\documentclass[%
reprint,
superscriptaddress,
groupedaddress,
nofootinbib,
amsmath,
amssymb,
aps,
prd,
floatfix,
]{revtex4-2}

\usepackage{graphicx}
\usepackage{dcolumn}
\usepackage{bm}
\usepackage{xcolor}
\usepackage{xspace}
\usepackage{tabularx}
\usepackage{ragged2e}
\usepackage{booktabs}
\usepackage{siunitx}
\usepackage[mathlines]{lineno}
\nolinenumbers
\usepackage{array}
\usepackage{caption}
\usepackage{subcaption}
\usepackage[nolist]{acronym}

\usepackage{hyperref}
\hypersetup{
    colorlinks,
    linkcolor={blue!50!black},
    citecolor={blue!50!black},
    urlcolor={blue!50!black}
}

\usepackage[capitalize]{cleveref}

\newcommand{\evidence}{\ensuremath{\mathcal{Z}}\xspace}
\newcommand{\likelihood}{\ensuremath{\mathcal{L}}\xspace}
\newcommand{\data}{\ensuremath{\vec{d}}\xspace}
\newcommand{\prior}{\ensuremath{\pi}\xspace}
\newcommand{\model}{\ensuremath{M}\xspace}
\newcommand{\parameters}{\ensuremath{\theta}\xspace}

\begin{document}

\title{RNLE: Residual neural likelihood estimation and its application to gravitational-wave astronomy}

\author{Mattia Emma and Gregory Ashton}
\email{mattia.emma@ligo.org}
\affiliation{%
Department of Physics, Royal Holloway University of London, Egham, TW20 0EX
}%

\date{\today}

\begin{abstract}
Simulation--based inference provides a powerful framework for Bayesian inference when the likelihood is analytically intractable or computationally prohibitive. By leveraging machine--learning techniques and neural density estimators, it enables flexible likelihood or posterior modeling directly from simulations.
We introduce Residual Neural Likelihood Estimation (RNLE), a modification of Neural Likelihood Estimation (NLE) that learns the likelihood of non-Gaussian noise in gravitational-wave detector data. Exploiting the additive structure of the signal and noise generation processes, RNLE directly models the noise distribution, substantially reducing the number of simulations required for accurate parameter estimation and improving robustness to realistic noise artifacts.
The performance of RNLE is demonstrated using a toy model, simulated gravitational-wave signals, and real detector noise from ground based interferometers. Even in the presence of loud non-Gaussian transients, glitches, we show that RNLE can achieve reliable parameter recovery when trained on appropriately constructed datasets. We further assess the stability of the method by quantifying the variability introduced by retraining the conditional density estimator on statistically identical datasets with different optimization seeds, referred to as training noise. This variability can be mitigated through an ensemble approach that combines multiple RNLE models using evidence-based weighting.
An implementation of RNLE is publicly available in the \texttt{sbilby} package, enabling its deployment within gravitational–wave astronomy and a broad range of scientific applications requiring flexible, simulation–based likelihood estimation.
\end{abstract}

\maketitle

\protected\def\protectedacused{\acused}

\begin{acronym}
\acro{LIGO}[LIGO]{Laser Interferometer Gravitational-Wave Observatory}
\acro{LHO}[LHO]{\ac{LIGO} Hanford observatory}
\acro{LLO}[LLO]{\ac{LIGO} Livingston observatory}
\acro{KAGRA}[KAGRA]{KAGRA}\acused{KAGRA}
\acro{iKAGRA}[iKAGRA]{initial-phase \ac{KAGRA}}
\acro{bKAGRA}[bKAGRA]{baseline-design \ac{KAGRA}}
\acro{GEO}[GEO]{GEO\,600 \ac{GW} detector}
\acro{aLIGO}{Advanced \ac{LIGO}}
\acro{A+}{Advanced+ \ac{LIGO}}
\acro{Asharp}[\ensuremath{\text{A}^\sharp}]{\ac{LIGO} \acs{Asharp}}
\acro{AdV}{Advanced \acl{Virgo}}
\acro{AdV+}{Advanced \acl{Virgo}+}
\acro{Virgo}{Virgo}\acused{Virgo}
\acro{VirgoNEXT}[Virgo\_nEXT]{Virgo\_nEXT}\acused{VirgoNEXT}

\acro{LSC}[LSC]{\acs{LIGO} Scientific Collaboration}
\acro{LV}[LV]{\acs{LIGO}--\acs{Virgo} Collaboration\protect\protectedacused{LVC}}
\acro{LVC}[LV]{\acs{LIGO}--\acs{Virgo} Collaboration\protect\protectedacused{LV}}
\acro{LVK}[LVK]{\acs{LIGO}--Virgo--KAGRA}
\acro{IGWN}[IGWN]{International \ac{GWH} Observatory Network}

\acro{O1}[O1]{first observing run}
\acro{O2}[O2]{second observing run}
\acro{O3}[O3]{third observing run}
\acro{O3a}[O3a]{first half of the third observing run}
\acro{O3b}[O3b]{second half of the third observing run}
\acro{O3GK}[O3GK]{observing run}
\acro{O4}[O4]{fourth observing run}
\acro{O4a}[O4a]{first part of the fourth observing run}
\acro{O4b}[O4b]{second part of the fourth observing run}
\acro{O4c}[O4c]{third part of the fourth observing run}
\acro{O5}[O5]{fifth observing run}

\acro{BH}[BH]{black hole}
\acro{BBH}[BBH]{binary black hole}
\acro{BNS}[BNS]{binary neutron star}
\acro{IMBH}[IMBH]{intermediate-mass black hole}
\acro{NS}[NS]{neutron star}
\acro{BHNS}[BHNS]{black hole--neutron star binaries}
\acro{NSBH}[NSBH]{neutron star--black hole binary}
\acro{PBH}[PBH]{primordial \ac{BH}}
\acro{CBC}[CBC]{compact binary coalescence}


\acro{IFO}[IFO]{interferometer}
\acro{FAR}[FAR]{false alarm rate}
\acro{IFAR}[IFAR]{inverse false alarm rate}
\acro{FAP}[FAP]{false alarm probability}
\acro{GR}[GR]{general relativity}
\acro{NR}[NR]{numerical relativity}
\acro{PN}[PN]{post-Newtonian}
\acro{EOB}[EOB]{effective-one-body}
\acro{ROM}[ROM]{reduced-order model}
\acro{IMR}[IMR]{inspiral--merger--ringdown}
\acro{PDF}[PDF]{probability density function}
\acro{PE}[PE]{parameter estimation}
\acro{CL}[CL]{credible level}
\acro{EOS}[EoS]{equation of state}
\acro{KLD}[KLD]{Kullback--Leibler divergence}
\acro{JSD}[JSD]{Jensen--Shannon divergence}
\acro{GCN}[GCN]{general coordinates network}
\acro{GWTC}[GWTC]{Gravitational-Wave Transient Catalog}
\acro{GWOSC}[GWOSC]{Gravitational Wave Open Science Center}

\acro{CWB}[cWB]{coherent WaveBurst}
\acro{LAL}[LAL]{\ac{LIGO} algorithm library}

\acro{CHRoCC}{central heating radius of curvature correction}
\acro{NonSENS}{non-stationary estimation and noise subtraction}

\acro{PTA}{Pulsar Timing Array}

\acro{MCMC}{Markov chain Monte Carlo}
\acro{ESS}{effective sample size}

\acro{RS}{rejection sampling}
\acro{IS}{importance sampling}
\acro{PSIS}{Pareto-smoothed \ac{IS}}
\acro{PSRS}{Pareto-smoothed \ac{RS}}

\acro{PP}{probability probability}
\acro{ASD}{amplitude spectral density}
\acro{IID}{independent and identically distributed}
\acro{KDE}{kernel density estimate}

\acro{ARNN}{autoregressive neural network}
\acro{PSD}{Power Spectral Density}
\acro{SBI}{Simulation Based Inference}
\acro{NLE}{Neural Likelihood Estimation}
\acro{ML}{Machine Learning}
\acro{SNR}{signal-to-noise ratio}
\acro{GW}{gravitational wave}
\acro{RNLE}{Residual Neural Likelihood Estimation}
\acro{NPE}{Neural Posterior Estimation}
\acro{NRE}{Neural Ratio Estimation}
\acro{MAF}{Masked Autoregressive Flow}
\acro{JS}{Jensen-Shannon}
\acro{UTC}{Universal Coordinated Time}

\end{acronym}
 
\acresetall
\section{Introduction\label{sec:introduction}}
The first direct detection of a \ac{GW} signal by the \ac{LIGO} occurred on the 14th September 2015~\cite{LIGOScientific:2016aoc}. Since then, the LIGO-Virgo-KAGRA (LVK) Collaboration~\cite{LIGOScientific:2014pky, VIRGO:2014yos, KAGRA:2020tym} has reported the observation of hundreds of additional \ac{CBC} events~\cite{ LIGOScientific:2025slb}. The analysis of these \ac{GW} signals has enabled unprecedented tests of Einstein’s theory of general relativity~\citep{LIGOScientific:2016lio, LIGOScientific:2019fpa, LIGOScientific:2020tif, LIGOScientific:2021sio}, insights into the evolutionary history of the Universe~\citep{LIGOScientific:2017adf,GWTC-4-cosmology}, and the nature of compact objects and matter at extreme densities~\citep{LIGOScientific:2018hze, Margalit:2017dij, LIGOScientific:2018cki, Nair:2019iur}. With the increasing number of events, it will become possible to infer the properties of the cosmological population of black holes and the formation pathways of compact binary systems with increasing precision~\cite{LIGOScientific:2025pvj}. To robustly carry out these studies, it is crucial to accurately infer the parameters of the \ac{GW} sources~\cite{Christensen:2022bxb}. 

The first algorithms for parameter estimation were developed in the late 1980s and employed maximum likelihood estimation techniques~\cite{Echeverria:1989hg}. With the application of the Metropolis-Hastings algorithm~\cite{Metropolis:1953am, Hastings:1970aa} to the calculation of posterior probabilities and the increased performance of hardware and computing power, in the mid 1990's it became feasible to use Bayesian inference in the multidimensional parameter space of \acp{GW}. The first to apply Markov chain Monte Carlo (MCMC) to \acp{GW} from compact binary mergers were~\citet{Christensen:1998gf}. The strength of the Bayesian approach lies in its ability to incorporate our theoretical knowledge of the \ac{GW} waveform model to obtain both a point estimate and a credible interval for each inferred parameter. MCMC algorithms were further developed by the LVK collaboration, leading to the \texttt{LALInference} adaptive MCMC routine~\cite{Veitch:2014wba}. This made it possible to estimate all 15 parameters describing the first \ac{GW} signal from a spinning binary black hole merger~\cite{LIGOScientific:2016vlm, Meyer:2020ijd}. Skilling’s work~\cite{Skilling:2004pqw, Skilling:2006gxv} independently introduced nested sampling as a general Bayesian inference algorithm. This method was adapted to gravitational-wave parameter estimation~\cite{Veitch:2014wba}, where posterior samples are obtained as a by-product of evidence evaluation. This method and further refined MCMC techniques were merged into the \texttt{Bilby} software~\cite{Ashton:2018jfp, Romero-Shaw:2020owr}. This is now the main tool for the production of parameter estimation results in the LVK oberving runs~\cite{KAGRA:2020tym, LIGOScientific:2025yae}. An alternative algorithm, \texttt{RIFT}~\cite{Lange:2018pyp, Wysocki:2019grj, Wofford:2022ykb}, uses Gaussian process interpolation with a highly-parallelizable grid-based parameter estimation approach~\cite{Pankow:2015cra}, and has also been successfully employed for parameter estimation in the \ac{O3}~\cite{LIGOScientific:2020ibl, KAGRA:2021vkt}. \\
While Bayesian parameter estimation algorithms have been instrumental in extracting the physical properties of the first hundreds of detected \acp{GW} events and continue to be actively refined~\cite{Williams:2023ppp, Prathaban:2024rmu, Pathak:2025bdi, Yallup:2025sty, Williams:2025szm}, their accuracy is limited by a number of underlying assumptions. In frequency-domain analyses, one such limitation arises from the use of the Whittle likelihood~\cite{Whittle:1951, ContrerasCristan:2006whittle, Thrane:2018qnx, Rao:2020reconciling}, which approximates the exact likelihood by assuming that the detector strain noise is stationary and Gaussian. These assumptions are violated in the presence of transient noise artifacts, commonly referred to as glitches~\cite{LIGOScientific:2019hgc, LIGO:2021ppb, LIGO:2024kkz}. Due to their frequent occurrence and potentially high \ac{SNR}, glitches have been shown to contaminate nearly 20\% of detected binary black hole signals and essentially all binary neutron star signals~\cite{LIGOScientific:2018kdd, Powell:2018csz, Edy:2021par, Davis:2022ird}. More broadly, glitches represent only one manifestation of the non-stationary behavior of interferometric gravitational-wave detectors, which also arises from time-dependent variations in the noise spectrum and environmental disturbances. Such non-stationarities introduce correlations between Fourier components and temporal variations in the noise power spectral density, leading to biased posterior distributions, underestimated uncertainties, and inconsistencies across data segments~\cite{Kumar:2022tto}.  \\
The most prominent signal affected by a glitch is GW170817~\cite{LIGOScientific:2017vwq}, the first observed \ac{BNS} merger. The simultaneous detection of an electromagnetic counterpart made it stand out as the first, and to date only, confirmed multimessenger event~\cite{LIGOScientific:2017ync, LIGOScientific:2017zic, Troja:2017nqp}. The coincidence of GW170817 with an extremely loud glitch in the Livingston detector~\cite{Pankow:2018qpo} demonstrated the importance of developing methods to perform accurate and fast parameter estimation in the presence of transient noise artifacts. Different techniques have been developed to achieve this goal~\cite{LIGOScientific:2018kdd}. Some aim to subtract known and modeled non-transient noise directly from the strain data using knowledge provided by the interferometer's auxiliary channels, e.g., \texttt{gwsubtract}~\cite{Davis:2018yrz, Vajente:2019ycy, Vajente:2022dme}. Other algorithms aim to model and subtract glitches from the data before analyzing the signal with the Whittle likelihood~\cite{Merritt:2021xwh, Chowdhury:2024jdx} or to model glitch and signal simultaneously~\cite{Plunkett:2022zmx, Ashton:2022ztk, Malz:2025xdg}. Among these, the \texttt{BayesWave} algorithm is the primary method used by the \ac{LVK}~\cite{Cornish:2014kda}. \texttt{BayesWave} employs linear combinations of wavelets to model non-Gaussianities and also allows for the simultaneous modeling of glitches and signals~\cite{ Chatziioannou:2021ezd, Hourihane:2022doe, Ghonge:2023ksb}. It is used by the \ac{LVK} collaboration to model and subtract glitches from data containing \ac{GW} signals~\cite{Cornish:2020dwh}. However,~\citet{Udall:2025bts} showed that the glitch subtraction process is inherently limited, leaving behind residual glitch power with an \ac{SNR} of 3–7 as an unavoidable consequence. They further demonstrated that this residual power is enough to introduce biases in parameter estimation results. \\
Concurrently, different \ac{ML} algorithms~\cite{Cuoco:2020ogp} have also been developed for the classification of non-transient noise~\cite{Powell:2015tta, Razzano:2018fxb, Soni:2021cjy}, the de-glitching of gravitational wave data, and to distinguish between glitch and signal power~\cite{Biswas:2013wfa, Cavaglia:2018xjq, Cavaglia:2020qzp}. While these methods introduce significant improvements with respect to the Whittle likelihood and have been successfully applied in the analysis of gravitational wave events~\cite{Davis:2022ird, Payne:2022spz, Macas:2023wiw, Udall:2024ovp}, none of them can yet be reliably applied in every setting to fully unbias \ac{GW} parameter estimation results~\cite{Ray:2025rtt}. 
More recently, novel approaches have been proposed to directly address this limitation by reducing or bypassing assumptions about the noise realization. For instance,~\citet{Chatterjee:2024obg} introduce a framework that reconstructs the gravitational wave signal independently of the underlying noise, while~\citet{Legin:2024gid} presents a score-based diffusion model that learns the non-Gaussian noise distribution directly from detector data, enabling accurate parameter recovery even in the presence of loud glitches. These developments represent important steps toward robust inference in glitch-contaminated data. \\
\ac{SBI} has established itself as a powerful \ac{ML}-based framework to tackle complex inference problems that are computationally difficult for traditional Bayesian methods or for which an analytical expression of the likelihood function is unavailable~\cite{Tavare:1997, Pritchard:1999, Chen:2023ILF, Deistler:2025sbiGuide}.
The central idea of \ac{SBI} is to exploit a forward model capable of generating synthetic data given some model parameters. 
These simulations are then used to train a conditional density estimator that approximates either the likelihood, the posterior, or a related quantity, thereby enabling efficient Bayesian inference without requiring an explicit likelihood function. 
Modern \ac{SBI} methods are typically classified into three main categories, depending on the quantity that is learned: \ac{NLE}~\cite{Papamakarios:2018zoy, Durkan:2018SNLE, Papamakarios:2019xvg, Dirmeier:2025SSNLE}, \ac{NPE}~\cite{Papamakarios:2016ctj, Greenberg:2019APT}, and \ac{NRE}~\cite{Miller:2021hys, Miller:2022shs}. 
In \ac{NLE}, a neural network is trained to approximate the likelihood function, and inference is performed via standard Bayesian techniques using the learned likelihood. 
In contrast, \ac{NPE} directly models the posterior distribution, and so samples can be directly drawn. 
Finally, \ac{NRE} estimates the likelihood-to-evidence ratio, which can be used to construct the posterior via importance weighting or rejection sampling. 
Each approach offers distinct advantages in terms of training stability, computational efficiency, and flexibility, and the choice between them depends on the structure of the simulator and the inference task at hand.
In the past decade, \ac{SBI} has found wider application in \ac{GW} astronomy and astrophysics~\cite{Cranmer:2019eaq, Gabbard:2019rde, Crisostomi:2023tle, Wang:2023sbi, AnauMontel:2024bos, Liang:2025yjw}. It has been employed for the rapid generation of sky-maps for public alerts~\cite{Marx:2025ioo}, and algorithms have been developed for the current~\cite{Graff:2011gv, Negri:2025cyc} and next generation of \ac{GW} detectors~\cite{Bhardwaj:2023xph, Alvey:2023naa, Vilchez:2024qnw}. A specific application of \ac{SBI} to gravitational-wave parameter estimation is the \texttt{DINGO} software package~\cite{Dax:2021tsq, Dax:2022pxd, Dax:2024mcn}. \texttt{DINGO} employs real detector data with simulated gravitational-wave signal injections to train a neural posterior estimation (\ac{NPE}) model, yielding an approximate posterior distribution. This approximation is subsequently refined through rejection sampling~\cite{Thrane:2018qnx, Ashton:2025xba}, and has been shown to reproduce the results obtained with standard Bayesian inference pipelines such as \texttt{Bilby}~\cite{Dax:2022pxd}. However, because the rejection sampling step relies on a Gaussian likelihood approximation, the resulting inference inherits the associated assumptions of stationarity and Gaussianity, and may therefore be biased in the presence of non-Gaussian noise transients.
 \\
In this work, we employ a modification of \ac{NLE} to address the challenges posed by non-Gaussian noise in gravitational-wave data analysis. We adopt \ac{NLE} rather than \ac{NPE} for two key reasons. First, \ac{NLE} has been shown to exhibit higher simulation efficiency, enabling it to learn more complex distributions with a fixed number of simulations~\cite{SpurioMancini:2022vcy}. Second, unlike \ac{NPE}, \ac{NLE} allows the conditional density estimator to be trained solely on detector noise while still yielding posterior distributions for the signal parameters, a property that is essential for our approach.
We introduce \ac{RNLE}, which combines the \ac{NLE} algorithm implemented in the \texttt{sbi} package~\cite{TejeroCantero:2020sbi, Papamakarios:2018zoy} with the additive structure of gravitational-wave signal and noise, with the goal of learning a likelihood for real detector data that captures non-stationarities and transient non-Gaussianities. The likelihood is obtained by training the conditional density estimator exclusively on noise realizations. For each likelihood evaluation, a waveform approximant realization is subtracted from the observed data, and the neural likelihood is evaluated on the resulting residuals.
This strategy is conceptually related to the approach introduced in~\citet{Legin:2024gid}, which also exploits the additivity of signal and noise, and learns the noise distribution directly from detector data to produce an unbiased estimate of the likelihood function. In contrast to~\citeauthor{Legin:2024gid}, which employ score-based diffusion models and learns the score of the noise distribution, our method uses neural likelihood estimation within the \texttt{sbi} framework and learns the likelihood directly from the noise distribution. We can therefore integrate the learned likelihood into the \texttt{Bilby} framework, and perform Bayesian parameter estimation using the nested sampling implementation of \texttt{dynesty}~\cite{Speagle:2019ivv}. \\
The paper is structured as follows. In Section~\ref{sec:methodology}, we give a summary of Bayesian parameter estimation and discuss the theoretical validity of the \ac{RNLE} approach. In Section~\ref{subsec:setup_RNLE}, we explain the setup of the \ac{RNLE} algorithm and the specifics of the employed estimator. In Section~\ref{sec:toy_model} we compare results obtained with \ac{NLE}, \ac{RNLE} and the Whittle likelihood for a simple toy model. In Section~\ref{sec:gw_setup} and~\ref{sec:simulated_GW}, we benchmark \ac{RNLE} against the Whittle likelihood using simulated detector noise for the training and a signal from a \ac{BBH} merger injected into simulated noise as an observation. After demonstrating the validity of our algorithm, in Section~\ref{sec:real_GW_quasi}, we analyze \ac{BBH} merger signals injected in quasi-Gaussian detector noise. In Section~\ref{sec:glitches_extreme}, we analyze \ac{BBH} injections in the proximity of very loud glitches to study the capability of our algorithm to recover accurate posteriors when the Whittle likelihood is biased. To follow up on this analysis, in Section~\ref{sec:blip_glitch} we analyze \ac{BBH} injections around a specific glitch with \ac{RNLE} likelihoods trained on varying datasets, and in Section~\ref{sec:training_noise} we assess the training noise by retraining on the same dataset with a different machine learning seed. We conclude in Section~\ref{sec:conclusion}, summarizing our results and outlining future development and usage possibilities for the \texttt{sbilby} software package~\cite{emma_sbilby_2026}.

\section{Methodology\label{sec:methodology}} 
Given observation data \data and a predictive model \model with associated parameters \parameters, Bayesian inference centres around constructing a likelihood $\likelihood(\data| \parameters, \model)$ and prior $\prior(\parameters | \model)$, then seeking to estimate the posterior probability density
\begin{equation}
    p(\theta | \data, \model) \propto \likelihood(\data | \parameters, \model)\prior(\theta| \model)\,,
\end{equation}
and the normalising evidence
\begin{equation}
    \evidence(\model | \data) = \int \, \likelihood(\data | \parameters, \model)\prior(\theta| \model) d\parameters\,.
\end{equation}
Calculation of the posterior and evidence can only be done in closed form for a limited set of problems (typically, those with so-called conjugate priors \cite{Raiffa:1961}). 
Therefore, computational Bayesian inference approaches have been developed\footnote{
Grid-based approaches can be applied in low-dimensional problems, but quickly become intractable for more than a few dimensions, therefore stochastic sampling methods such as MCMC and Nested Sampling are generally applied in higher-dimensional problems.}
enabling approximation with induced numerical error.

Computational approaches typically require an explicit construction of the likelihood in closed form.
This is suitable for many astrophysical data analysis problems where the likelihood takes a simple form following standard probability distributions.
Most often, this is the Gaussian distribution thanks to the central limit theorem~\cite{Cowan:1998ji}.
However, in some instances, the likelihood is intractable (i.e. difficult or impossible to write down in closed form) or computationally costly~\cite{Alsing:2019xrx, Jo:2022rln, Christensen:2022bxb, Lehman:2024vyl}.
Over the last few years, a new approach has been developed to address such problems: \ac{SBI}.
While there are multiple variants of \ac{SBI}, in this work we consider \ac{NLE} in which a neural density estimator $q_{\phi}$, with weights $\phi$, is trained to approximate $\likelihood(\data | \parameters)$ using training data consisting of $N$ pairs $\{\parameters_i, \data_i\}$.
To construct the training data, samples $\parameters_i$ are drawn from a generating distribution (generally the prior).
The samples are then passed into the simulator $s(\parameters_i)$ to produce a realisation of the $\data_i$.
Note that the simulator may, in general, also depend on a set of latent variables $z$ that are randomly sampled during the simulation.

Let us now restrict ourselves to the set of cases where the forward model described by the simulator consists of a deterministic model $m(\vartheta)$ and an additive noise process $n(\lambda)$ where we have defined two subsets $\vartheta$, the model parameters and $\lambda$, the parameters describing the noise process with $\theta=\vartheta \cup \lambda$.
That is, we consider only models where $s_i = m_i + n_i$ where $m_i = m(\vartheta)$ is deterministic while $n_i = n(\lambda)$ is a random sample drawn from a noise distribution.

Standard \ac{NLE} proceeds by taking as input the full simulator $s(\theta)$ and proceeding to approximate $\likelihood(\data | \theta)$.
However, a wide variety of astrophysical problems are well described by a deterministic model with additive noise. Under this assumption, it is appropriate to instead approximate the residual likelihood $\likelihood(\data - m(\vartheta) | \lambda)$ which is simpler due to the lower dimensionality.

To better understand the motivation let us consider cases where the likelihood is known in closed form. For additive white Gaussian noise with standard deviation $\sigma$, for a single observation $d_i$, the log-likelihood is given by 
\begin{equation}
    \log\likelihood(d_i | \theta) = 
    -\frac{1}{2}\left(
    \frac{\left(d_i - m(\vartheta)\right)^2}{\sigma^2} - \log(2\pi\sigma^2)\right)\,,
    \label{eq:Gaussian_likelihood}
\end{equation}
and if the set of data \data is independent and identically distributed (i.i.d.), then
\begin{equation}
    \log\likelihood(\data | \theta) = \sum_{i} \log\likelihood(d_i | \theta).
    \label{eq:multi_Gaussian_likelihood}
\end{equation}
However, given that the data only enters the likelihood through the residual $r_i=d_i - m(\vartheta)$, we can introduce the \emph{residual likelihood} as
\begin{equation}
    \log\likelihood(r_i | \lambda) = 
    -\frac{1}{2}\left(
    \frac{\left(r_i\right)^2}{\sigma^2} - \log(2\pi\sigma^2)\right)\,,
\end{equation}
which is identical to the standard likelihood function. For standard applications of this likelihood, it would be odd to differentiate between the two likelihoods, but if one where to apply \ac{NLE}, learning the residual likelihood is easier as the density estimate that must be learned is only that of the noise distribution and not the full joint distribution of the data and \parameters.

More related to \ac{GW} is the case of coloured Gaussian noise. For interferometric gravitational-wave data, in the absence of a signal (which we refer to as $\model_n$), the detector data is well-described by stationary coloured Gaussian noise.
Assuming we know a priori the \ac{PSD} that describes the noise, if $\data$ is a regular real-valued time-series with a sampling frequency $f_s$ and duration $T$, then the likelihood is known and can be approximated by the so-called Whittle likelihood:
\begin{equation}\label{eqn:loglGWN}
    \log\likelihood(\data| \model_n) \propto -\frac{1}{2}\langle \tilde{d}, \tilde{d} \rangle \,,
\end{equation}
where $\tilde{d}=\texttt{fft}(\data)/f_s$ is the scaled complex-valued Fourier-transformed data, the noise-weighted inner product is defined as
\begin{equation}
    \langle a, b \rangle := \frac{4}{T} \sum_{j} \mathcal{R}\left(\frac{a^*_j b_j}{P_j}\right)\,,
\end{equation}
and $P_j$ is the real-valued \ac{PSD}.
If a signal is present in the data, then the generate model $\model_s$ predicts an additive sum of the signal and noise such that
\begin{equation}\label{eqn:loglGWN_2}
    \log\likelihood(\data| \vartheta, \model_s) = -\frac{1}{2}\langle \tilde{d}-m(\vartheta), \tilde{d}-m(\vartheta) \rangle - \sum_j \log\left(2\pi P_j\right) \,.
\end{equation}
Again, we can rewrite this as a residual likelihood
\begin{equation}\label{eqn:loglGWN_3}
    \log\likelihood(\vec{r}~| \model_s) \propto -\frac{1}{2}\langle \tilde{r}, \tilde{r} \rangle - \sum_j \log\left(2\pi P_j\right) \,.
\end{equation}
In this instance, the noise has no parameters (i.e. $\lambda$ is the empty set). However, one could also consider parameterising the \ac{PSD} $P_j\rightarrow P_j(\lambda)$ in which case
\begin{equation}\label{eqn:loglGWN_4}
    \log\likelihood(\vec{r}~| \lambda, \model_s) =
    -\frac{1}{2}\langle \tilde{r}, \tilde{r} \rangle_{\lambda}
    - \sum_i\log\left(2\pi P_j(\lambda)\right)
    \,,
\end{equation}
where we add the subscript $\lambda$ to the noise-weighted inner product to denote that these parameters are required by the \ac{PSD}. \\
This can straightforwardly be generalized for a multi-detector analysis. Since the detector noise is assumed to be independent, the joint likelihood is obtained by combining the individual detector likelihoods multiplicatively, which corresponds to summing their logarithms. Thus, for detectors $d \in \{1, \dots, D\}$, the total log-likelihood is  
\begin{equation}
\log \mathcal{L}_{\mathrm{tot}}(\vec{r}~| \lambda, \model_s) = \sum_{d=1}^{D} \log \mathcal{L}_{d}(\lambda),
\label{eq:multidet_like}
\end{equation}
where $\mathcal{L}_{d}(\vec{r}~| \lambda, \model_s)$ denotes the RNLE-learned likelihood for detector $d$ evaluated at parameters $\lambda$.

For these cases, where the likelihood is known in closed form, \ac{NLE} is not required. However, our hypothesis is that when the assumptions underlying the likelihood construction are no longer valid (e.g. the noise is non-Gaussian, non-stationary, or contains transient noise artefacts), provided the noise and signal remain additive, the residual likelihood is easier to approximate with \ac{NLE} than the full likelihood. 
\subsection{Simulation Based Inference setup \label{subsec:setup_RNLE}}
We employ the \texttt{sbi} package's implementation of \ac{NLE} to construct the \ac{RNLE} algorithm, which we implement within the newly developed \texttt{sbilby} framework~\cite{emma_sbilby_2026}. Following the approach described in Section~\ref{sec:methodology}, the package comprises two main components for the simulation and analysis of \ac{GW} data. 

The first component is specific to gravitational-wave parameter estimation and provides functionality to simulate Gaussian noise realizations from prescribed power spectral densities, as well as to access real detector noise. In this work, we always use pieces of data of 4 s, and the analysis is performed in the time domain. The pieces of data are whitened in the frequency domain and subsequently cut in the time domain to obtain training data for the conditional density estimators (see Section~\ref{sec:gw_setup} for the details). This training data, consisting of pairs of noise parameters and data $\{ \lambda_i, \data_i \}$, is passed to the \texttt{sbi} package to train the conditional density estimator through the second component of our algorithm. This second component is independent of the particular inference problem and can be employed in any application for which separate generative models of the noise and signal processes are implemented. Throughout this work, we employ the \ac{MAF} as the conditional density estimator~\cite{Papamakarios:2017tec}. 

A \ac{MAF} models the joint probability density of the parameters by decomposing it into a product of one-dimensional conditional densities, each parameterized by an \ac{ARNN}~\cite{Uria:2016, Rezende:2015}. The \acp{ARNN} provide a flexible and tractable parameterization of conditional probability distributions. In \acp{ARNN}, the joint probability of a multivariate variable $\mathbf{x} = (x_1, \ldots, x_D)$ is factorized as
\begin{equation}
p(\mathbf{x}) = \prod_{i=1}^{D} p(x_i \,|\, x_{1:i-1}),
\end{equation}
where each conditional distribution $p(x_i \,|\, x_{1:i-1})$ is represented by a neural network whose inputs depend only on the preceding components $x_{1:i-1}$. This autoregressive structure enables the model to efficiently capture statistical dependencies among parameters while maintaining exact likelihood evaluation. In our framework, the ARNN sequentially transforms samples from a base distribution into complex likelihood surfaces $\likelihood(\data - m(\vartheta) \,|\, \lambda, \model_s)$. To ensure reproducibility, we employ the default architecture provided in \texttt{sbi} v0.22.0. In this configuration, the likelihood $p(x \mid \theta)$ is modeled using a stack of five autoregressive flow transforms, each parameterized by a masked autoencoder for distribution estimation network~\cite{Germain:2015yft} with two hidden layers of 50 units and rectified linear unit activation functions~\cite{Glorot:2011relu}. 

The resulting conditional density estimator integrates directly with the \texttt{Bilby} framework, as the NLE implementation is designed to be fully compatible with \texttt{Bilby}’s inference interface. For sampling, we employ \texttt{dynesty}. At each likelihood evaluation, we subtract a realization of the model $m(\vartheta)$, with $\vartheta$ drawn from the prior, from the analyzed data, and evaluate the likelihood on the resulting residuals $r(\theta) = \tilde{d}(\theta) - m(\vartheta)$. Although the likelihood directly depends only on the residual parameters $\lambda$, we obtain samples for the model parameters as a natural byproduct of this subtraction process. The modular design of our algorithm allows the user to specify custom signal and noise generation functions, ensuring its applicability to a broad class of inference problems beyond gravitational-wave analyses.
 \\

\section{Results: toy model \label{sec:toy_model}}
To test the validity of \ac{RNLE}, we begin with the analysis of a simple timeseries problem consisting of a sine-Gaussian toy model following the equation:
\begin{equation}
    \mathcal{F}(t,\alpha,f)= A \exp(-(t/\alpha)^2)\times \sin(2 \pi f t),
    \label{eq:sine_gaussian}
\end{equation}
where $t$ is the time, and $f$, $A$, and $\alpha$ are free parameters that we will refer to as frequency, amplitude, and standard deviation. We simulate observational data by adding white Gaussian noise sampled from a normal Gaussian distribution with standard deviation $\sigma$ to a realization of Eq.~\ref{eq:sine_gaussian} for specific values of the free parameters. The value of $\sigma$ is drawn from a uniform prior $\mathcal{U}[0,2]$. We aim to analyze this observation using likelihoods generated with both the canonical \ac{NLE} algorithm and our newly implemented \ac{RNLE} algorithm, and compare the results with those obtained using the correct Gaussian likelihood, following Eq.s~\ref{eq:Gaussian_likelihood} and \ref{eq:multi_Gaussian_likelihood}. We employ the \texttt{dynesty} nested sampling package implemented in \texttt{Bilby} for the posterior sampling. For the training of the \ac{NLE} conditional density estimator, we employ simulations of white Gaussian noise added to simulations of the signal, employing the priors in Table~\ref{tab:priors_toy_model} in the appendix. To train \ac{RNLE}, we employ simulations of white Gaussian noise only, varying the $\sigma$ parameter.

We perform benchmarking against the Gaussian likelihood first, using a restricted parameter space where $f$ and $A$ are fixed, and then sampling in the full three-dimensional signal space. To assess how well the conditional density estimator recovers the true likelihood, we compute the \ac{JS} divergence between the posterior probability density functions obtained with the (R)NLE and the Gaussian likelihood~\citep{Lin:1991zzm}. For each comparison, we draw 3000 samples from both posteriors, fit a kernel density estimator to each sample set, and evaluate the corresponding probability densities. The \ac{JS} divergence is computed using the \texttt{scipy} implementation~\citep{Virtanen:2019joe}, which returns the JS distance measured in bits, i.e., using a base-2 logarithm so that $\max(\mathrm{JS})=1$. For our analysis, we use the corresponding \ac{JS} divergence, obtained by squaring the returned distance. This procedure is repeated one hundred times, and the reported values are averages over these trials. We consider two posteriors indistinguishable if their \ac{JS} divergence is smaller than $10/N$, where $N$ is the number of samples used for the evaluation, e.g., with $N=3000$ the threshold is $0.003$~\cite{Romero-Shaw:2020owr}. \par
\begin{figure*}[t]
    \centering
    \includegraphics[width=0.47\linewidth]{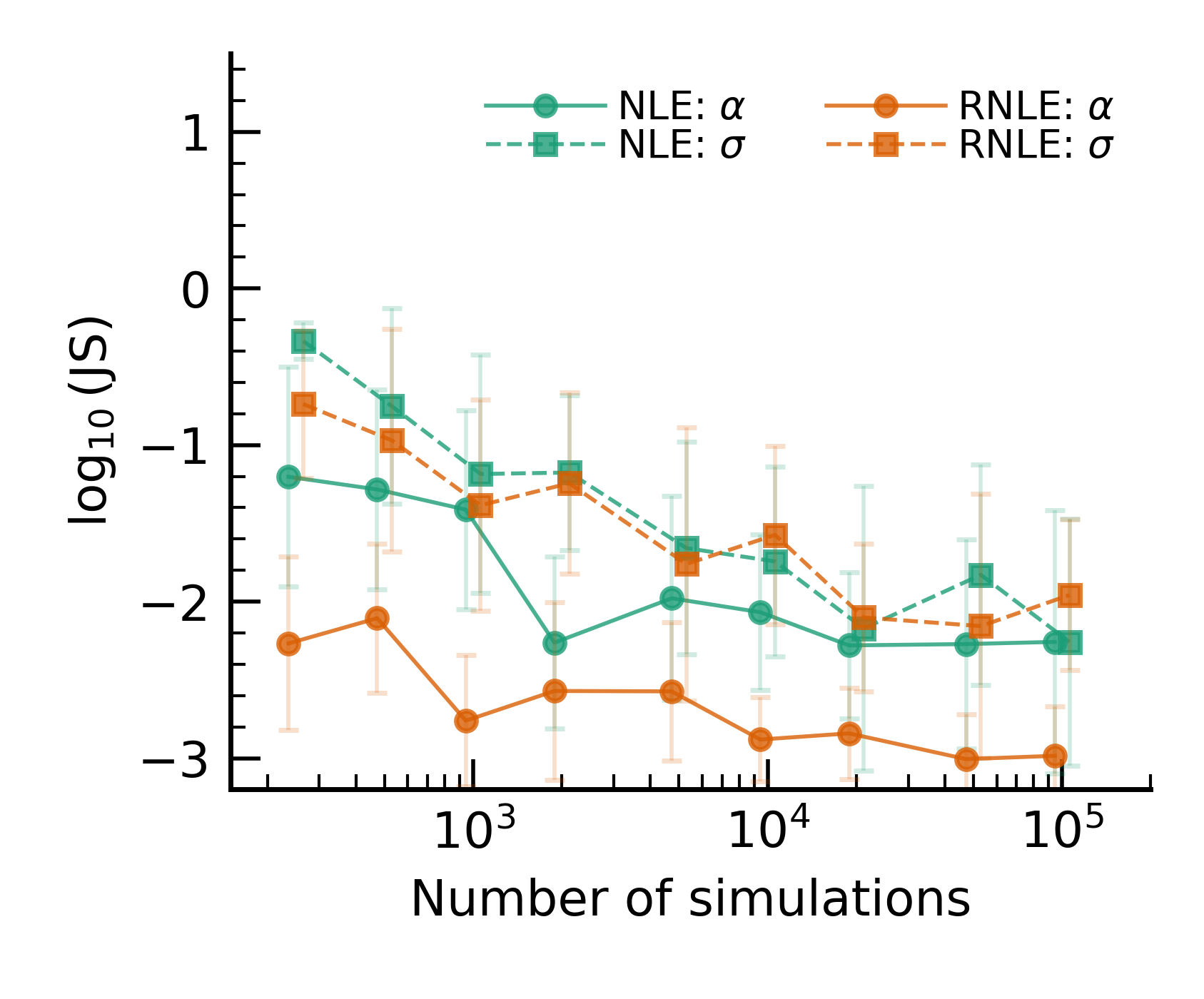}
    \includegraphics[width=0.47\linewidth]{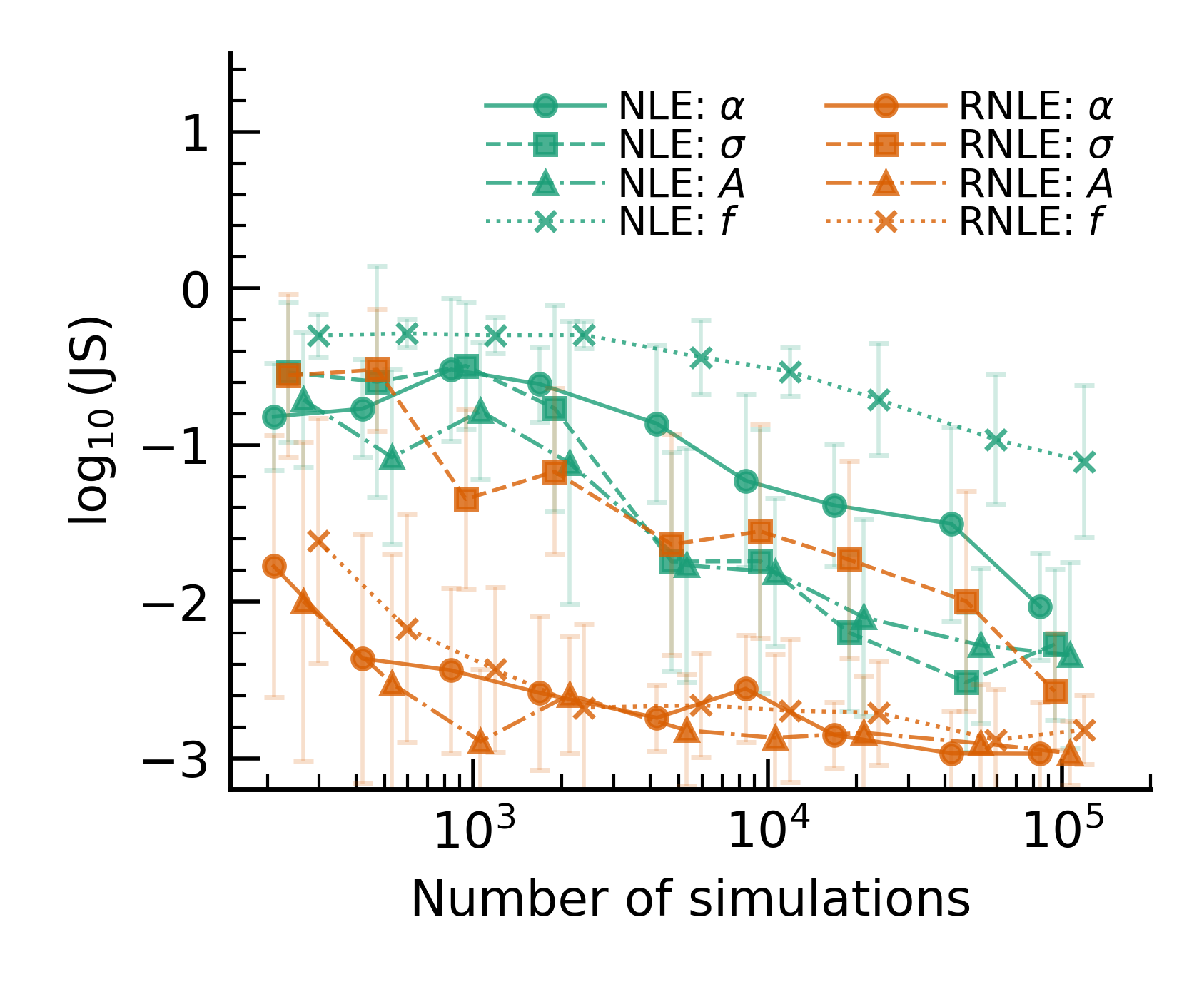}
    
    \caption{Logarithm of the Jensen Shannon divergence value for the posterior distributions obtained with the likelihoods trained with \ac{NLE}, in green, and \ac{RNLE}, in orange. The squares stand for the noise related parameter $\sigma$, the circles, triangles, and crosses for the variance, frequency, and amplitude of the sine-Gaussian signal, respectively. The 2-dimensional results are shown in the left panel, while the right panel shows the 4-dimensional results.}
    \label{fig:benchmark_gaussian}
\end{figure*}

The results of the benchmarking are shown in Figure~\ref{fig:benchmark_gaussian}, for two dimensions in the left panel and four dimensions in the right panel. The plots show the variation of the \ac{JS} divergence value with the number of simulations used in the training for \ac{NLE} and \ac{RNLE} against the Gaussian likelihood in green and orange, respectively. 
From the plots, we see that the \ac{JS} values for the signal parameter posteriors recovered with \ac{RNLE} are systematically lower than the \ac{NLE} ones. The same is not true for the noise parameter $\sigma$, for which the values are consistent in the two approaches. The better performance in the \ac{RNLE} case is related to the lower dimensionality of the training parameter space, one for \ac{RNLE} and two (four in the right panel) for \ac{NLE}. We would expect the \ac{JS} values for the \ac{NLE} results to eventually converge to the \ac{RNLE} ones when further increasing the number of simulations. Those results show that \ac{RNLE} is a viable method to produce posterior probabilities and requires fewer simulations to achieve the same accuracy as \ac{NLE}. The results also confirm the statistical equivalence of the posterior distributions obtained with the Gaussian likelihood and the one obtained with the RNLE likelihood.\\

\begin{figure*}[t]
    \centering
    \includegraphics[width=0.45\linewidth]{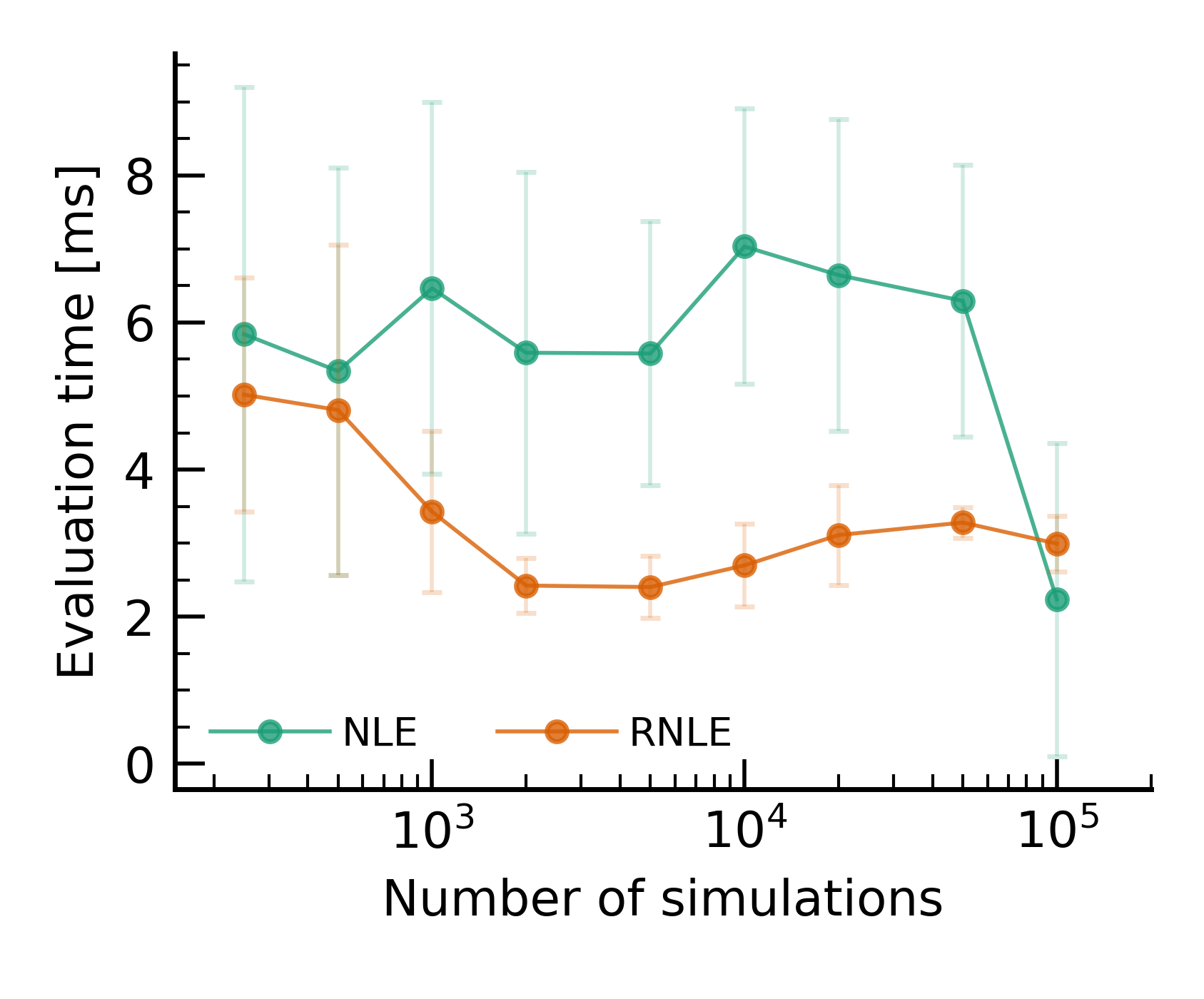}
    \includegraphics[width=0.45\linewidth]{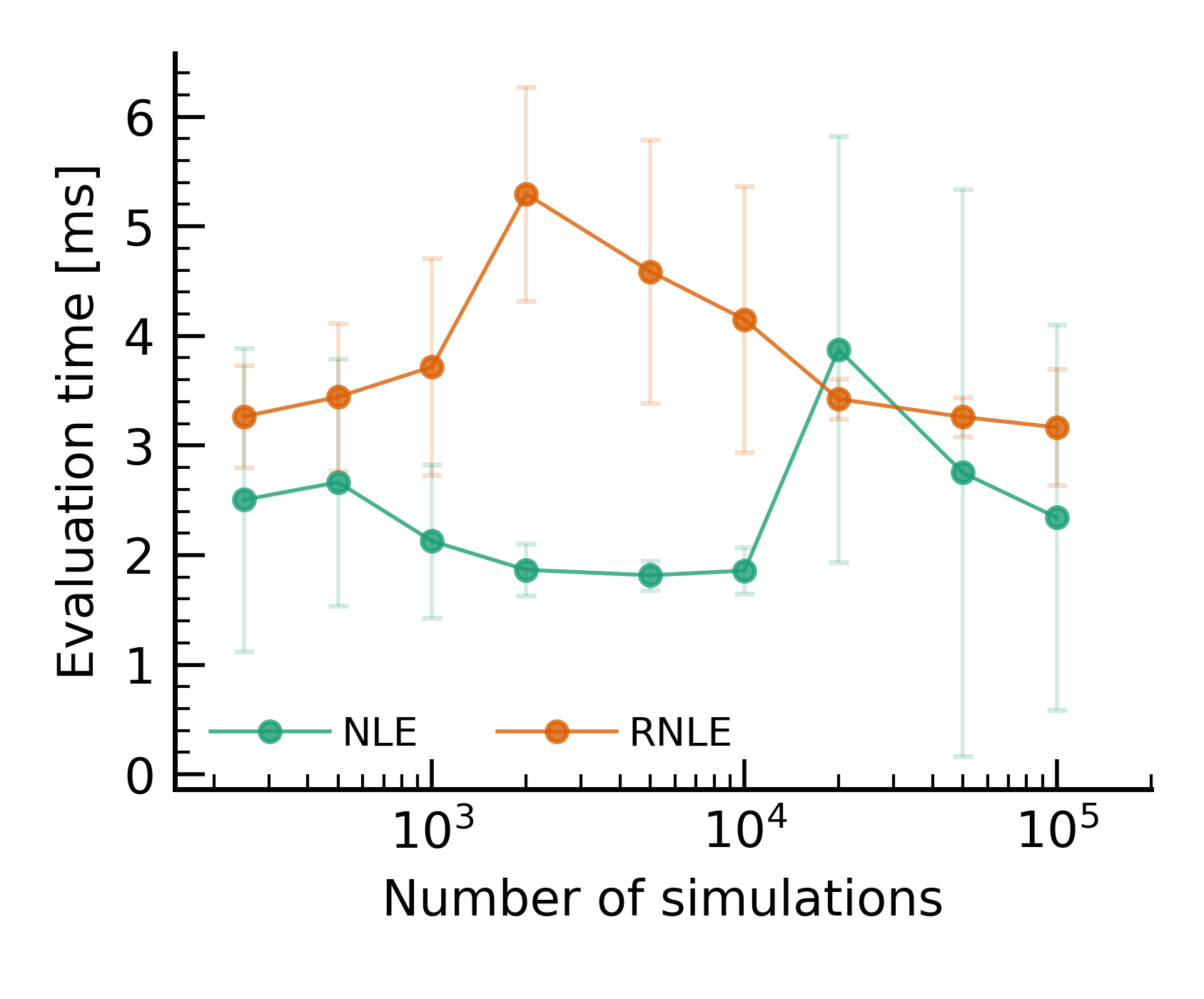}
    
    \caption{Likelihood evaluation time in milliseconds for \ac{NLE}, green, and \ac{RNLE}, orange, versus the number of simulations used in the training. We employ two free parameters in the left panel and four in the right one.}
    \label{fig:benchmark_gaussian_2}
\end{figure*}

In Figure~\ref{fig:benchmark_gaussian_2}, we show the time required for a single likelihood evaluation for the \ac{NLE}, in green, and the \ac{RNLE}, in orange, trained likelihoods. Similarly to Figure~\ref{fig:benchmark_gaussian}, for the left panel we employed two free parameters, and for the right panel we employed four. We observe that the computational cost of \ac{RNLE} likelihood evaluations is comparable to that of standard \ac{NLE}. While \ac{RNLE} involves an additional operation—namely the explicit subtraction of a signal realization from the data prior to likelihood evaluation—this overhead does not translate into a systematic increase in total runtime for the toy model considered here, where the forward model evaluation is computationally inexpensive. For gravitational-wave analyses, however, the model evaluation, i.e., the waveform generation, typically constitutes the dominant contribution to the computational cost in standard Bayesian sampling. In this context, \ac{RNLE} retains the cost of waveform evaluation and introduces an additional overhead associated with the neural likelihood evaluation, making it more expensive than regular sampling. In contrast, once trained, \ac{NLE}-based inference does not require waveform evaluations during sampling and therefore can be substantially faster. In practice, for our toy-model the overall computational performance depends on several factors, including the hardware architecture, the total number of likelihood evaluations performed by the sampler, the effective sampling efficiency, and the average evaluation time per likelihood call. When accounting for these effects, we find that the total wall time of \ac{RNLE}-based analyses is broadly similar to that of \ac{NLE}, with run-to-run fluctuations that are comparable in magnitude for the two methods. As a result, the implementation of \ac{RNLE} does not incur a significant computational penalty.

\section{Configuring RNLE for efficient GW analysis \label{sec:gw_setup}}

\begin{figure*}[t]
    \centering
    \includegraphics[width=\linewidth]{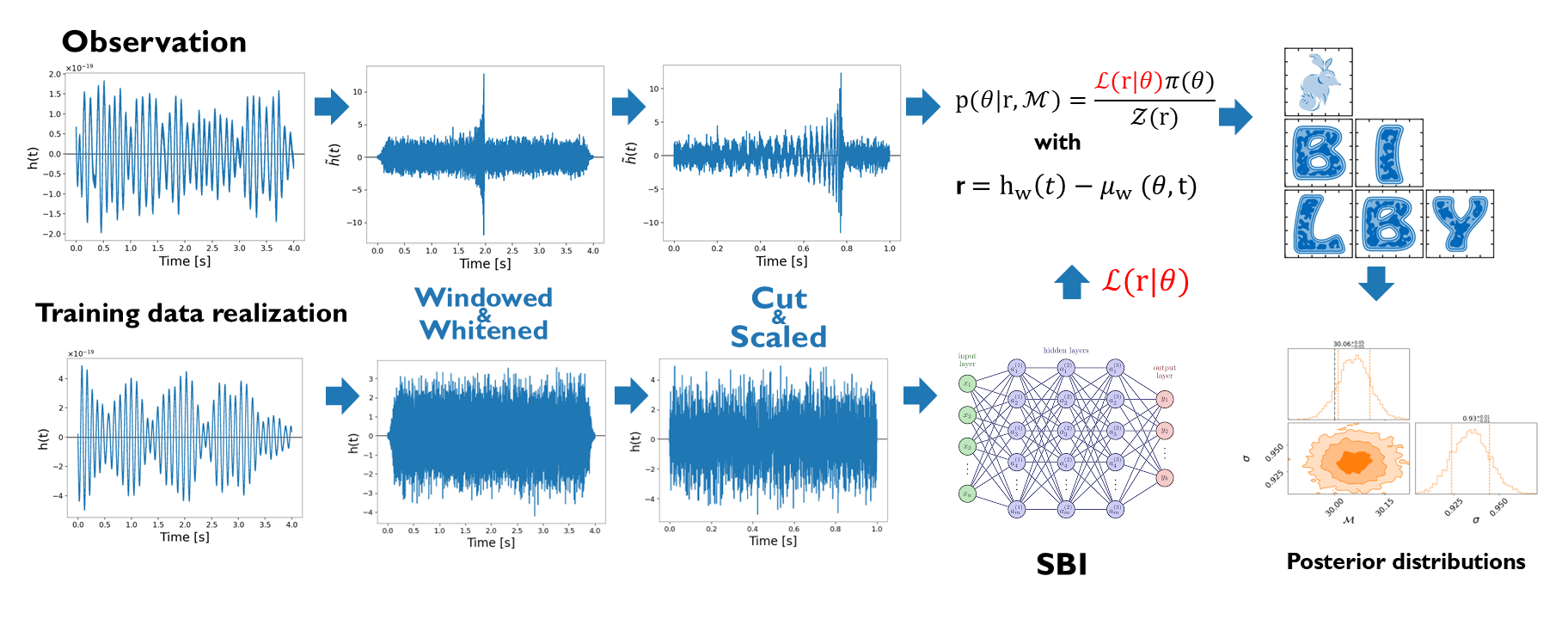}
    \caption{Schematic overview of the workflow we employ to produce posterior distributions from a set of training data realizations and an observation using RNLE. The pre-processing for the observations, top row, and the training data, bottom row, is the same.  We input the training data into the \ac{SBI} conditional density estimator, which trains a likelihood. This is used in \texttt{sbilby} with the \texttt{dynesty} sampler to perform inference and obtain posterior distributions. At every evaluation of the likelihood, before passing the data to the \ac{RNLE} likelihood object we subtract a whitened \ac{GW} signal realization from it.}
    \label{fig:schematic_processing}
\end{figure*}

To apply the \ac{RNLE} algorithm to the analysis of real gravitational wave events, we start by testing it on synthetic data. We create an observation by combining coloured-Gaussian noise, generated from a fixed PSD, and an injected \ac{GW} waveform produced with the \texttt{IMRPhenomPv2} waveform approximant~\cite{Hannam:2013oca, Santamaria:2010yb, Husa:2015iqa, Khan:2015jqa}. The injection values for the signal parameters are listed in Table~\ref{tab:simulation}. We simulate the Gaussian noise to ensure that the assumptions for the validity of the Whittle likelihood approximation are satisfied, using the publicly available high-sensitivity \ac{PSD} curve presented in~\citet{KAGRA:2013rdx} with a horizon distance of 180 Mpc. To effectively compare the results between our \ac{RNLE} implementation and \texttt{Bilby}, we ensure that we train the conditional density estimator on data processed as done internally by the \texttt{Bilby} algorithm~\cite{Talbot:2025vth}. In practice, this means that we ensure that our Gaussian noise simulations are properly windowed. Different from a standard \texttt{Bilby} analysis, we are operating in the time domain and therefore need to whiten the data simulated in the frequency domain too. We fix the sampling frequency of the data to 4096 Hz. Since we are operating in the time domain, the whitening procedure for each simulation comprises three main steps~\citep{LIGOScientific:2019hgc}:
\begin{enumerate}
    \item A Fast-Fourier transform to obtain the frequency domain representation, $h(f)$, of our time domain data, $h(t)$ , after properly windowing it using a Tukey window~\cite{Tukey:1967Spectrum} given by Eq. 11 in~\citet{Talbot:2025vth}. For a data length of 4 s, we employ a roll-off of 0.2 s and a tapering paramater value $\alpha=0.1$ following the \texttt{Bilby} default in version 2.6.0;
\item Whitening the frequency domain signal, after bandpassing the data with a frequency mask $f_{\mathrm{mask}}$, between $20$ and $2048$ Hz, that selects only the frequencies retained in the analysis:
\begin{equation}
    \tilde{h}(f)=\frac{h(f)}{\sqrt{P_j\times\frac{1}{N} \sum_{n=0}^{N-1} w[n]^2\times\frac{d}{4}}}~,
\end{equation}
where $d$ is the duration of the whitened data in seconds, $N$ is the number of data samples, the index $n$ runs over the elements of the window and $P$ is the \ac{PSD}.
\item An inverse Fast-Fourier transform is used to recover the whitened time domain signal:
\begin{equation}
h_w(t) = \mathrm{IFFT}\!\left( \tilde{h}(f) \right) \times 
\frac{N_f}{\sqrt{\sum f_{\mathrm{mask}}}}~,
\label{eq:inverse_fourier}
\end{equation}
where $N_f$ is the number of frequency bins. The normalization factor 
$\left(\sum f_{\mathrm{mask}}\right)^{-1/2}$ compensates for the fact that the mask sets excluded frequencies to zero, thereby ensuring that the overall amplitude of the whitened time-domain signal remains properly normalized.
\end{enumerate}
The whitened time-domain strain $h_w(t)$ is scaled by an auxiliary ``mock'' parameter $\sigma$, introduced to enable \ac{SBI} to operate correctly.  
Simulation--based inference requires that every simulated dataset be associated with at least one latent parameter drawn from a prior, so that the conditional density estimator can learn a mapping between parameters and data.  
In our case, the simulations contain only noise and therefore do not naturally carry any parameters. We introduce an auxiliary scale parameter $\sigma$, with a broad prior $\mathcal{U}[0,2]$, to ensure that the training procedure remains well posed. This choice rescales the noise realizations without altering their relative structure, such that prominent features, e.g., transients, are preserved. \\
To mitigate numerical artifacts introduced by the whitening procedure at the edges of the time series and to reduce data dimensionality, we truncate the data. As we have to accordingly crop the observation when analyzing it with the \ac{RNLE} likelihood, we make sure that we are not cutting away any part of the time series that includes the signal and contributes information to the inference. \\
In Figure~\ref{fig:schematic_processing}, we present a schematic overview of the workflow used to obtain posterior distributions with \ac{RNLE} from a set of training data realizations—either simulations or real detector data—and a given observation.  For validation purposes, the observation consists of a simulated \ac{GW} signal injected into coloured Gaussian noise, whereas in the following sections it corresponds to a \ac{GW} signal injected into real detector noise. Both the training data and the observations are pre-processed using an identical procedure.
The training data realizations are passed to the conditional density estimator, with the settings described in Section~\ref{subsec:setup_RNLE}, yielding a likelihood model.
For each evaluation of the \ac{RNLE} likelihood, the \ac{GW} parameters are sampled from the priors listed in Table~\ref{tab:prior_gw}, and a corresponding frequency-domain waveform is generated using the same waveform model employed for the observation generation. The waveform is whitened and transformed into the time domain, and subsequently subtracted from the whitened time-domain observation. The resulting residual time series is then passed to the likelihood for evaluation. This procedure is repeated at each likelihood evaluation within the nested sampling algorithm, which ultimately produces posterior samples for the signal parameters and the noise residual parameter $\sigma$.
 \\

\subsection{Analysis of 1D GW signal\label{subsec:gw_1D_test}}
\begin{figure*}[t]
    \centering
    \includegraphics[width=0.45\linewidth]{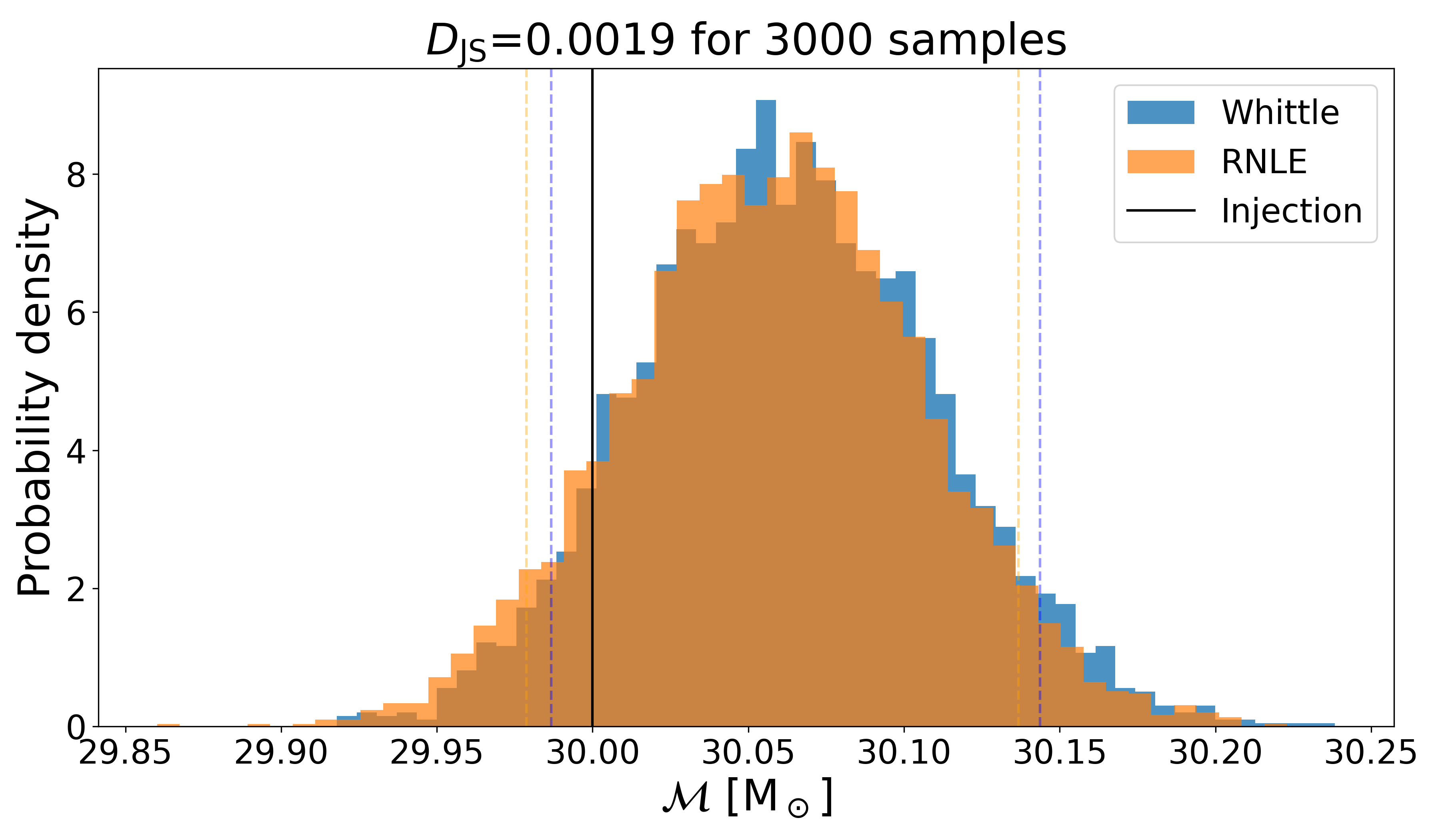}
    \includegraphics[width=0.45\linewidth]{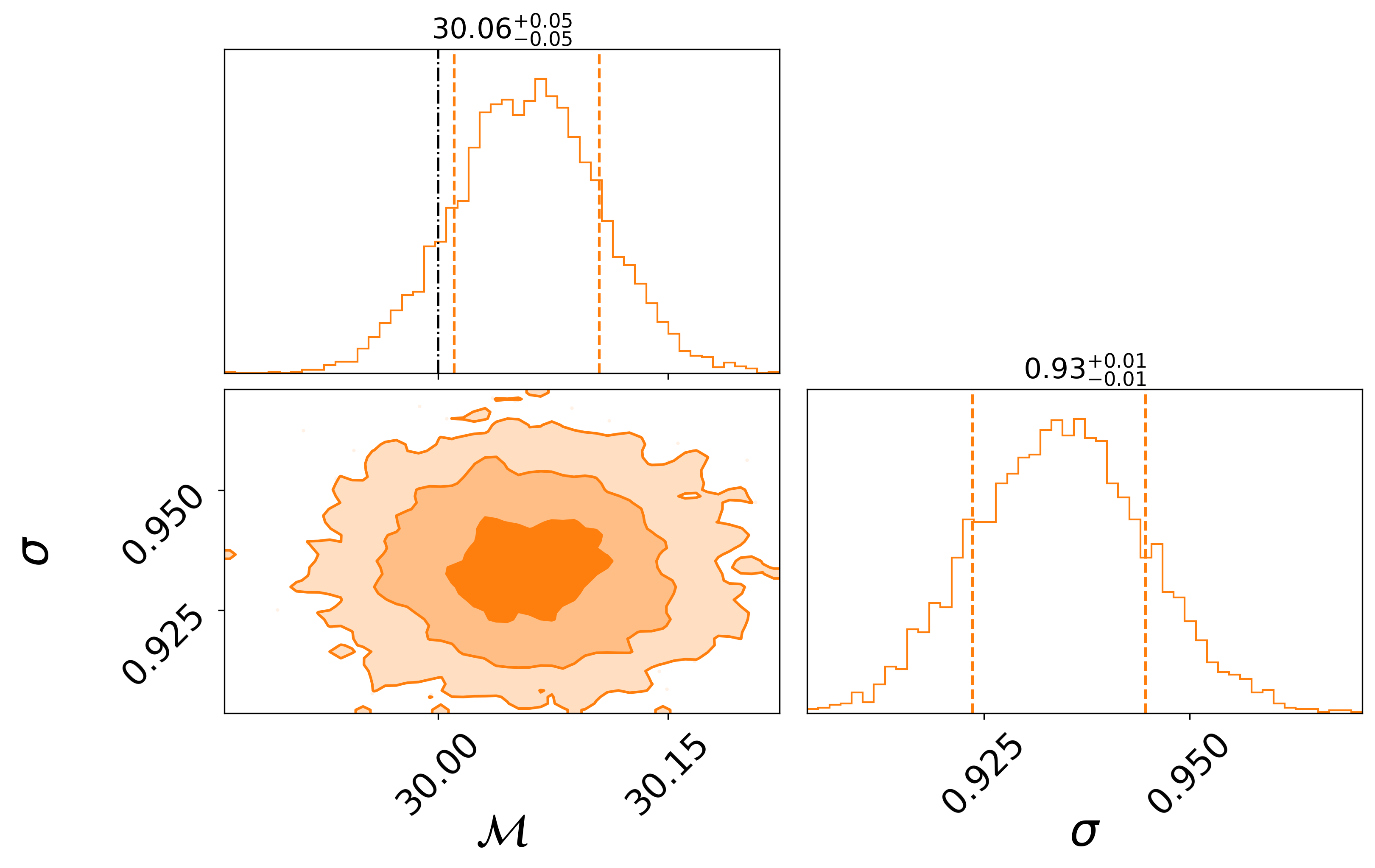}
    \caption{\textbf{Left}: Comparison of the posterior distribution of the chirp mass $\mathcal{M}$ obtained using the \ac{RNLE} likelihood trained using 3000 simulations of coloured-Gaussian noise, in orange, and the Whittle likelihood, in blue. The \ac{JS} divergence value is reported in the title, confirming the agreement between the two results. \textbf{Right}: Full posterior of the \ac{RNLE} analysis. The histograms show the posterior distributions for the noise scaling parameter $\sigma$ and the chirp mass. The dash-dotted black line in the histogram shows the true value of the parameter.}
    \label{fig:benchmark_gw_1D}
\end{figure*}

To validate the RNLE framework for gravitational-wave parameter estimation, we first perform a simplified one-dimensional analysis, focusing on a single signal parameter. In this controlled setup, we test whether RNLE can accurately recover posterior distributions consistent with those obtained using the standard Whittle likelihood. Restricting the analysis to the detector-frame chirp mass allows us to systematically explore the algorithm’s performance, identify optimal configuration settings, and significantly reduce computational cost. \\
In the left panel of Figure~\ref{fig:benchmark_gw_1D}, we show the comparison between the posterior distribution of the chirp mass~\cite{LIGOScientific:2025hdt} obtained with \ac{RNLE} and that obtained using the Whittle likelihood under identical sampler settings, observation, and priors. We perform a \ac{JS} divergence test, following the specifics outlined in Section~\ref{sec:toy_model}. The resulting value is below the chosen threshold, confirming that the two posterior distributions are statistically consistent.\\
In the right panel of Figure~\ref{fig:benchmark_gw_1D}, we show a corner plot for the \ac{RNLE} likelihood analysis. The posterior distributions for the noise scaling parameter $\sigma$ and the chirp mass are uncorrelated, with the true value of $\sigma$ equal to 1.

\subsection{Dependence on Training Dataset Size\label{subsec:gw_1D_simulation_number}}
\begin{figure*}[htp!]
    \centering
    \includegraphics[width=0.7\linewidth]{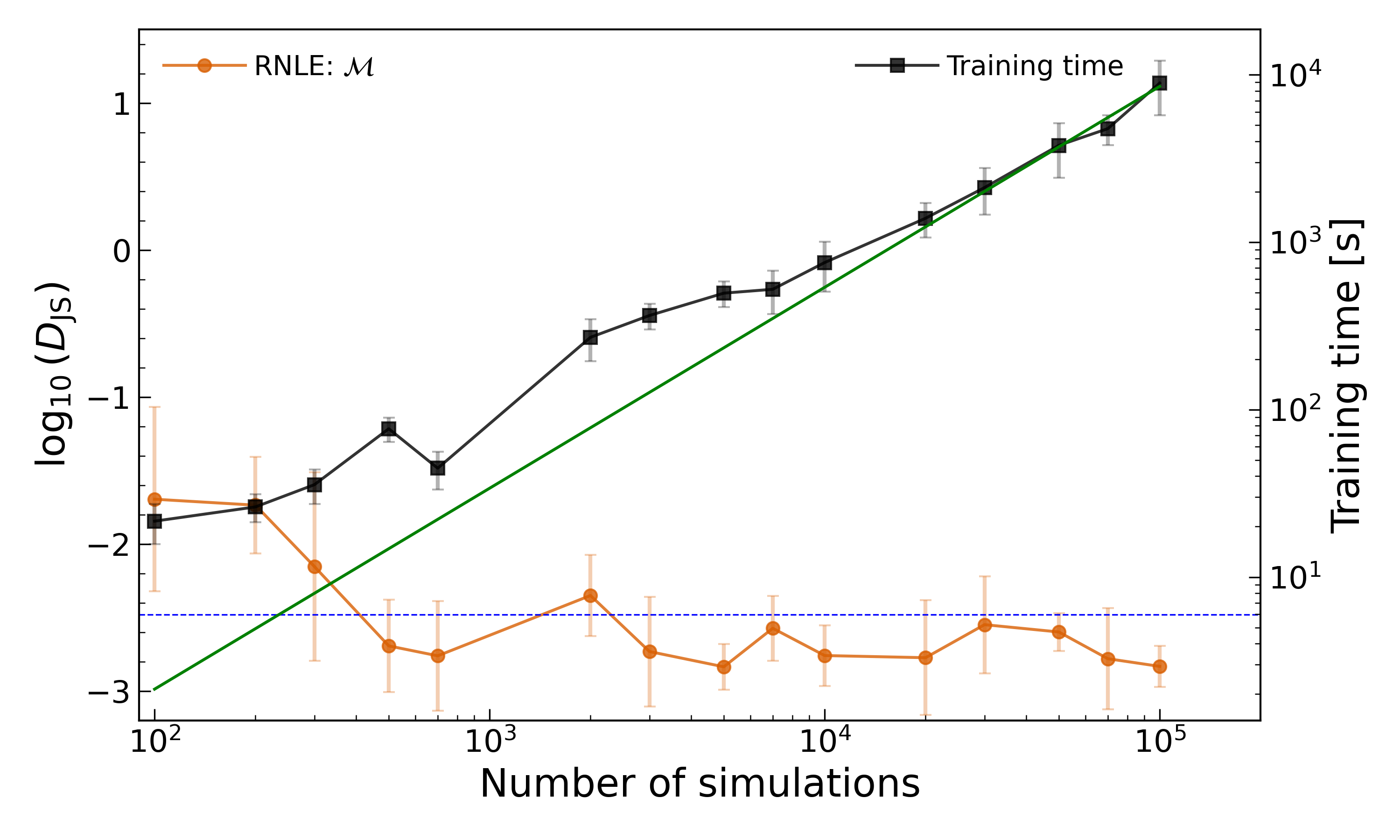}
    \caption{In orange, the logarithm base ten of the \ac{JS} divergence value for the chirp mass posteriors, computed between the Whittle and \ac{RNLE} results, against the number of simulations employed to train the conditional density estimator. The blue dotted line indicates the \ac{JS} divergence threshold signifying the statistical equivalence between the two distributions. The black squares show the \ac{RNLE} likelihood training time against the number of simulations. We use a logarithmic scale for the $x$-axis and the right $y$-axis. The errorbars are computed over five realizations of re-training and sampling. The green line is obtained using the optimal values from a linear fit in log-log space for the training time versus the number of simulations.}
    \label{fig:benchmark_gw_initial}
\end{figure*}

To assess how the accuracy of the \ac{RNLE} likelihood depends on the size of the training dataset, we investigate the convergence of the inferred posterior distributions as a function of the number of training data realizations. In particular, we quantify how the agreement between the \ac{RNLE} and Whittle likelihood results improves as additional training data are provided to the conditional density estimator. \\
In Figure~\ref{fig:benchmark_gw_initial}, we show the \ac{JS} divergence between the chirp mass posterior distributions obtained with \ac{RNLE} and with the Whittle likelihood as a function of the number of simulations used for training. The blue dotted line indicates the \ac{JS} divergence threshold corresponding to statistical equivalence between the two distributions. The black squares denote the training time required for the \ac{RNLE} likelihood as a function of the number of simulations. A linear fit in log--log space is performed to illustrate the scaling behavior, showing that the training time approaches linear scaling for $N_{\mathrm{sim}} \gtrsim 10^4$. \\
From these results, we observe that the \ac{JS} divergence stabilizes below the equivalence threshold once more than approximately $2000$ simulations are used for training, indicating that the \ac{RNLE} likelihood can reliably reproduce the Whittle likelihood results beyond this point. At the same time, the training time increases rapidly with the size of the dataset, ranging from below one hour for $\sim2000$ simulations to several hours for $10^5$ simulations. Based on this trade-off between accuracy and computational cost, we adopt $3000$ simulations for all subsequent analyses involving a one-dimensional signal prior, unless stated otherwise.

\subsection{ Analysis of training noise \label{subsec:gw_1D_training_noise}}
\begin{figure*}[t]
    \centering
    \includegraphics[width=0.8\linewidth]{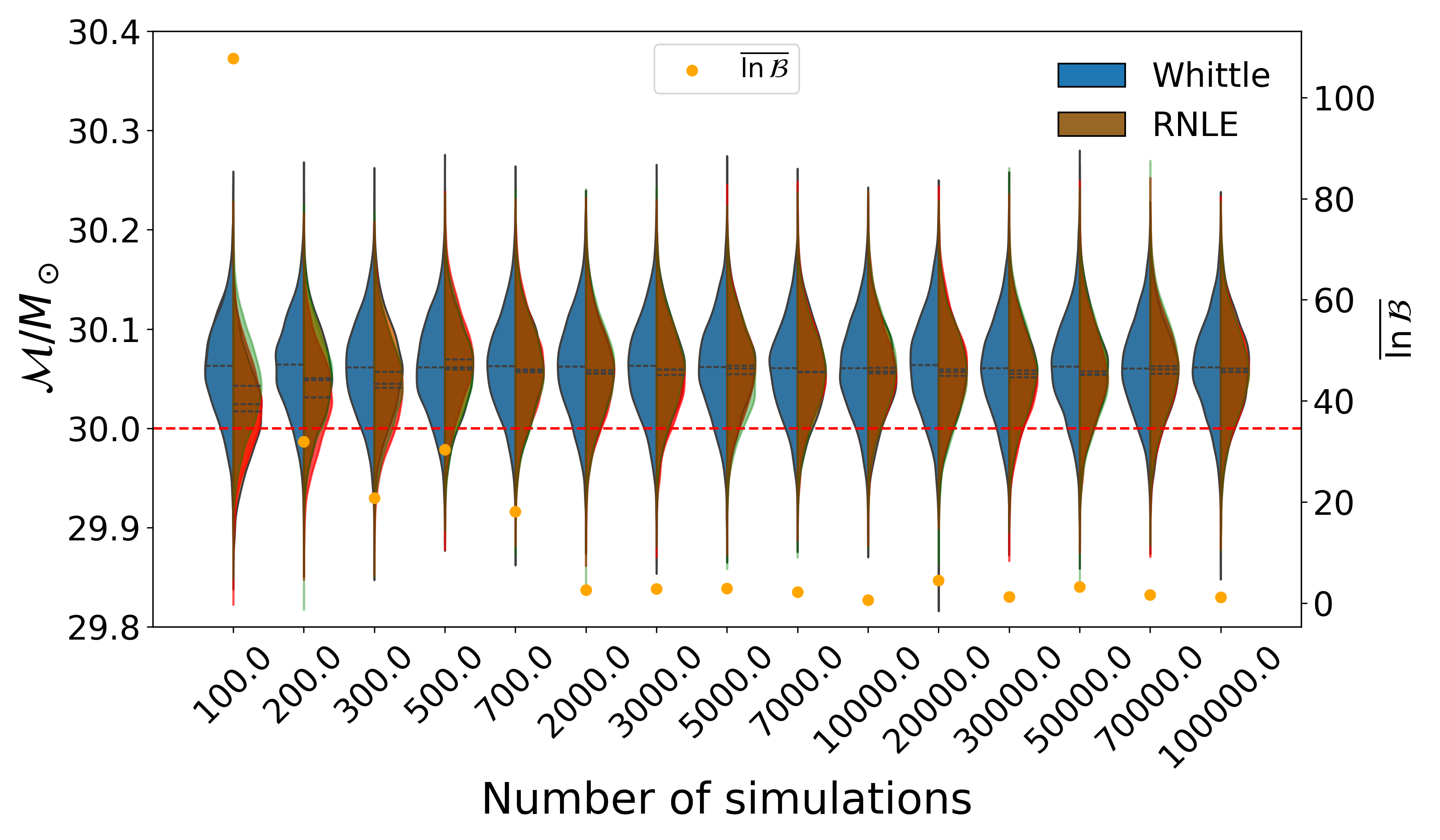}
    \caption{Violin plot showing the posterior distributions for the chirp mass varying the number of simulations used to train the conditional density estimator. For each violin, the left side in blue, shows the posterior obtained with the Whittle likelihood, while the right side shows the RNLE posteriors obtained with different likelihoods trained on the same number of simulations. Each likelihood realization is trained with a different optimization seed and a different, but statistically equivalent, set of simulations. The red dashed line indicates the true value of the chirp mass for the injected signal. The orange dots represent the mean of the log-Bayes factor computed for the different realizations of the RNLE likelihood against the number of simulations used in the training.}
    \label{fig:benchmark_gw_violins}
\end{figure*}

To investigate the stability of the conditional density estimator and to quantify the training noise, i.e., the uncertainty associated with the training process itself, we perform an ensemble analysis in which multiple \ac{RNLE} likelihoods are trained independently. For each chosen number of simulations, we construct three training datasets by drawing independent realizations of coloured Gaussian noise from the same \ac{PSD}. Each dataset is used to retrain the conditional density estimator with different optimization seeds, allowing us to isolate variability arising from both the finite training set and the stochastic nature of the training procedure. Since all training datasets are drawn from the same underlying noise-generating process, we expect that, in the limit of sufficiently large training sets, variations in the specific realizations should not affect the learned likelihood in a systematic way. Any residual variability can therefore be attributed to the stochasticity of the training procedure itself. \\
In Figure~\ref{fig:benchmark_gw_violins}, we show violin plots comparing the posterior distribution of the chirp mass obtained with the Whittle likelihood to those obtained from the different realizations of the \ac{RNLE} likelihood, for increasing numbers of training simulations. The red dotted line indicates the true chirp mass value of the injected \ac{GW} signal. Each set of violins corresponds to three independently trained \ac{RNLE} likelihoods using the same number of simulations. \\
To quantify the spread among the ensemble members, we compute the log-Bayes factor for each \ac{RNLE} likelihood with respect to the Whittle likelihood when analyzing the same observation. The orange points in Figure~\ref{fig:benchmark_gw_violins} show the mean of the log-Bayes factors across the ensemble for each training size. We observe that this quantity decreases as the number of simulations increases and stabilizes for training sets with more than approximately $2000$ simulations. This behavior suggests that the variability associated with training noise decreases as the size of the training dataset increases, indicating a progressive stabilization of the learned likelihood. This trend is consistent with the results shown in Figure~\ref{fig:benchmark_gw_initial}, supporting the choice of $\mathcal{O}(10^3)$ simulations as a reasonable compromise between accuracy and computational cost. A complementary discussion of training noise is presented in Section~\ref{sec:training_noise}, where we also explore a strategy to combine the posteriors obtained from different likelihood realizations, weighting each posterior by its associated Bayesian evidence.

\subsection{Dependence on analysis window size\label{subsec:gw_1D_cutting_time}}
\begin{figure*}[htp!]
    \centering
    \includegraphics[width=0.46\linewidth]{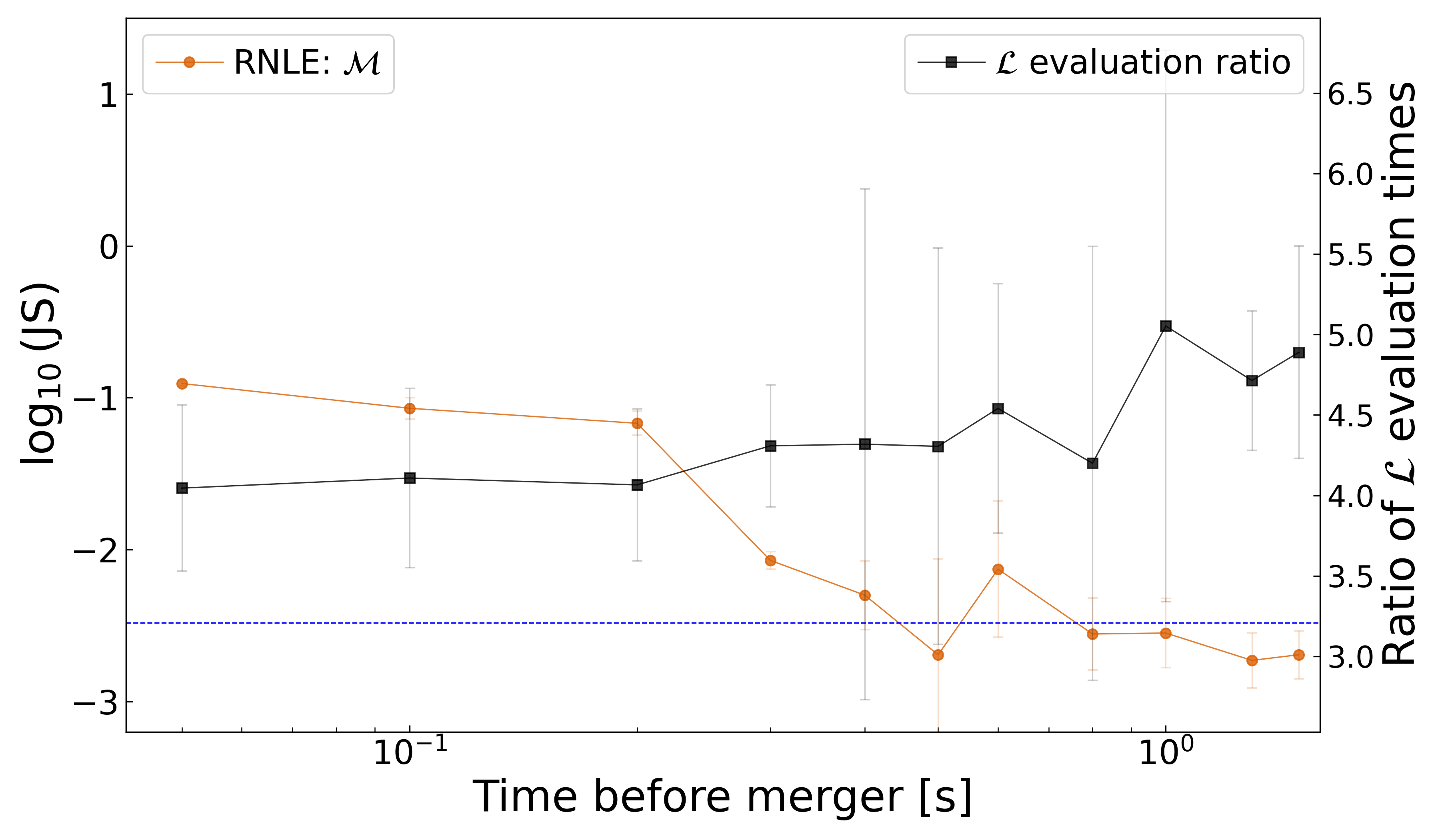}
    \includegraphics[width=0.46\linewidth]{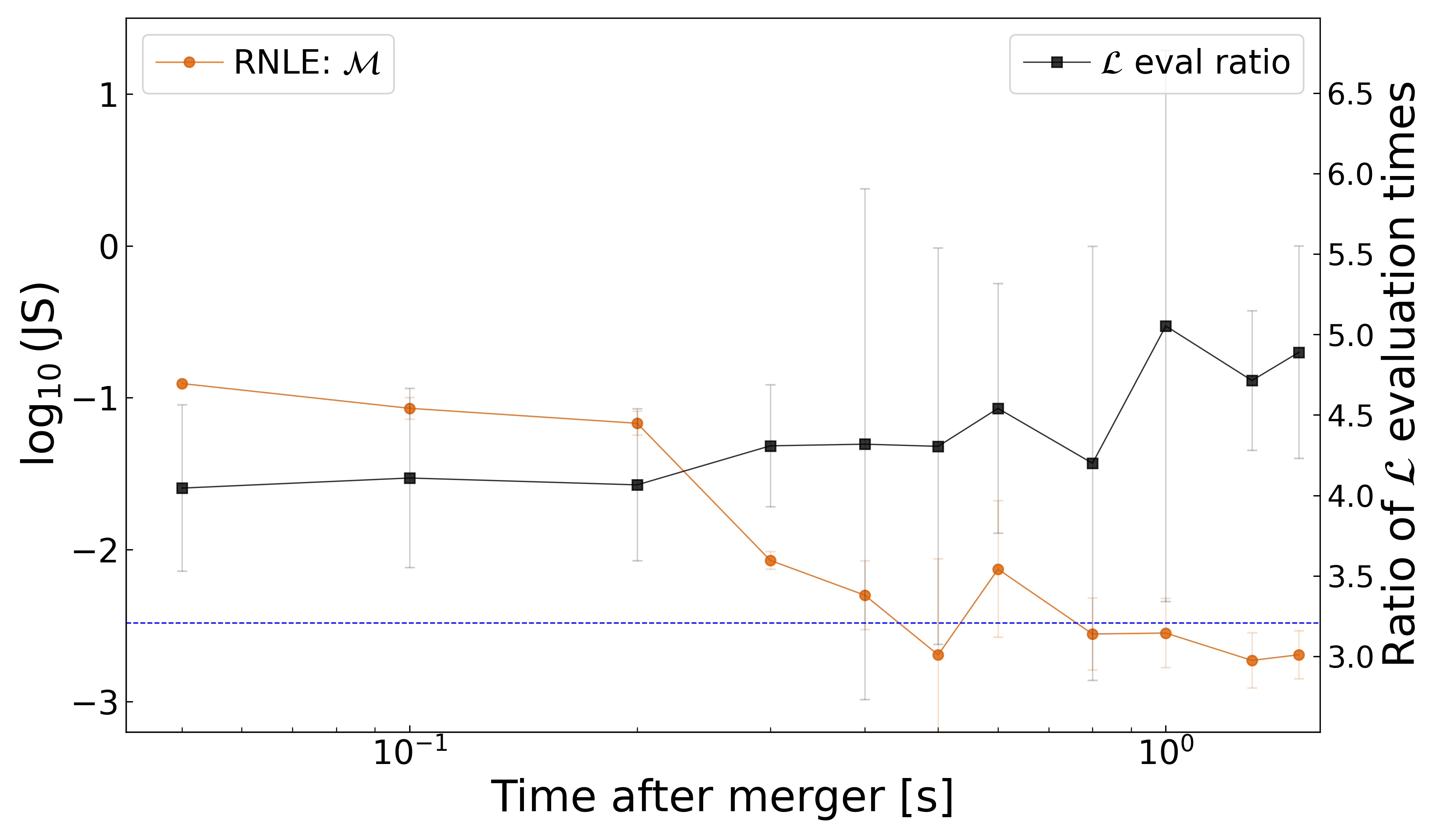}
    \caption{In orange, the logarithm base ten of the \ac{JS} divergence value for the chirp mass posteriors against the time included before the merger, left panel, and after the merger, right panel. The blue dotted line indicates the \ac{JS} divergence threshold signifying the statistical equivalence between the two distributions. The black squares show the ratio between the time needed for a single likelihood evaluation with  \ac{RNLE} and the Whittle likelihood. We use a logarithmic scale for the x-axis.}
    \label{fig:benchmark_gw_time}
\end{figure*}

To quantify how the accuracy of the \ac{RNLE} likelihood depends on the size of the analyzed data segment, we study the sensitivity of the recovered posterior distributions to variations in the analysis window. In particular, we assess how reducing the amount of data before or after the merger affects the agreement between \ac{RNLE} and the Whittle likelihood, as measured by the \ac{JS} divergence of the chirp mass posteriors. \\
In Figure~\ref{fig:benchmark_gw_time}, we show the \ac{JS} divergence as a function of the time interval retained before the merger (left panel) and after the merger (right panel) for the \ac{RNLE} analysis. In each case, the complementary portion of the data segment is held fixed: when varying the pre-merger duration, we include a fixed $0.3\,\mathrm{s}$ of post-merger data, and conversely, when varying the post-merger duration, we include $0.8\,\mathrm{s}$ of pre-merger data. The benchmarking is performed against Whittle likelihood results obtained using a four-second data segment, corresponding to two seconds before and two seconds after the merger. The dotted horizontal line indicates the \ac{JS} divergence threshold below which the two posterior distributions are considered statistically equivalent. The square markers denote the ratio between the computational time required for a single likelihood evaluation using \ac{RNLE} and that required by the Whittle likelihood. \\
While the per-evaluation cost of the \ac{RNLE} likelihood is only moderately higher— by a factor of $\sim 3$--$5$—this does not fully capture the total computational overhead of an end-to-end inference run. In practice, \ac{RNLE}-based analyses require a larger number of effective likelihood evaluations, owing to the increased dimensionality of the parameter space resulting from the inclusion of an additional noise parameter, which in turn impacts sampling efficiency and convergence. When accounting for both the per-evaluation cost and the total number of likelihood calls performed by the sampler, we find that the overall wall time of \ac{RNLE} analyses is approximately an order of magnitude larger than that of corresponding Whittle-likelihood runs. We emphasize, however, that this increased computational cost reflects the added flexibility and robustness of the learned likelihood, rather than an inherent inefficiency of the evaluation itself. \\
From the left panel of Figure~\ref{fig:benchmark_gw_time}, we observe that truncating the data too close to the merger removes essential signal information, preventing recovery of posterior distributions consistent with those obtained using the Whittle likelihood. In particular, retaining at least $0.8\,\mathrm{s}$ of data prior to the merger is required to recover statistically equivalent posteriors for this injection. This requirement depends on the signal parameters, most notably the chirp mass, as the time spent by the \ac{GW} signal in the ground based interferometers' detection band scales as $t_c - t(f) \propto \mathcal{M}^{-5/3}$, where $t_c$ denotes the coalescence time~\cite{Maggiore:2007ulw}.

\section{Analysing simulated BBHs in coloured Gaussian noise \label{sec:simulated_GW}}

\begin{figure*}[t]
    \centering
    \includegraphics[width=0.44\linewidth]{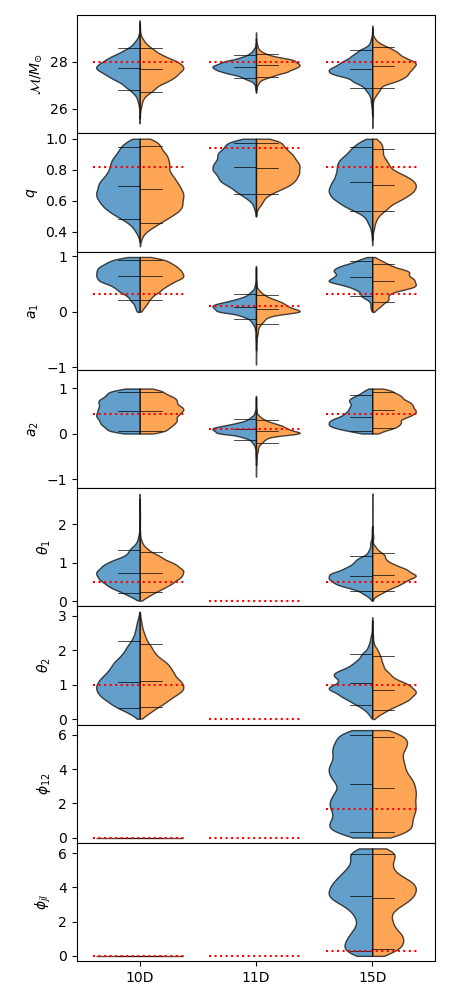}
    \includegraphics[width=0.44\linewidth]{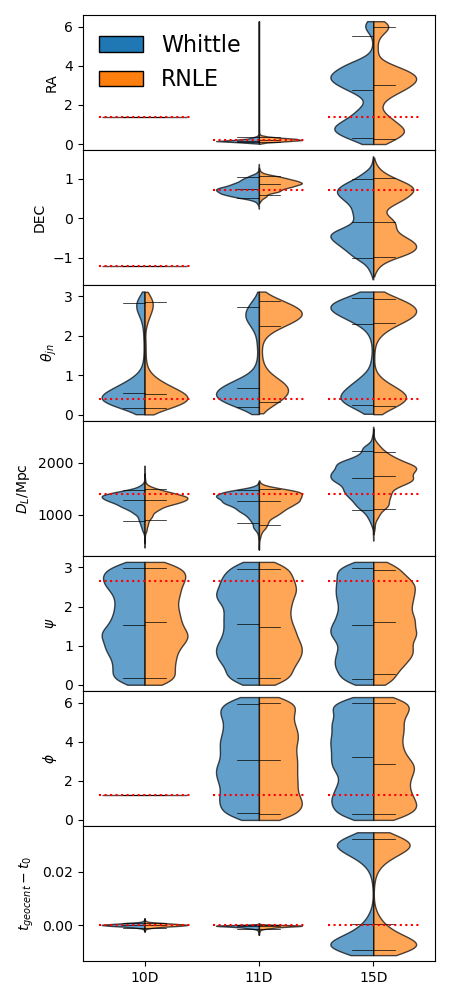}
    \caption{Comparison of violin plots for the 10-,11-, and 15-dimensional posterior distributions obtained with the \ac{RNLE} and Whittle likelihood, in orange and blue, respectively. These were obtained from a precessing \ac{BBH} merger with a resticted sampling parameter space, an aligned-spin \ac{BBH} merger signal, and a precessing \ac{BBH} signal sampling in the full parameter space, respectively. In all cases the \ac{GW} signal was injected into coloured Gaussian noise. The injection values of the signal parameters can be found in Table~\ref{tab:simulation} and are overplotted as red dotted lines on the violins. \ac{JS} divergence values for all the parameters are listed in the same Table. We show the mean and 90\% credibility intervals as black lines for each violin half. For fixed parameters we plot the posterior as a Dirac delta on the true value. The left panel shows the posteriors for the intrinsic parameters, while the right panel shows the extrinsic parameters.}
    \label{fig:benchmark_gw_all_violin}
\end{figure*}

Having identified empirically well-performing settings for the training of the conditional density estimator in the one-dimensional case, we now extend our analysis to the full parameter space of gravitational-wave waveform injections. This step allows us to assess the performance and robustness of the \ac{RNLE} framework in realistic, high-dimensional inference scenarios. We adopt the injection parameters and prior distributions listed in Tables~\ref{tab:simulation} and~\ref{tab:prior_gw} throughout this section. \\
For all \ac{RNLE} analyses presented here, we use a one-second data segment comprising $0.8\,\mathrm{s}$ preceding the trigger time and $0.2\,\mathrm{s}$ following it, extracted from the four-second stretch employed for whitening. To improve the numerical stability of the learned likelihood in these higher-dimensional settings, we increase the number of training simulations to $3\times10^4$. We emphasize that this choice has not been systematically optimized: while a smaller training set may be sufficient, we have not explored the minimum number of simulations required to ensure stable performance. Importantly, the training procedure remains fully independent of the underlying signal parameters. 
Figure~\ref{fig:benchmark_gw_all_violin} summarizes the results of three benchmark configurations with increasing complexity. These include: (i) a 10-dimensional (10D) parameter space for a precessing binary black hole (BBH) signal with a subset of extrinsic parameters fixed; (ii) an 11-dimensional (11D) aligned-spin BBH signal analyzed using data from two detectors; and (iii) a 15-dimensional (15D) precessing BBH signal. \\
Each panel in Figure~\ref{fig:benchmark_gw_all_violin} shows violin plots of the posterior distributions for the sampled parameters. For each parameter, we compare the posterior obtained using the Whittle likelihood, with the posterior obtained with \ac{RNLE}. From left to right, the violins represent the 10D, 11D, and 15D analyses. The left panel of the figure displays the intrinsic source parameters: the detector-frame chirp mass $\mathcal{M}$, the mass ratio $q$, the spin magnitudes $a_1$ and $a_2$, the tilt angles $\theta_1$ and $\theta_2$ between each black hole’s spin and the orbital angular momentum, the azimuthal angle between the spin vectors $\phi_{12}$, and the azimuthal angle between the total and orbital angular momenta $\phi_{jl}$~\cite{Veitch:2014wba, LIGOScientific:2016vlm}. The right panel shows the extrinsic parameters: right ascension $\mathrm{RA}$, declination $\mathrm{DEC}$, the viewing angle $\theta_{JN}$, the luminosity distance $D_L$, the geocenter time $t_{\mathrm{geocent}}$, the polarization angle $\psi$, and the phase $\phi$. The \ac{JS} divergence values for all the parameters and analyses are reported in Table~\ref{tab:simulation} in Appendix~\ref{app:tables}. \\
In the 10D analysis, the posterior distributions obtained with \ac{RNLE} are statistically identical from those obtained with the Whittle likelihood across all parameters, as demonstrated by the \ac{JS} values. This strong agreement demonstrates that, in moderately high-dimensional settings, \ac{RNLE} can accurately reproduce the results of traditional likelihood-based inference. \\
For the 15D precessing BBH system, small discrepancies are visible in a subset of parameters. These differences are consistent with stochastic sampling variability and reflect the well-known degeneracies and complex posterior structure characteristic of high-dimensional precessing BBH parameter spaces~\cite{Baird:2012cu, Usman:2018imj}, rather than a systematic bias introduced by the \ac{RNLE} method. 

To further disentangle potential algorithmic effects from model complexity, we repeat the analysis for an 11D aligned-spin BBH system using simulated data from the Hanford and Livingston detectors. In this configuration, separate conditional density estimators are trained for each detector using disjoint simulation sets, while identical \acp{PSD} are employed to ensure consistency in the whitening procedure. The total likelihood is constructed following Eq.~\ref{eq:multidet_like}. This modular approach allows the RNLE framework to scale naturally to larger detector configurations. As shown by the 11D violins, the agreement between the Whittle and \ac{RNLE} likelihoods remains strong for the intrinsic parameters. This confirms that the deviations observed in the 15D case are not indicative of a deficiency in the \ac{RNLE} framework.
Some residual differences persist in the extrinsic parameters—most notably in right ascension, declination, and the viewing angle $\theta_{JN}$—which exhibit multimodal posterior structures. These features are consistent with the expected sampling variability in complex, multimodal inference problems and do not suggest a systematic bias. Overall, as is expected, these results validate the accuracy and reliability of the \ac{RNLE} algorithm across a range of signal dimensionalities and detector configurations.

\section{Analysing simulated BBHs in quasi-Gaussian observed data}
\label{sec:real_GW_quasi}
Having verified that the \ac{RNLE} algorithm can reproduce the results obtained with the Whittle likelihood under the assumptions of stationarity and Gaussianity, we next extend the method to data observed by the \ac{GW} detectors. We construct a pipeline that downloads publicly available strain data from the \ac{GW} detectors~\cite{Trovato:2019liz, LIGOScientific:2025snk}, whitens it, and partitions it into one-second segments, following the procedure described in Section~\ref{sec:simulated_GW} and shown in Figure~\ref{fig:schematic_processing}. For each segment, whitening is performed with a \ac{PSD} estimated from the preceding 16 seconds of data using a Hann\footnote{We acknowledge that, for consistency with the data's whitening procedure, a Tukey window should be used. As this inconsistency was found in an advanced stage of the analysis, we tested that the effects on the results are negligible, see Appendix~\ref{app:psd_window}, without further modifications.} window and employing the median method with the \texttt{gwpy} package~\cite{gwpy}. These segments are used as input data to train the conditional density estimator. When evaluating the \ac{RNLE} likelihood, at each sampling step, the subtracted signal realization is whitened using the same \ac{PSD} used to whiten the observation.

\subsection{Identification of quasi-Gaussian data segment}\label{subsec:quasi_guassian_data}

As an initial test, we consider a segment of data for which the noise is approximately Gaussian: the \ac{BNS} range~\cite{Schutz:2011tw,Chen:2017wpg} is stable, the Omicron trigger rate is low~\citep{Robinet:2015omicron,Robinet:2020lbf}, and no glitches with \ac{SNR}$>5$ are present. We arbitrarily select a 20-second data segment centered on the well-behaved data around 2{:}00~a.m.~\ac{UTC} on 10 August 2019, which is used to construct the observation and to estimate the \ac{PSD} for whitening. To generate the training data realizations, we then use 30 minutes of detector data on either side of this central segment, explicitly excluding the 20-second window used for the observation.
The data is divided into consecutive one-second segments, each paired with its corresponding \ac{PSD}. Segments are randomly sampled to construct the training set, ensuring that each segment is used only once.

\begin{figure*}[t]
    \centering
    \includegraphics[width=0.47\linewidth]{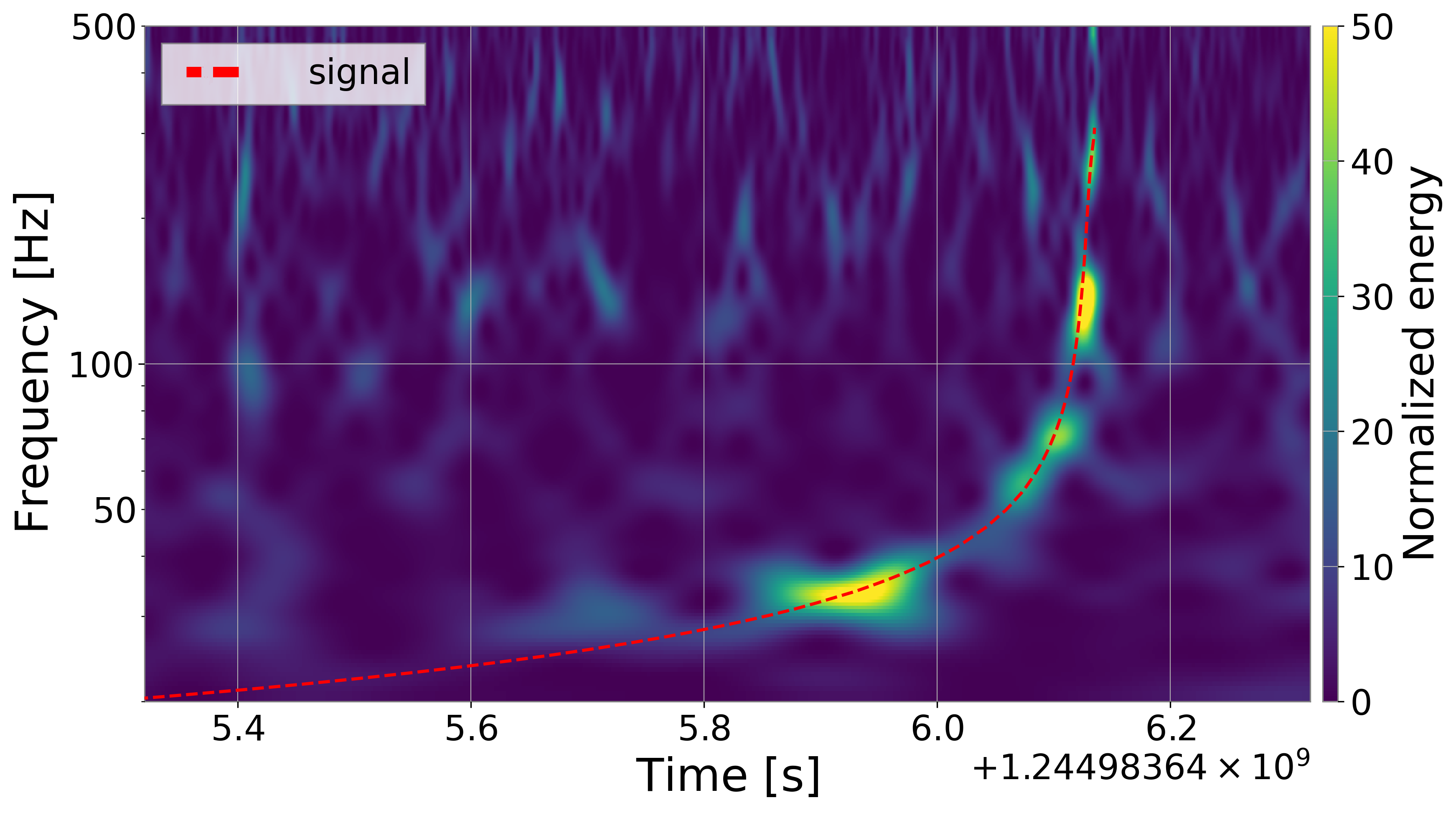}
     \includegraphics[width=0.47\linewidth]{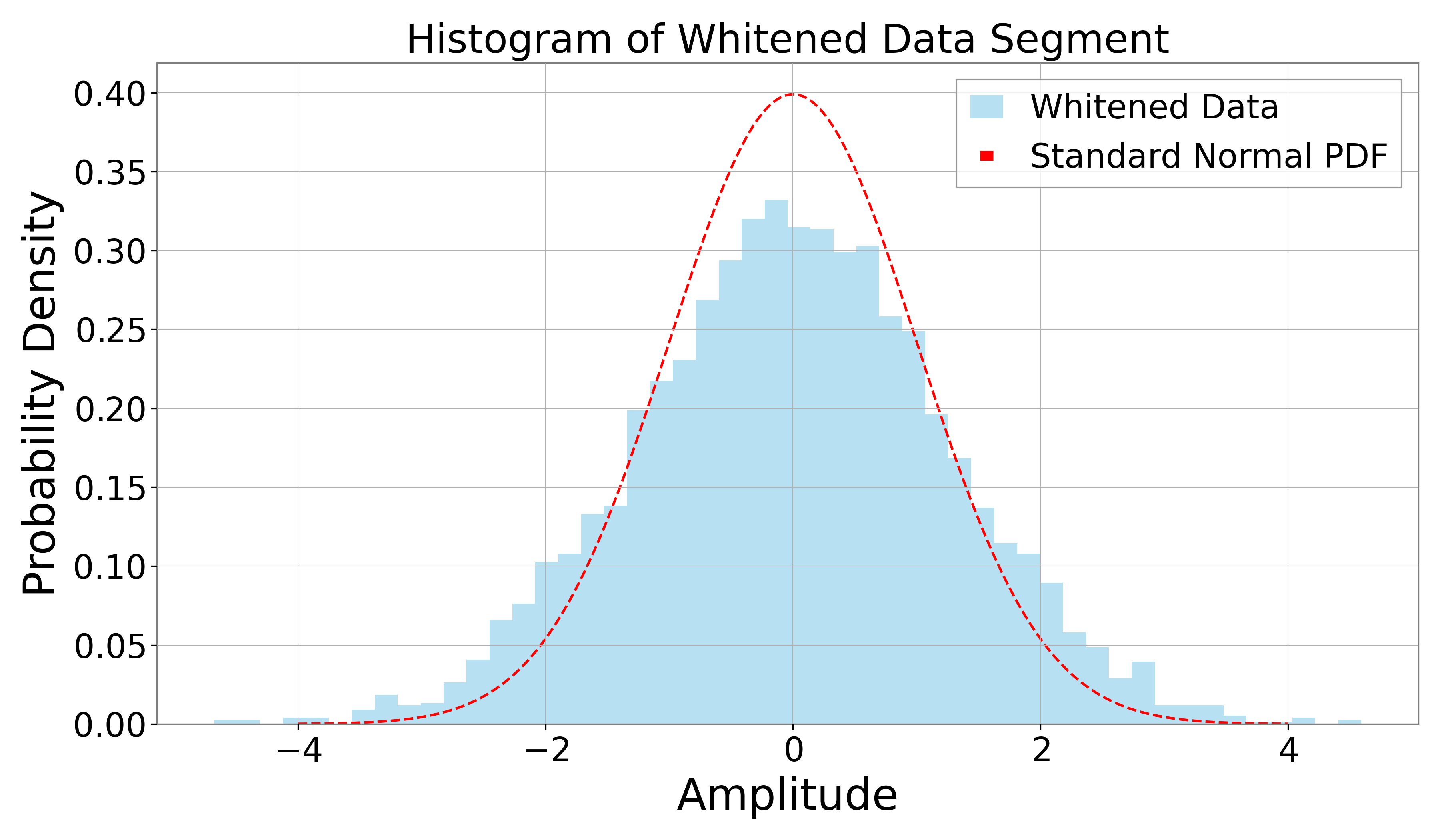}
    \caption{\textbf{Left}: Time-frequency plot of 1 seconds of data from the Hanford detector around 2:00 a.m. \ac{UTC} on 10 August 2019 used as the analysis window for \ac{RNLE}. The data was selected for its quasi-Gaussian noise properties. A red-dashed trace is overplotted to the injected \ac{GW} signal. The parameters of the injection are recorded in Table~\ref{tab:simulation}. \textbf{Right}: Gaussianity test for the selected data. We compare the amplitude of the whitened data segment without the signal to a standard normal distribution, red dashed line.}
    \label{fig:real_data_1D_data}
\end{figure*}

For the test observation, we inject the same aligned-spin \ac{BBH} merger used in Section~\ref{sec:simulated_GW} into a segment excluded from the training set. The resulting time–frequency representation is shown in the left panel of Figure~\ref{fig:real_data_1D_data}, where the injected signal track is overplotted as a red dotted line. The right panel displays a Gaussianity check of the data segment without the injection by comparing the histogram of whitened amplitudes to a normal distribution.

\subsection{Analysing a BBH injection into a quasi-Gaussian data segment}
\label{subsec:analysis_quasi_Gaussian}
\begin{figure*}[t]
    \centering
    \includegraphics[width=0.47\linewidth]{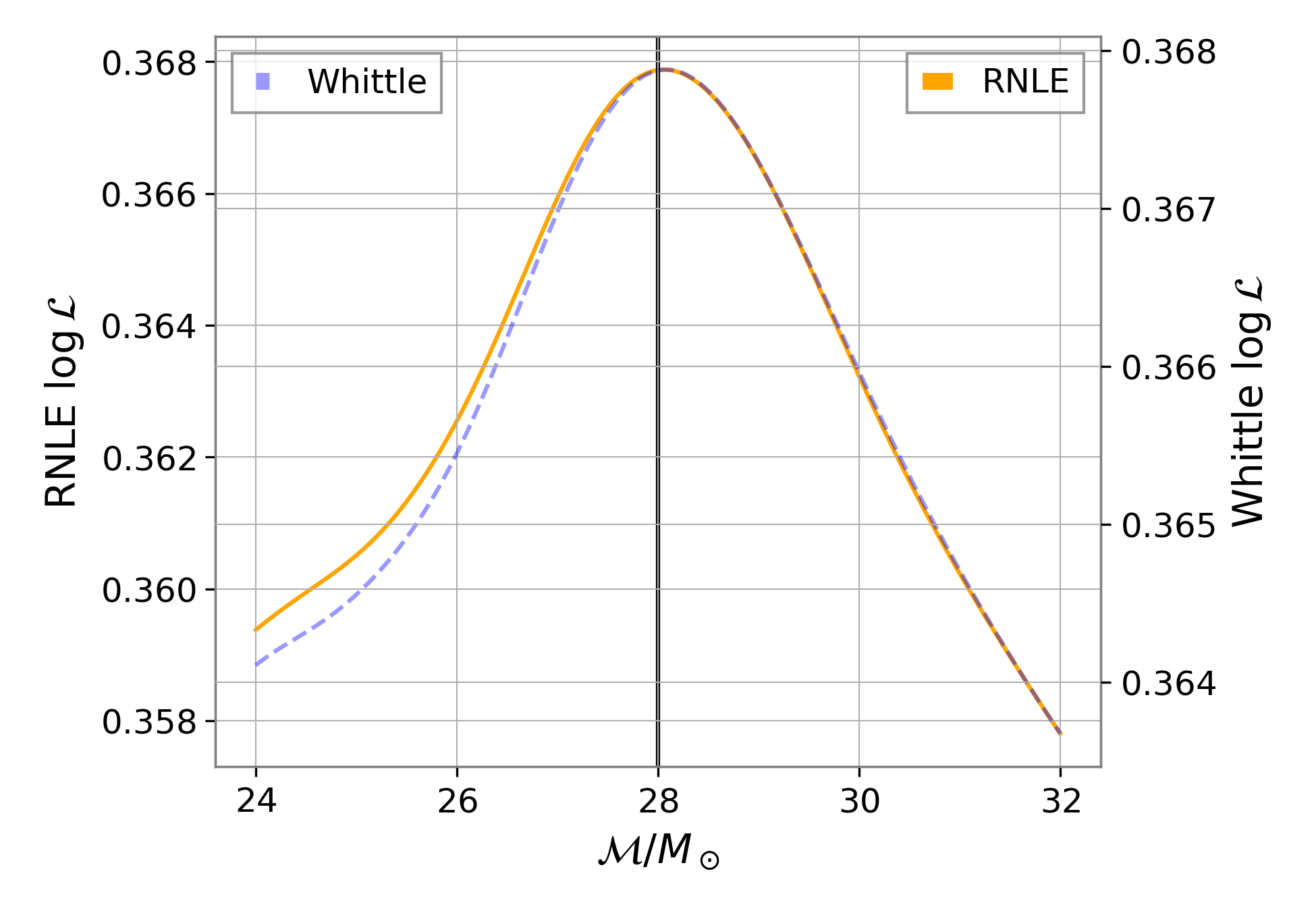}
    \includegraphics[width=0.47\linewidth]{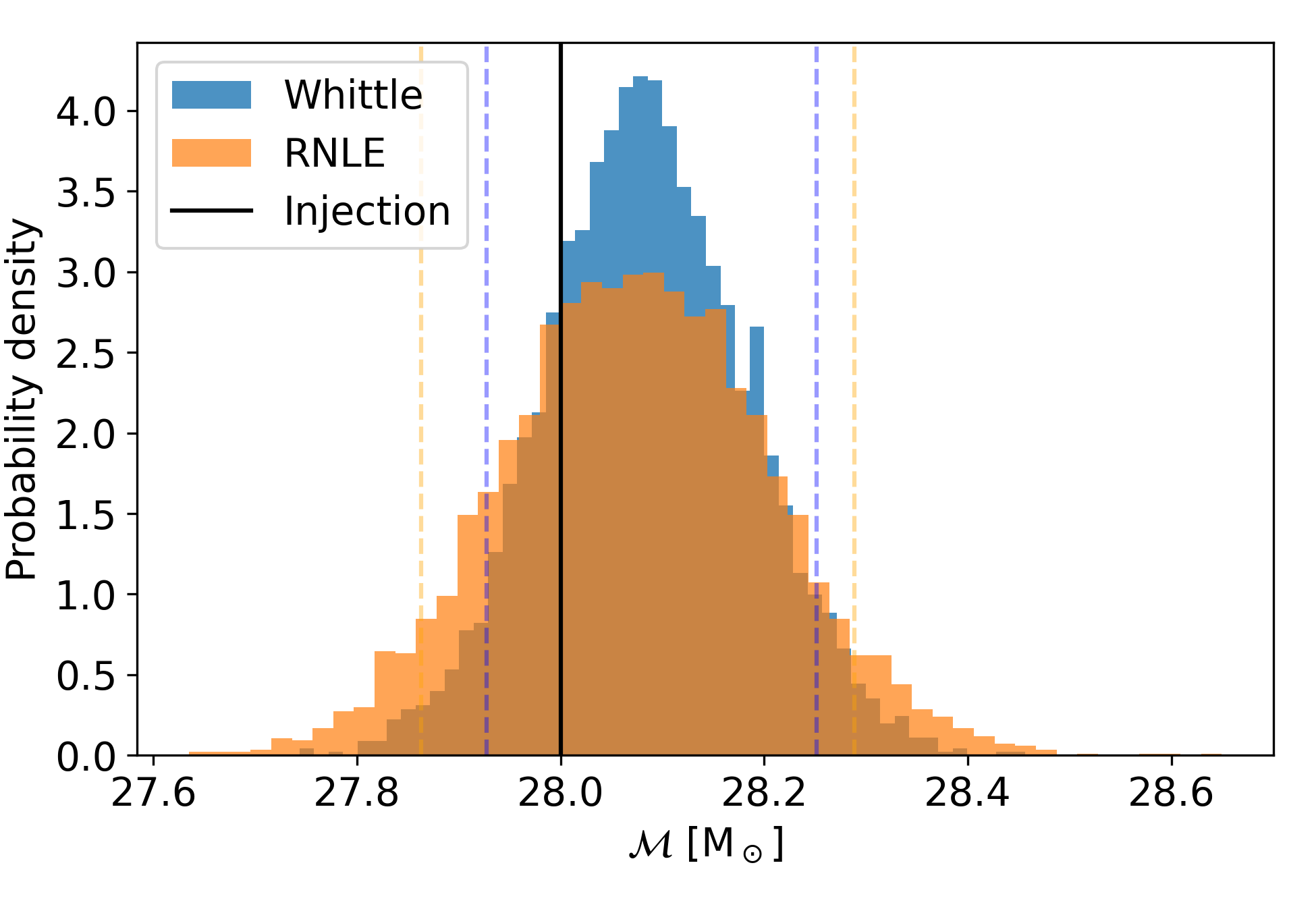}
    \caption{\textbf{Left}: Comparison between the Whittle likelihood, in blue, and the learnt \ac{RNLE} likelihood, in orange. for the selected data. \textbf{Right}: Comparison of the posterior probability distributions obtained with \texttt{dynesty} using the Whittle likelihood, in blue, and the \ac{RNLE} likelihood, in orange. The black vertical line indicates the injection value, while the dotted vertical lines represent the 90\% credible intervals.}
    \label{fig:real_data_1D_results}
\end{figure*}

After training the \ac{RNLE} conditional density estimator on 3000 data segments, we compare the resulting likelihood with the Whittle likelihood evaluated on the observation shown in Figure~\ref{fig:real_data_1D_data}. The left panel of Figure~\ref{fig:real_data_1D_results} presents the two likelihoods as a function of the chirp mass, holding all other parameters fixed. The \ac{RNLE} likelihood is broadly consistent with the Whittle likelihood, with minor differences expected given the inherent non-Gaussian and non-stationary features of observational data. This behaviour is mirrored in the one-dimensional posteriors obtained using \texttt{dynesty} within \texttt{Bilby}: the \ac{RNLE} posterior is marginally broader, and the Jensen–Shannon divergence remains above the statistical-equivalence threshold, taking a value of 0.01. Nonetheless, both methods recover comparable median parameter values, demonstrating that the \ac{RNLE} training and evaluation pipeline can operate reliably on observational data.

\begin{figure*}[htpb]
    \centering
    \includegraphics[width=0.44\linewidth]{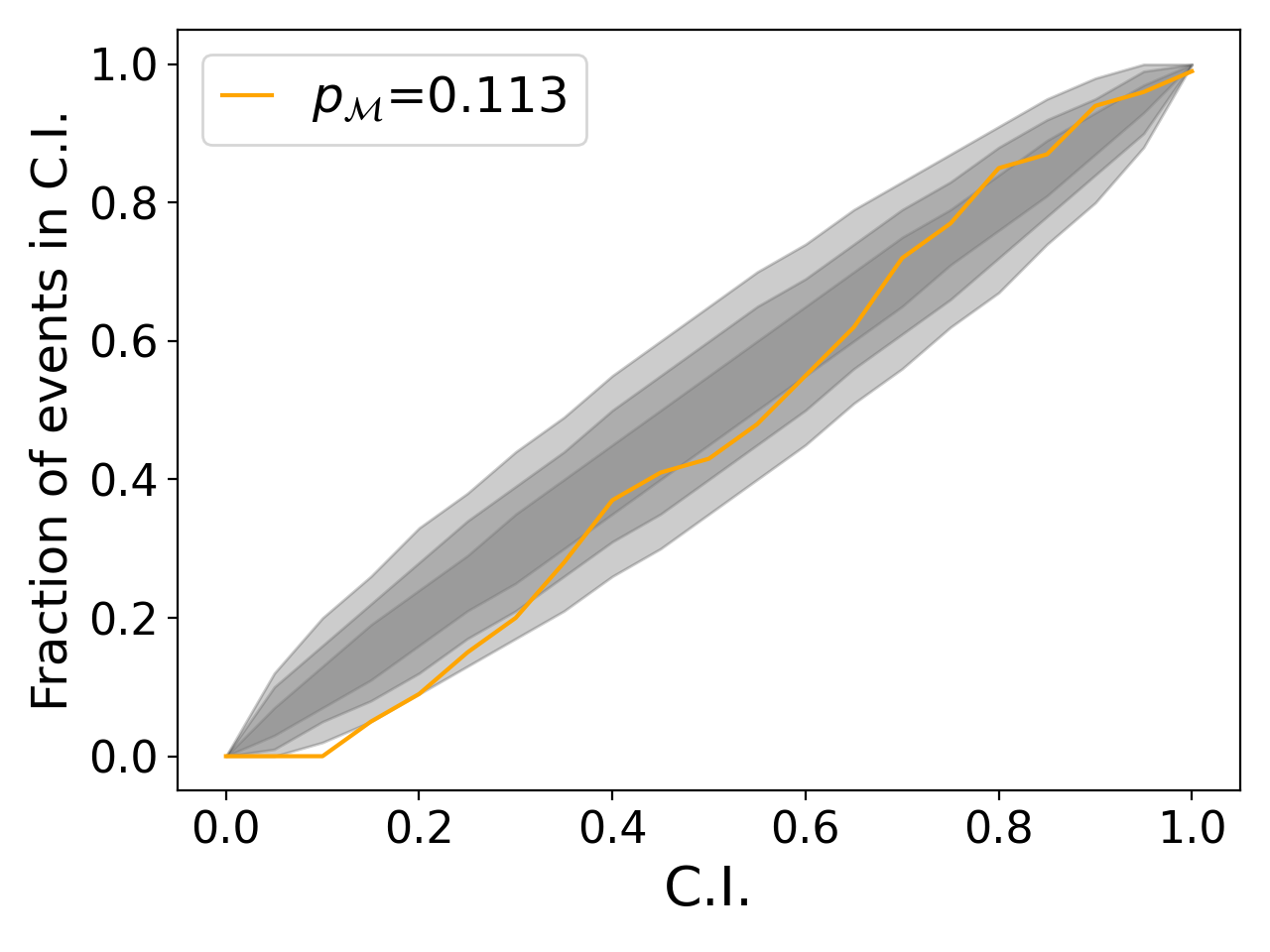}
    \includegraphics[width=0.44\linewidth]{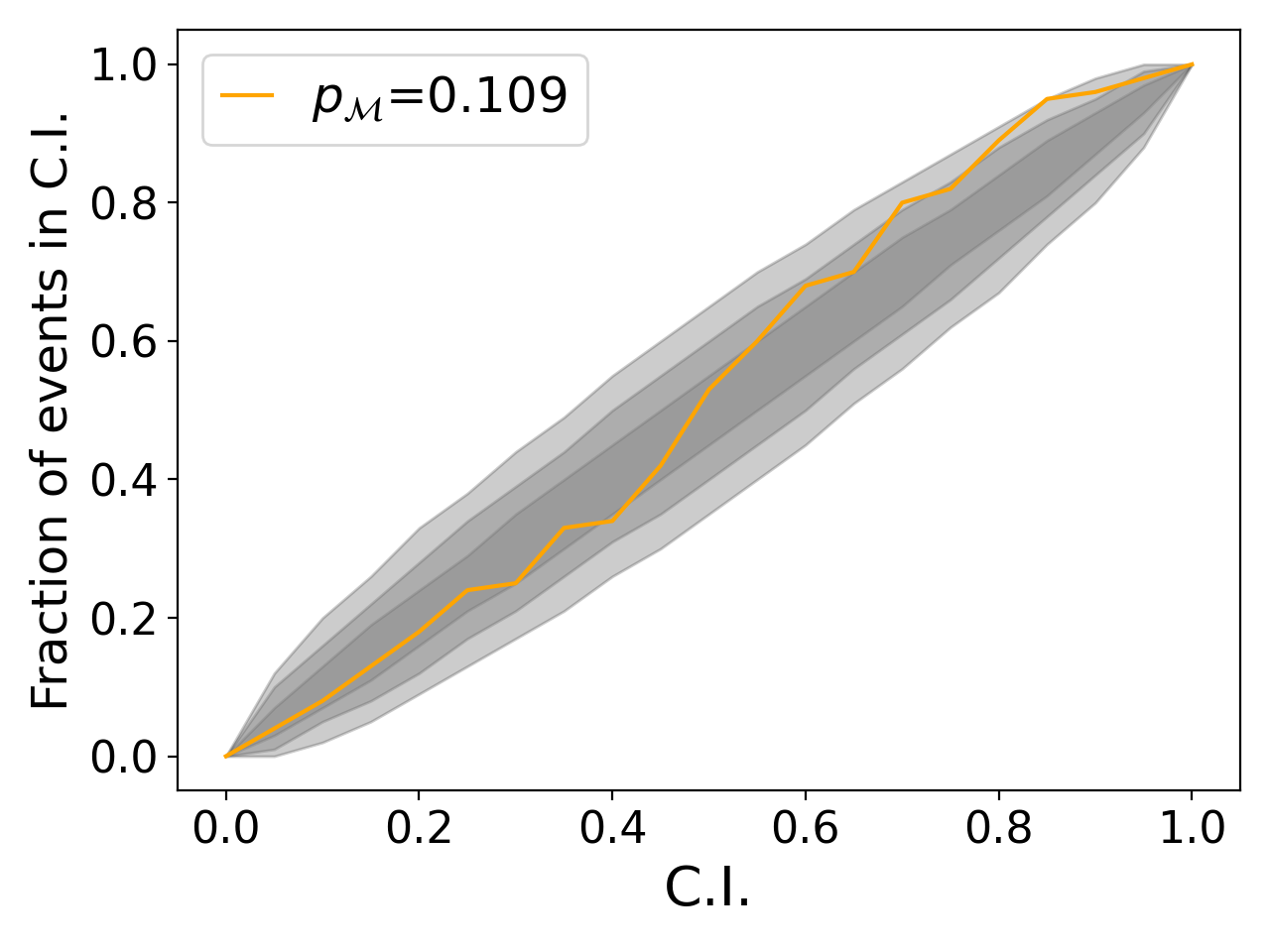}
    \caption{Probability–probability plot for BBH injections into observational data over 100 realizations. We use one RNLE likelihood model for all runs, drawing the chirp mass uniformly from $\mathcal{U}[26,30]$ for each injection and embedding each signal in a different detector-data segment. The results shown in the left panel are obtained with an \ac{RNLE} likelihood trained on 3000 data realizations, while for the results in the right panel we employed 5000 data pieces for the training.}
    \label{fig:real_data_ppplot}
\end{figure*}

To further assess the robustness and calibration of our implementation, we perform a probability–probability (PP) test~\cite{Romero-Shaw:2020owr}. We carry out one hundred independent analyses in which both the data segment used for the injection and the injected chirp-mass value are varied, while keeping the trained \ac{RNLE} likelihood fixed. The chirp mass values are drawn from a uniform prior $\mathcal{U}[26,30]$, and each injection is analyzed using the same inference settings. For this test, the RNLE likelihood is first trained on 3000 noise realizations drawn from data surrounding the segment used for the observation analyzed in Figure~\ref{fig:benchmark_gw_1D}. Figure~\ref{fig:real_data_ppplot} shows the resulting PP plot, where the $x$-axis denotes the nominal credibility level and the $y$-axis indicates the fraction of events for which the true parameter value lies within the corresponding posterior credible interval. For a statistically well-calibrated inference procedure, the PP curve is expected to follow the diagonal. \\
The PP plot obtained with 3000 training realizations passes the calibration test, yielding a p-value of $p_{\mathcal{M}} = 0.113$. However, the curve exhibits a mild S-shaped deviation from the diagonal, indicative of slightly underconstrained posteriors. Such behavior is a known feature of simulation-based inference methods when the learned likelihood is not fully converged. To investigate whether this effect is driven by limitations in the training dataset, we repeat the analysis using an RNLE likelihood trained on an expanded set of 5000 noise realizations. The corresponding PP plot also passes the calibration test, with a p-value of $p_{\mathcal{M}} = 0.109$, and shows a reduced S-shaped deviation. This improvement suggests that increasing the number of training realizations enhances the stability of the learned likelihood and leads to better-calibrated posterior distributions.

\section{Analysing BBH injections in extremely non-Gaussian data segments}
\label{sec:glitches_extreme}
We have demonstrated that \ac{RNLE} recovers results consistent with the Whittle likelihood in all scenarios where the assumptions of stationarity and Gaussianity remain valid. We now turn to testing the performance of our algorithm in regimes where the Whittle likelihood is known to produce biased results. To this end, we select two segments of detector data containing loud glitches (\ac{SNR}$>20$), inject a gravitational-wave signal on top of each glitch, and attempt to recover the injected chirp mass using \texttt{dynesty} in combination with a likelihood obtained via \ac{RNLE}, while keeping all other analysis settings fixed. As in the previous sections, we restrict the inference to a single varying parameter in order to isolate the effect of non-Gaussian noise and clearly demonstrate the viability of the RNLE approach. \\
Although we focus here on a one-dimensional parameter space for clarity, the method is in principle applicable to higher-dimensional analyses. This expectation is motivated by the fact that the \ac{RNLE} training procedure depends only on the noise properties of the data and is independent of the signal model, as discussed in the previous section. As a result, extending the inference to additional signal parameters would not require retraining the likelihood. Nevertheless, demonstrating the practical performance and scalability of the method in higher-dimensional signal parameter spaces remains an important direction for future work.

\subsection{Identification of a highly non-Gaussian data segment}\label{subsec:id_non_Gaussian}
\begin{figure*}[htp!]
    \centering
    \includegraphics[width=\textwidth]{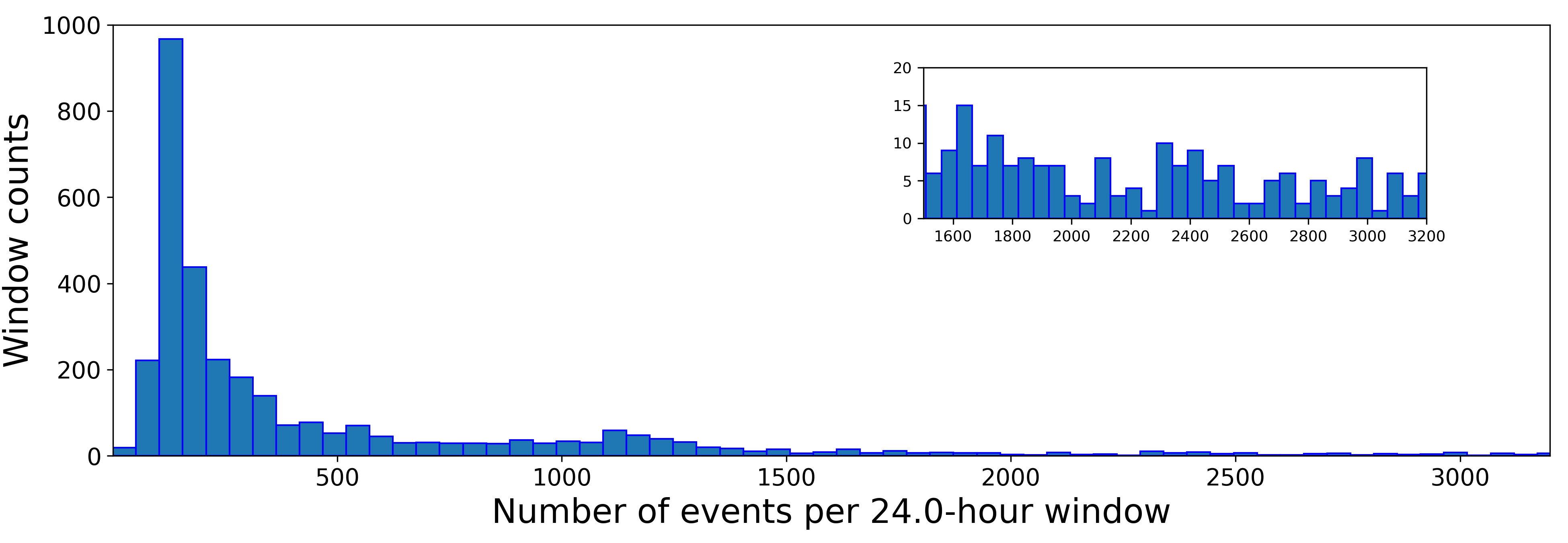}
    \caption{Histogram of the number of 24-hour windows for each glitch count value in \ac{LHO} during \ac{O3b}. Glitch counts are computed using a 24-hour sliding window with a one-hour step, based on the Gravity Spy database. The horizontal axis shows the number of glitches within a 24-hour window, while the vertical axis reports the number of windows exhibiting a given glitch count. The inset highlights the windows with the highest glitch rates, from which the analysis window used in this work is selected.}
    \label{fig:glitch_frequency_O3b}
\end{figure*}

To identify suitable periods of elevated glitch activity, we analyze data from the \ac{LHO} during the \ac{O3b} using a 24-hour sliding window, shifted forward by one hour at each step. For each window, we quantify glitch occurrence by extracting the number of entries recorded in the \texttt{GravitySpy} database~\cite{Zevin:2016qwy, Glanzer:2022avx, Zevin:2023rmt}. Figure~\ref{fig:glitch_frequency_O3b} shows the resulting glitch counts as a function of time across the \ac{O3b}.

\begin{figure*}[htp!]
    \centering
    \includegraphics[width=0.47\textwidth]{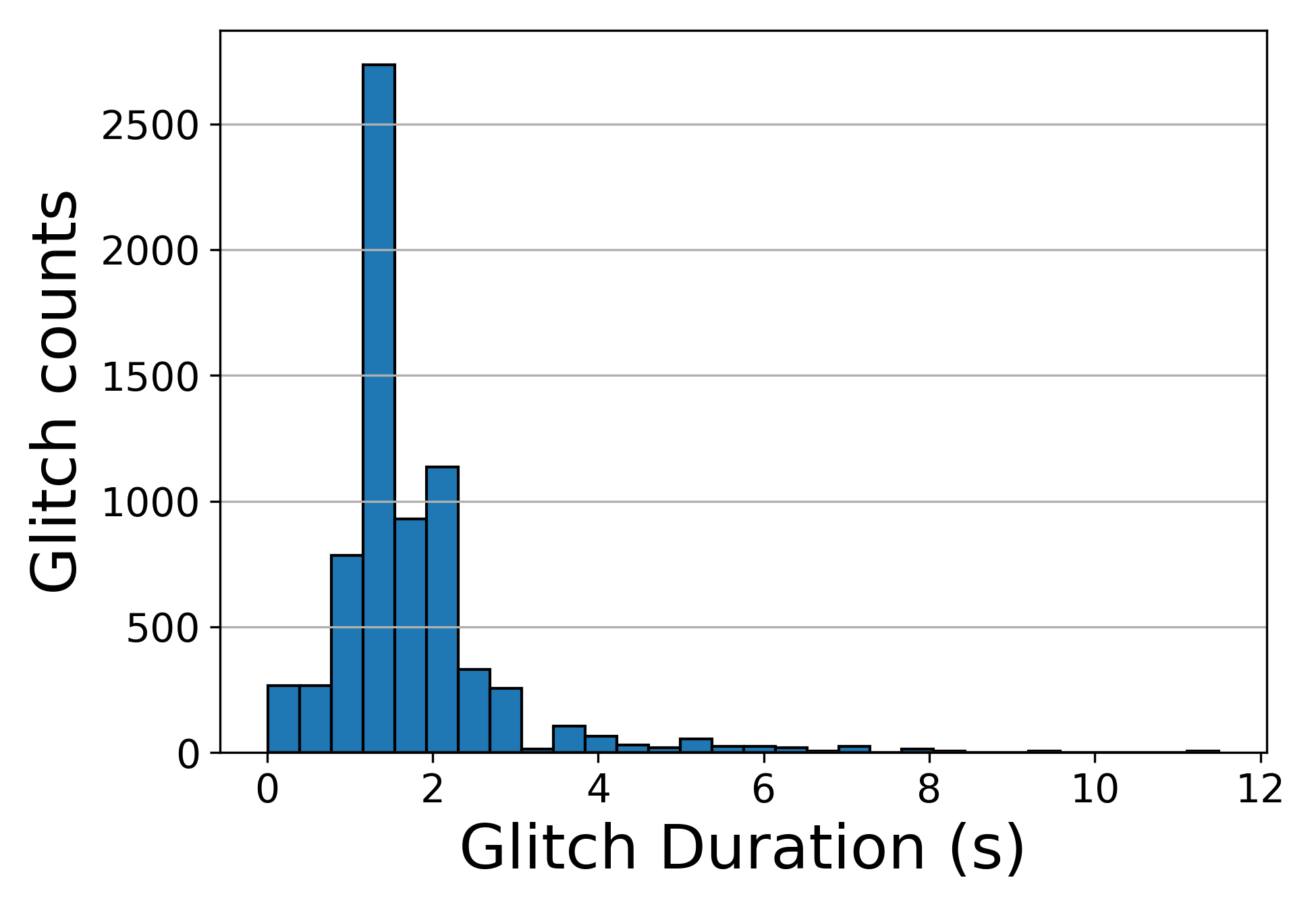}
    \includegraphics[width=0.47\textwidth]{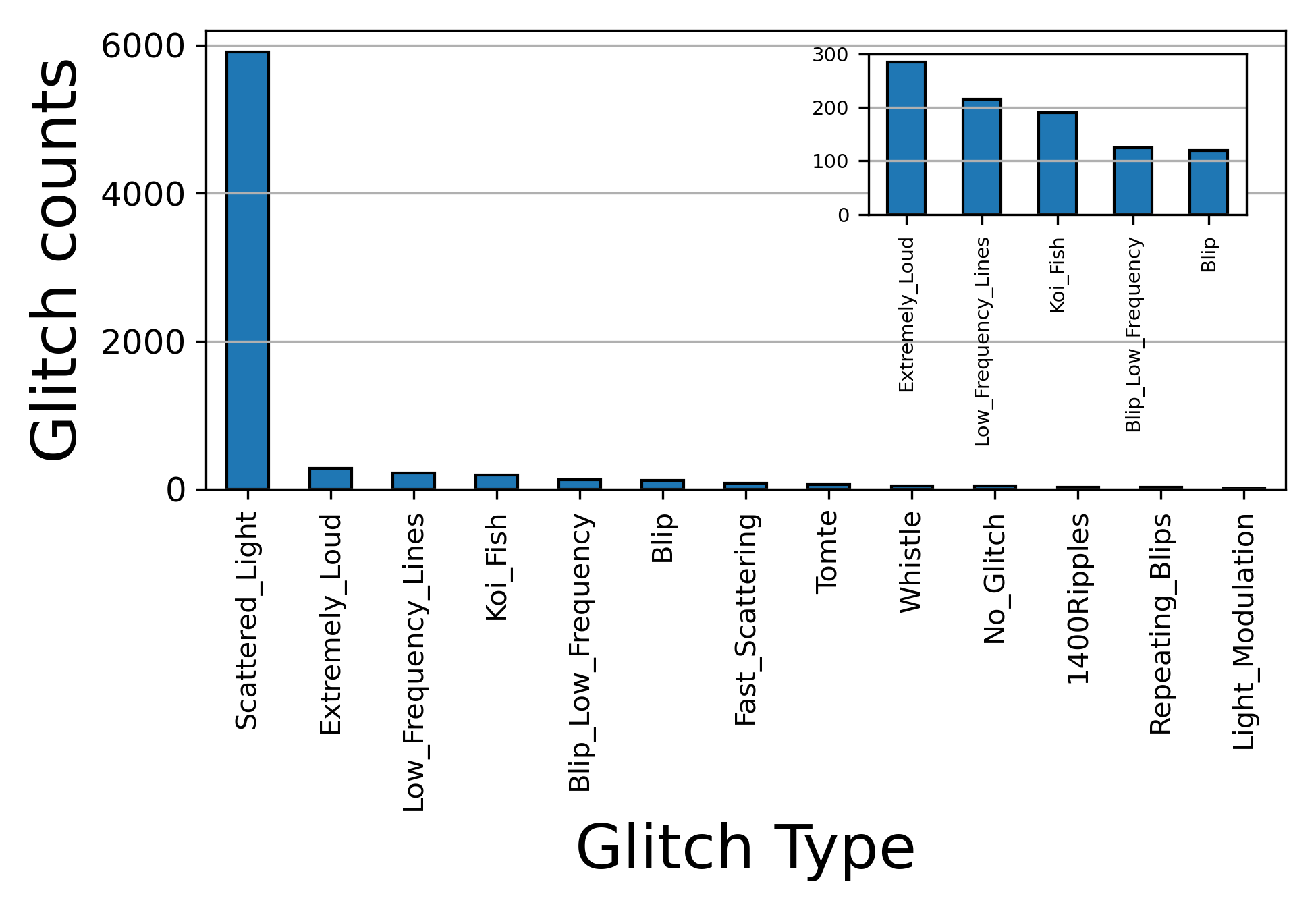}
    \includegraphics[width=0.47\linewidth]{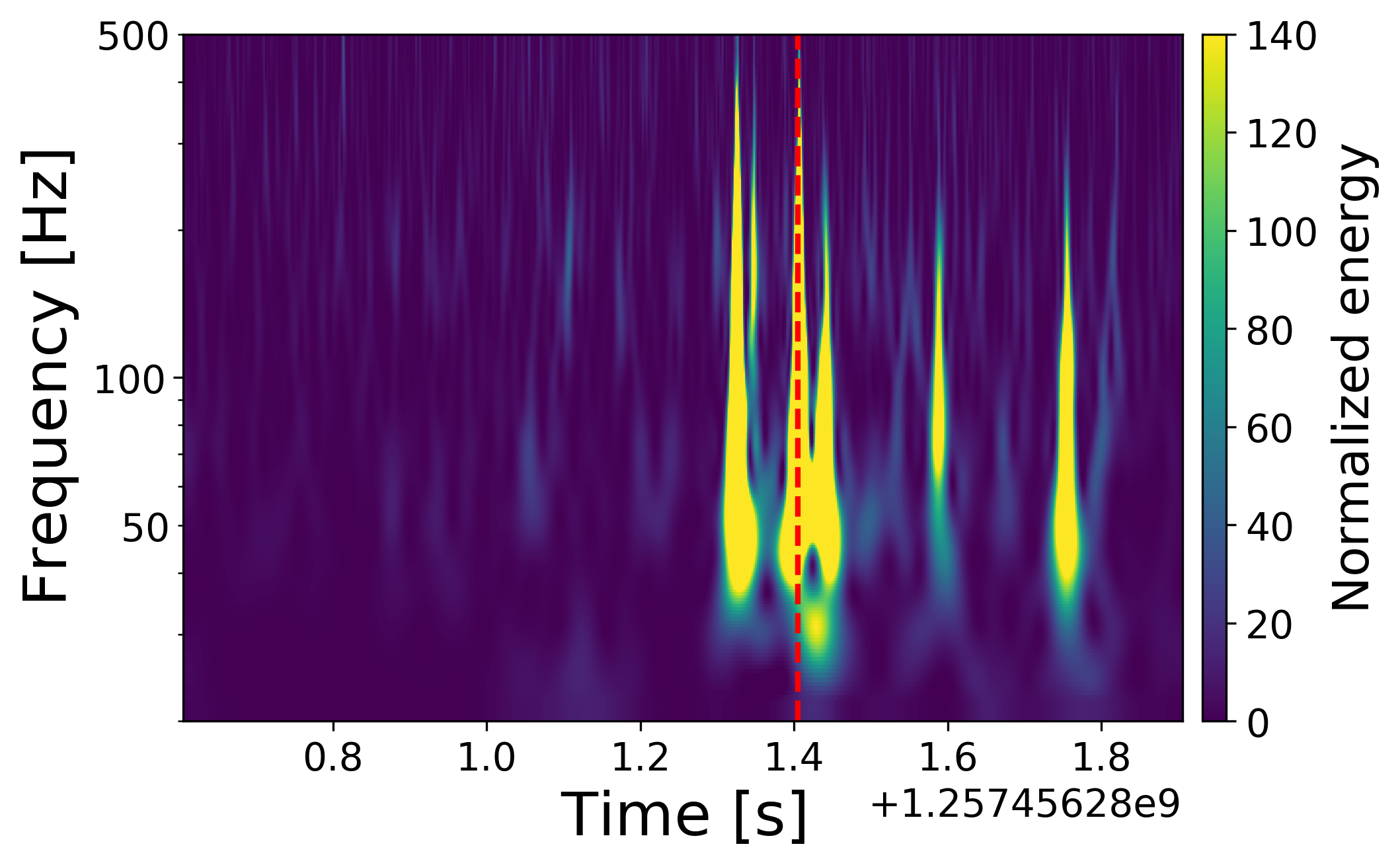}
    \includegraphics[width=0.47\linewidth]{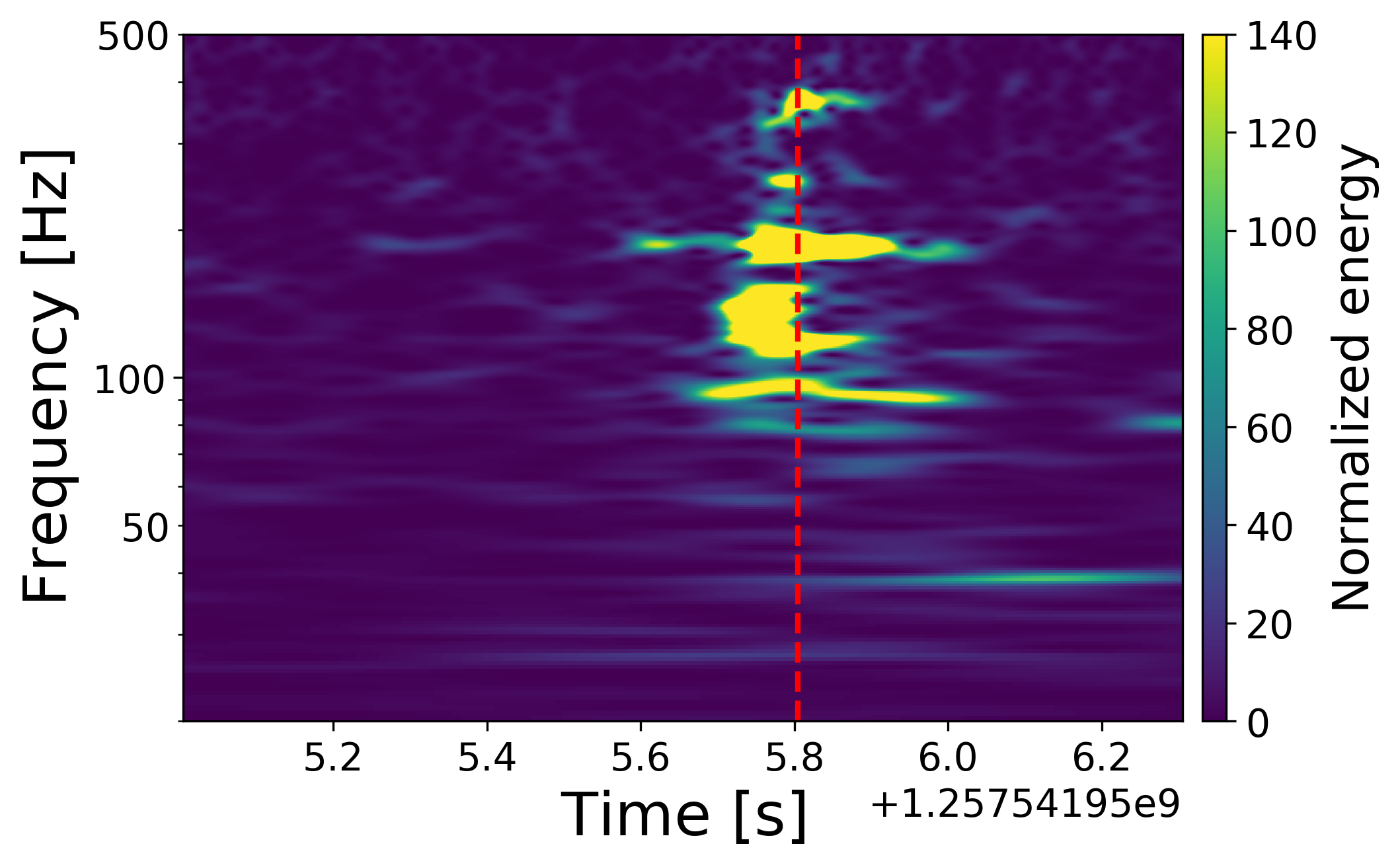}
    \caption{\textbf{Top panels}: Analysis of the properties of glitches in a 30-hour window in \ac{LHO} from \ac{O3b} around the two selected glitches. In the left panel the glitch duration versus the glitch count. In the right panel, a histogram of the glitch labels assigned by \texttt{GravitySpy}. The inset highlights the counts for the \textit{Koi fish} glitches, which correspond to the glitches analyzed in this study.  \textbf{Bottom panels}: Time-frequency plot of 1 second of data from the Hanford detector to visualize the glitches selected for the study. These glitches stand out for their long duration, high \ac{SNR}, and frequency content in the injected \ac{GW} signal's band. The red dotted vertical line indicates the trigger time of the injection, corresponding to the coalescence of the \ac{BBH} system and the moment when the $(2,2)$ mode of the waveform attains its maximum amplitude. }
    \label{fig:glitches_window_properties}
\end{figure*}

From this distribution, we investigate the properties of time windows associated with elevated glitch rates. Our goal is to analyze an extreme scenario in which the inferred chirp mass is severely biased. For this reason, the time window used in our study is selected based on both the overall glitch rate and the tendency of individual glitches within that period to bias the inference, ensuring that the sample is representative of a strongly non-Gaussian noise environment. We identify the period between 9–12 November 2019 as fitting our requirements. The characteristics of the glitches identified by \texttt{GravitySpy} during this interval are shown in the top panels of Figure~\ref{fig:glitches_window_properties}.

Within this window, we concentrate on two particularly loud glitches occurring at GPS times 1257456281.4 and 1257541955.8, corresponding to 21:24:23 UTC on 10 November 2019 and 21:12:17 UTC on 11 November 2019, respectively. For clarity, we refer to these as Glitch A and Glitch B. Glitch A has an \ac{SNR} of 39.94, a duration of 2 s, and a central frequency of 360 Hz, while Glitch B has an \ac{SNR} of 44.37, a duration of 1.5 s, and a central frequency of 448 Hz. Time–frequency representations of both glitches are shown in Figure~\ref{fig:first_glitch}. Although \texttt{GravitySpy} classifies both glitches as \textit{Koi fish}, their morphology is more complex than typical examples of this category. As seen in the top panels of Figure~\ref{fig:glitches_window_properties}, these events also correspond to unusually long durations compared to other glitches in this period, and nearly 200 \textit{Koi fish} glitches occur within the selected time window.

\subsection{Analysing a BBH injection into a highly non-Gaussian data segment}
\begin{figure*}[htp!]
    \centering
    \includegraphics[width=0.46\linewidth]{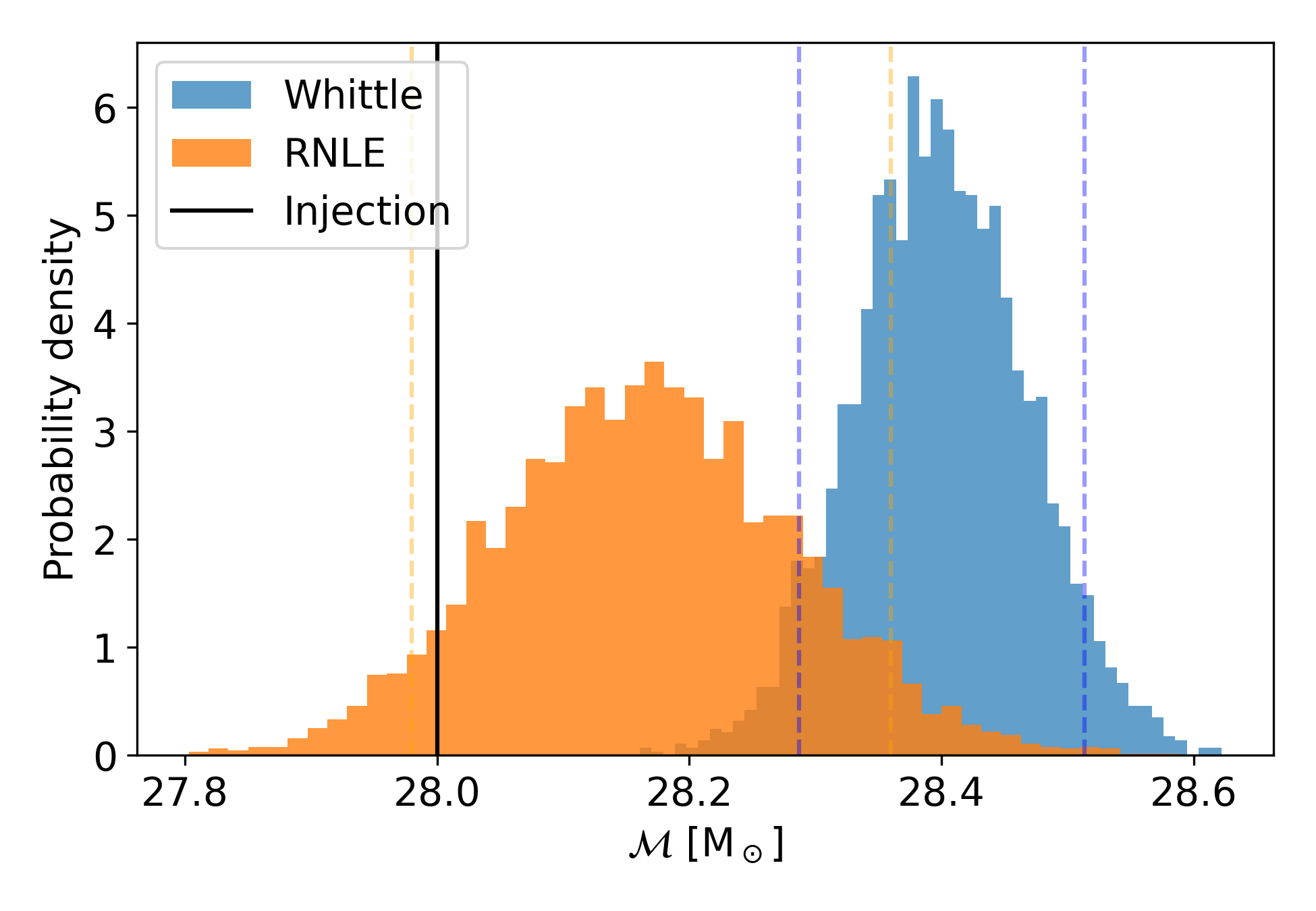}
    \includegraphics[width=0.47\linewidth]{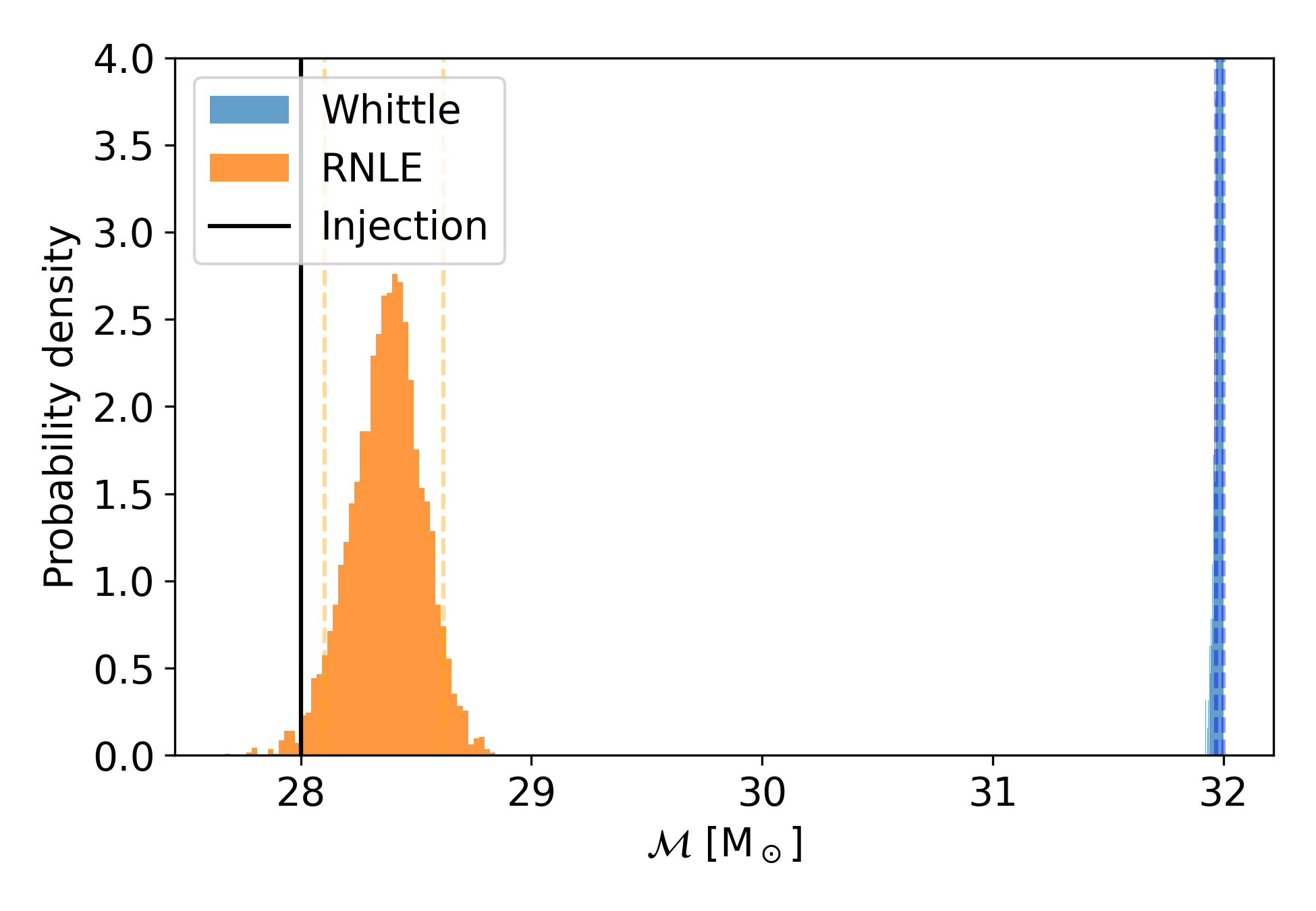}
    \caption{ Posterior probability distribution of the chirp mass obtained running \texttt{Bilby} with the \texttt{dynesty} sampler using the Whittle likelihood, in blue, and the \ac{RNLE} likelihood, in orange. The dotted lines indicate the 90\% contours and the black vertical line shows the injection value. The left panel shows the results for Glitch A and the right panel shows the same for Glitch B.}
    \label{fig:first_glitch}
\end{figure*}

We extract 3600 s of strain data around each glitch and process these segments following the procedure described in Section~\ref{sec:real_GW_quasi}. This yields one simulation per second of data, providing a dataset from which we train an \ac{RNLE} likelihood for each glitch independently. We then inject a \ac{GW} signal, using the same source parameters as in Section~\ref{sec:simulated_GW}, into the data segment containing the glitch. The trigger time of the injection is shown as a red dashed vertical line in the lower panels of Figure~\ref{fig:first_glitch}. Throughout this discussion, we define the offset as the relative temporal separation between the glitch and the injected \ac{GW} signal. A positive offset therefore, indicates that the glitch occurs before the signal, whereas a negative offset means that it occurs after the signal. We take the time of the glitch to be the event time provided by \texttt{GravitySpy}. The simulated signal is always placed at the trigger time, i.e., 0.8 seconds after the start of the analyzed segment and 0.2 seconds before its end. 

The posterior distributions shown in the bottom panels of Figure~\ref{fig:first_glitch} correspond to the case of zero offset, in which the signal and glitch are temporally coincident. We display the chirp-mass posterior recovered with the Whittle likelihood in blue and the posterior obtained using the \ac{RNLE} likelihood in orange, for Glitch A, left panel, and Glitch B, right panel. In this analysis, the chirp mass is the only free parameter. For both glitches, the \ac{RNLE} likelihood yields a posterior distribution that more accurately reflects the injected chirp mass compared to the Whittle likelihood. For Glitch A, the injected value, vertical black line, lies within the 90\% credible interval, dotted lines. For Glitch B, although the 90\% interval does not fully contain the injected value, the full posterior contains it, while the Whittle posterior rails against the upper prior bound. These results demonstrate that \ac{RNLE} can enable accurate parameter estimation in the presence of non-Gaussian noise. However, the performance depends on several factors, including the chosen dataset, the relative timing between the glitch and the signal, the signal-to-noise ratios of both, and the number of simulations used during training.

\section{Analysing BBH injections around a Blip glitch}
\label{sec:blip_glitch}
We want to study the performance of the \ac{RNLE} algorithm in the presence of glitches while varying the training dataset and the position of glitch vs signal and compare results to state-of-the-art deglitching algorithms, like \texttt{BayesWave}. Therefore, we follow a similar approach to~\citet{Ghonge:2023ksb}. They have injected gravitational wave signals on and around various types of glitches and performed parameter estimation analyses before and after deglitching the data using the \texttt{BayesWave} algorithm, finding that parameter estimation results are biased for injections within $<0.1$ s of the glitch. They analyzed injections at offsets between -0.075 and 0.075 s with a timestep of 0.025 s. We use the same offsets for our injections. We analyze one of the blip glitches from their study, for which \texttt{BayesWave} was not always able to unbias the posterior distributions. Blip glitches are characterized by their short duration, approximately $10\,\mathrm{ms}$, and broad frequency bandwidth, around $100\,\mathrm{Hz}$~\cite{Cabero:2019orq}. They are observed in both LIGO detectors, with an average occurrence rate of about two per hour. They are of particular interest because their morphology closely resembles the gravitational-wave signal from high-mass compact binary mergers, while their physical origin remains largely unknown.
\subsection{Identification of RNLE training datasets for the analysis of a blip glitch}\label{subsec:blip_glitch_datasets}
\begin{figure*}[t]
    \centering
    \includegraphics[width=0.44\linewidth]{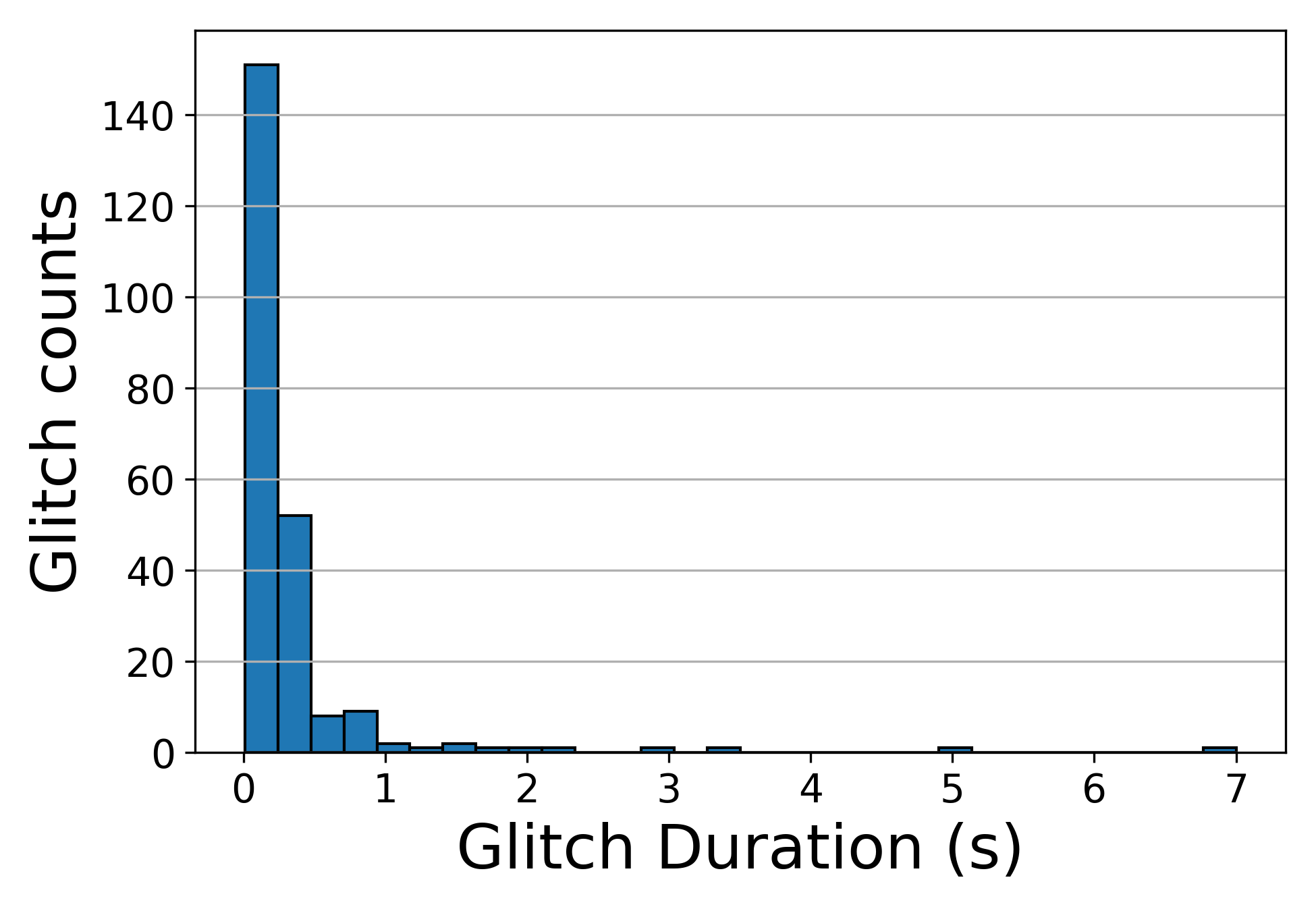}
    \includegraphics[width=0.44\linewidth]{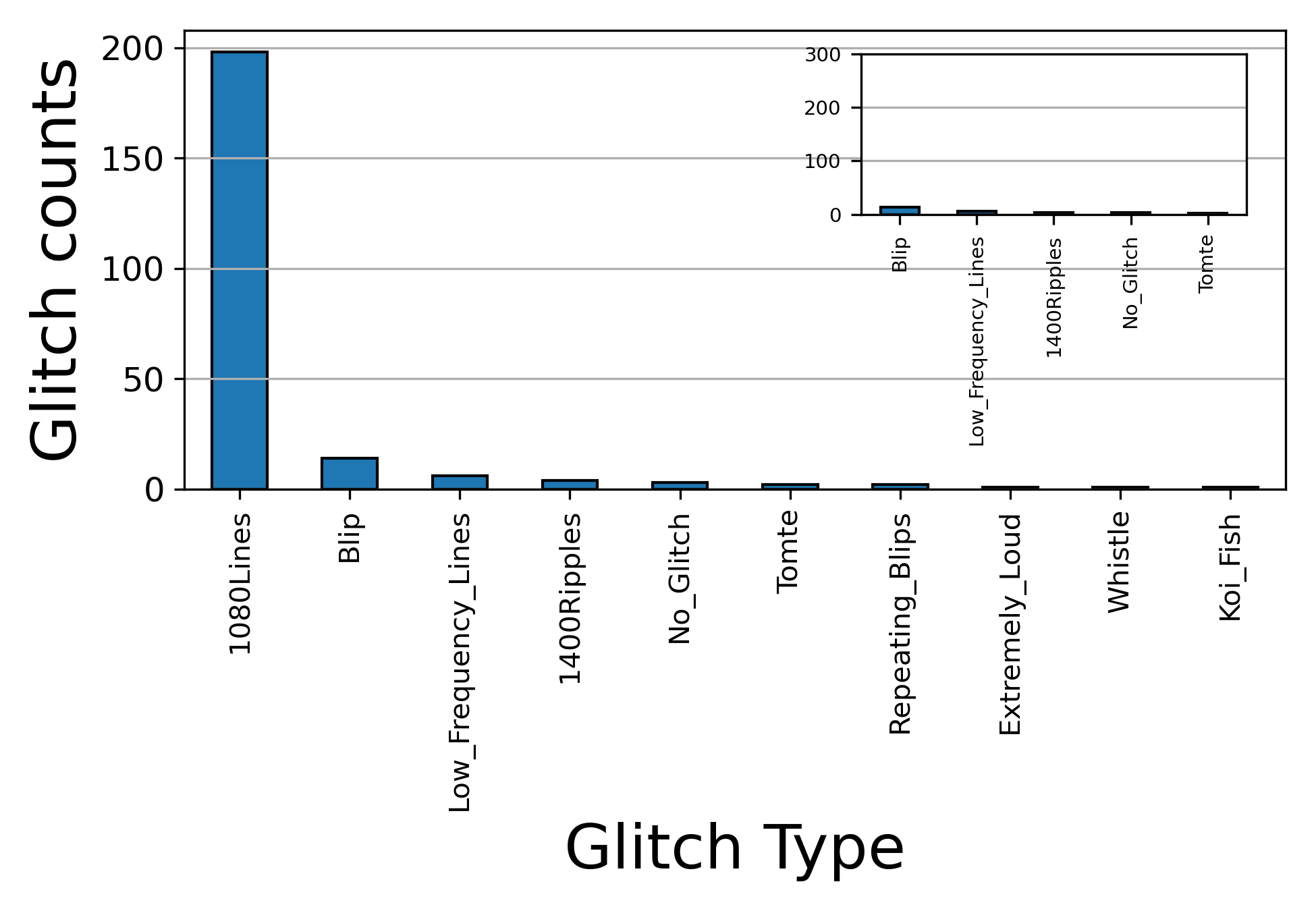}
    \includegraphics[width=0.44\linewidth]{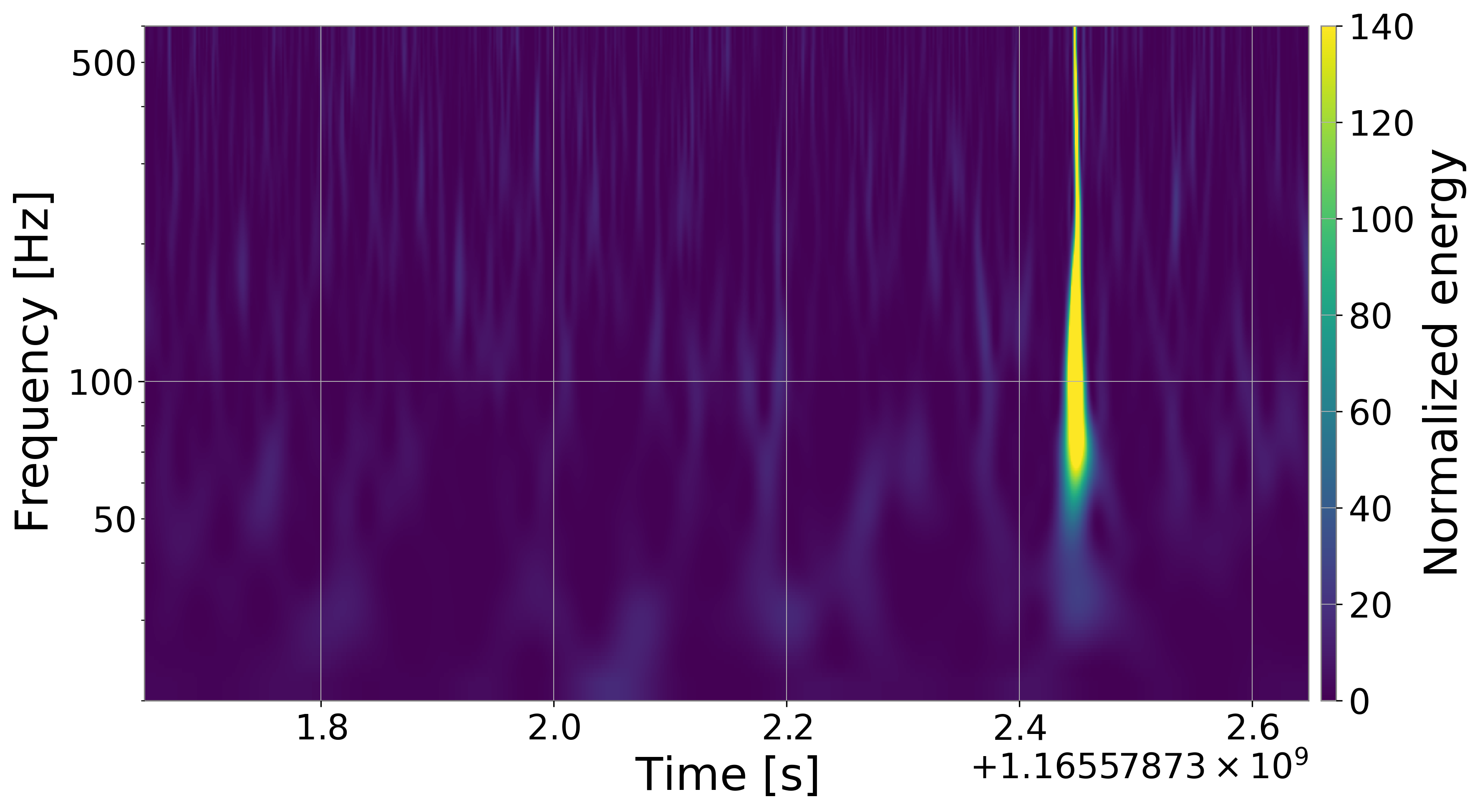}
    \includegraphics[width=0.44\linewidth]{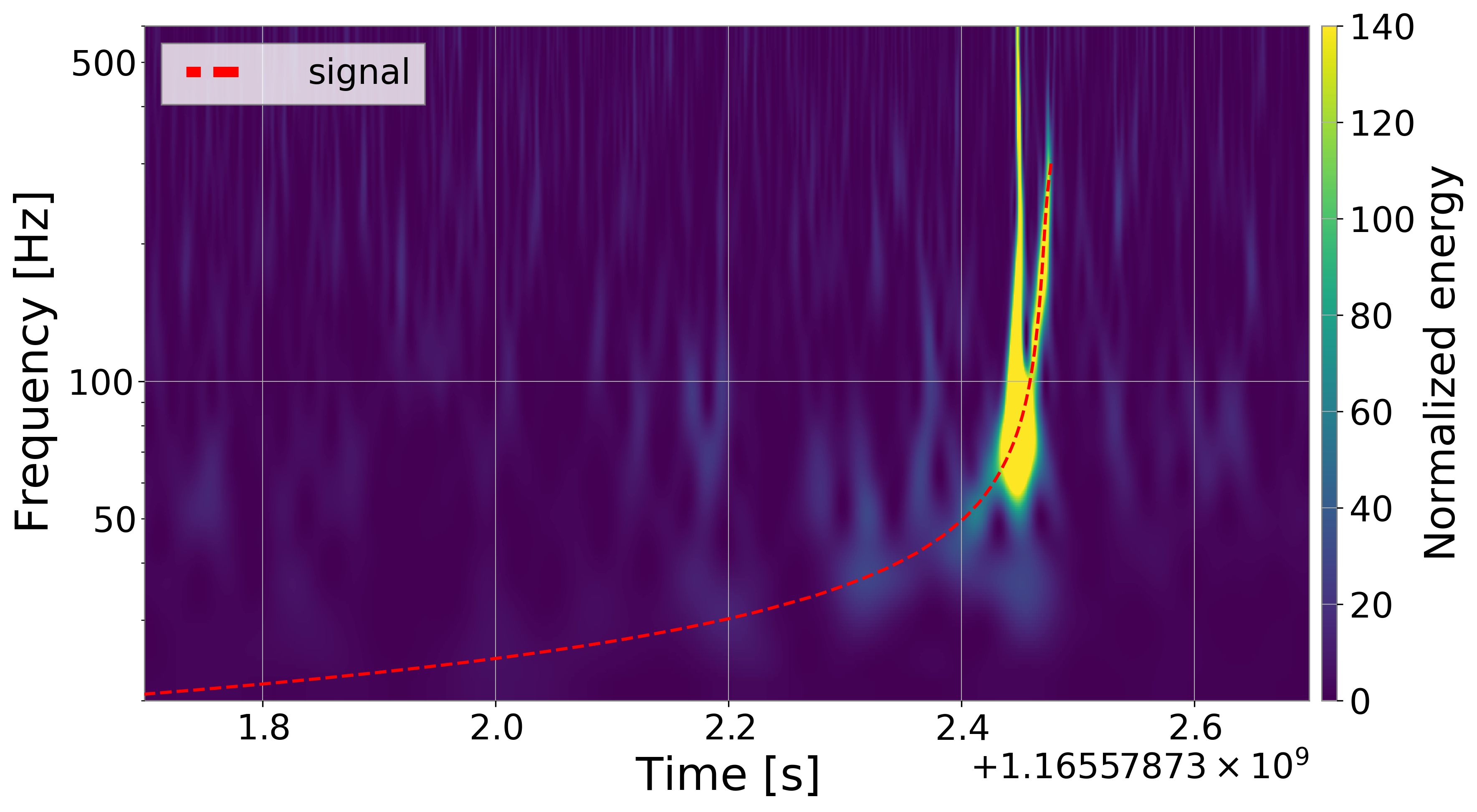}
    \caption{\textbf{Top panels}: Histograms of the glitch duration, left, and glitch type, right, for all glitches identified by \texttt{GravitySpy} in a 2-hour window centered on the selected blip glitch. \textbf{Bottom panels}: Time-frequency plots of the selected blip glitch, left, and the glitch and an injected \ac{GW} signal, right. In the latter, the red dotted line indicates the evolution of the signal. The offset for the glitch is 0.05 seconds and we double the signal's SNR with respect to the analysis so that it is clearly visible. }
    \label{fig:real_data_blip_glitch}
\end{figure*}

The analysis focuses on a blip glitch observed in the \ac{LHO} detector during the \ac{O2} observing run at GPS time 1165578732.45, corresponding to 11:51:55~UTC on 12 December 2016. The glitch exhibits a signal-to-noise ratio of 15.34, a duration of $0.25\,\mathrm{s}$, and a central frequency of $549\,\mathrm{Hz}$. The top panels of Figure~\ref{fig:real_data_blip_glitch} show, respectively, the distributions of glitch durations and glitch types obtained from \texttt{GravitySpy} within a 2-hour window centered on the selected event. The duration of our chosen glitch is representative of the local glitch population, and we note that only 
14 blip glitches are present in this interval. The bottom-left panel of Figure~\ref{fig:real_data_blip_glitch} displays the time–frequency representation of one second of data surrounding the glitch. The bottom-right panel shows the same data segment with an injected \ac{GW} signal, highlighted by the overplotted red dotted trajectory. {\hypersetup{hidelinks}\citeauthor{Ghonge:2023ksb}} found that this blip glitch can substantially bias parameter-estimation results and were unable to recover unbiased posteriors for all the tested offsets after glitch subtraction.

To study this behavior with the \ac{RNLE} likelihood, we analyze injections around the blip glitch at the same set of temporal offsets, with the goal of recovering accurate posterior distributions in all cases. For this purpose, we construct several datasets to train the conditional density estimator. These datasets, summarized in Table~\ref{tab:datasets}, differ in the number of simulations and in their defining characteristics. In contrast to previous analyses—where we restricted ourselves to simulations based solely on data adjacent to the analyzed segment—here we develop two classes of training sets: (i) datasets containing data drawn from the vicinity of the observed glitch time, and (ii) datasets in which each simulated example includes a blip glitch taken from the full \texttt{GravitySpy} repository for \ac{LHO} during \ac{O2}. This strategy is motivated by the fact that the small number of local blip glitches might be insufficient for the network to learn their variability, whereas \ac{O2} contains thousands of labeled blip glitches that can be leveraged to provide a more robust and representative training distribution. To construct the glitch datasets we selected glitches with \ac{SNR}$>10$, $f_{\mathrm{peak}}>60$, and duration$>0.2$ s to maximize the similarity to the analyzed glitch.
 
\begin{table*}[t]
\caption[Dataset characteristics]{\label{tab:datasets}
Characteristics of the different datasets employed to train the conditional density estimator for \protect\ac{RNLE}. 
Selected blip glitches satisfy the criteria: duration $>0.2\,\mathrm{s}$, \protect\ac{SNR}$>10$, and $f_{\mathrm{peak}}>60\,\mathrm{Hz}$. 
Unless otherwise specified, glitches are taken from \protect\ac{O2}.}
\begin{tabular}{lll}
\hline
\textbf{Dataset label} & \textbf{Number of simulations} & \textbf{Characteristics} \\
\hline
GD1.0 & 9240 & Selected blip glitches with offset 0.0\,s \\
GD1.1 & 9240 & Selected blip glitches with offset $-0.025$\,s \\
GD1.2 & 9240 & Selected blip glitches with offset $-0.05$\,s \\
GD1.3 & 9240 & Selected blip glitches with offset $-0.075$\,s \\
GD1.4 & 9240 & Selected blip glitches with offset $0.025$\,s \\
GD1.5 & 9240 & Selected blip glitches with offset $0.05$\,s \\
GD1.6 & 9240 & Selected blip glitches with offset $0.075$\,s \\
GD2.0 & 184000 & Selected blip glitches with offset $\mathcal{U}[-0.1,0.1]$, \\
& &each taken 100 times \\
GD2.1 & 394990 & Selected blip glitches from \protect\ac{O2} and \protect\ac{O3a} \\
& & with offset $\mathcal{U}[-0.1,0.1]$, each taken 100 times \\
GD3.0 & 36751 & Selected blip glitches with discrete offsets at \\ 
& &$[-0.075,-0.05,-0.025,0.025,0.05,0.075]$ \\
T1    & 14300  & Seconds of data around the trigger time \\
\hline
\end{tabular}
\end{table*}

\subsection{Analysing a BBH injection around a blip glitch}\label{subsec:blip_glitch_analysis}
\begin{figure*}[t]
    \centering

    \begin{subfigure}{0.47\linewidth}
        \centering
        \includegraphics[width=\linewidth]{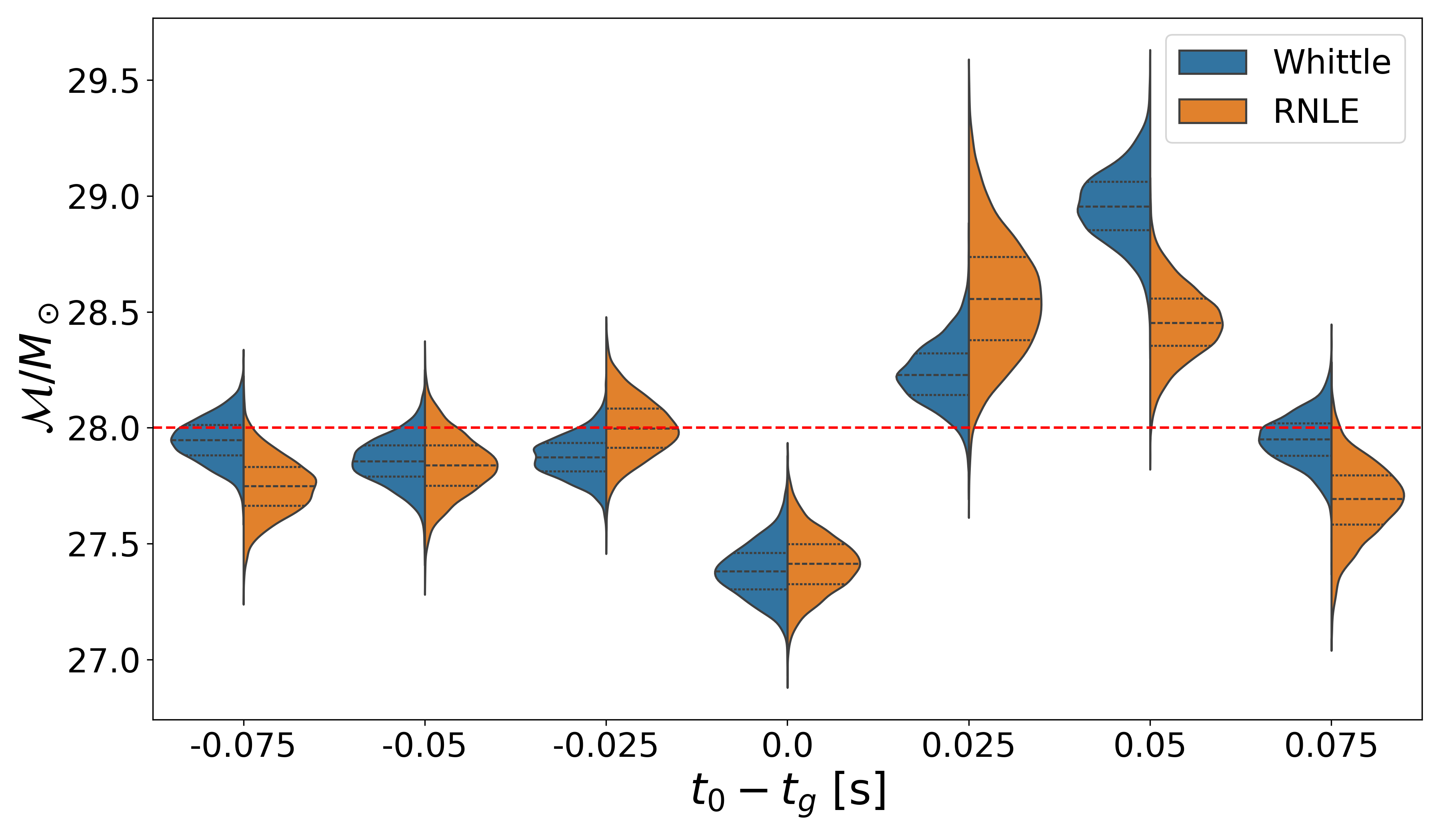}
        \caption{Results obtained using an \ac{RNLE} likelihood trained on the T1 dataset containing 14\,300 seconds of data around the blip glitch.}
        \label{fig:blip_A}
    \end{subfigure}
    \hfill
    \begin{subfigure}{0.47\linewidth}
        \centering
        \includegraphics[width=\linewidth]{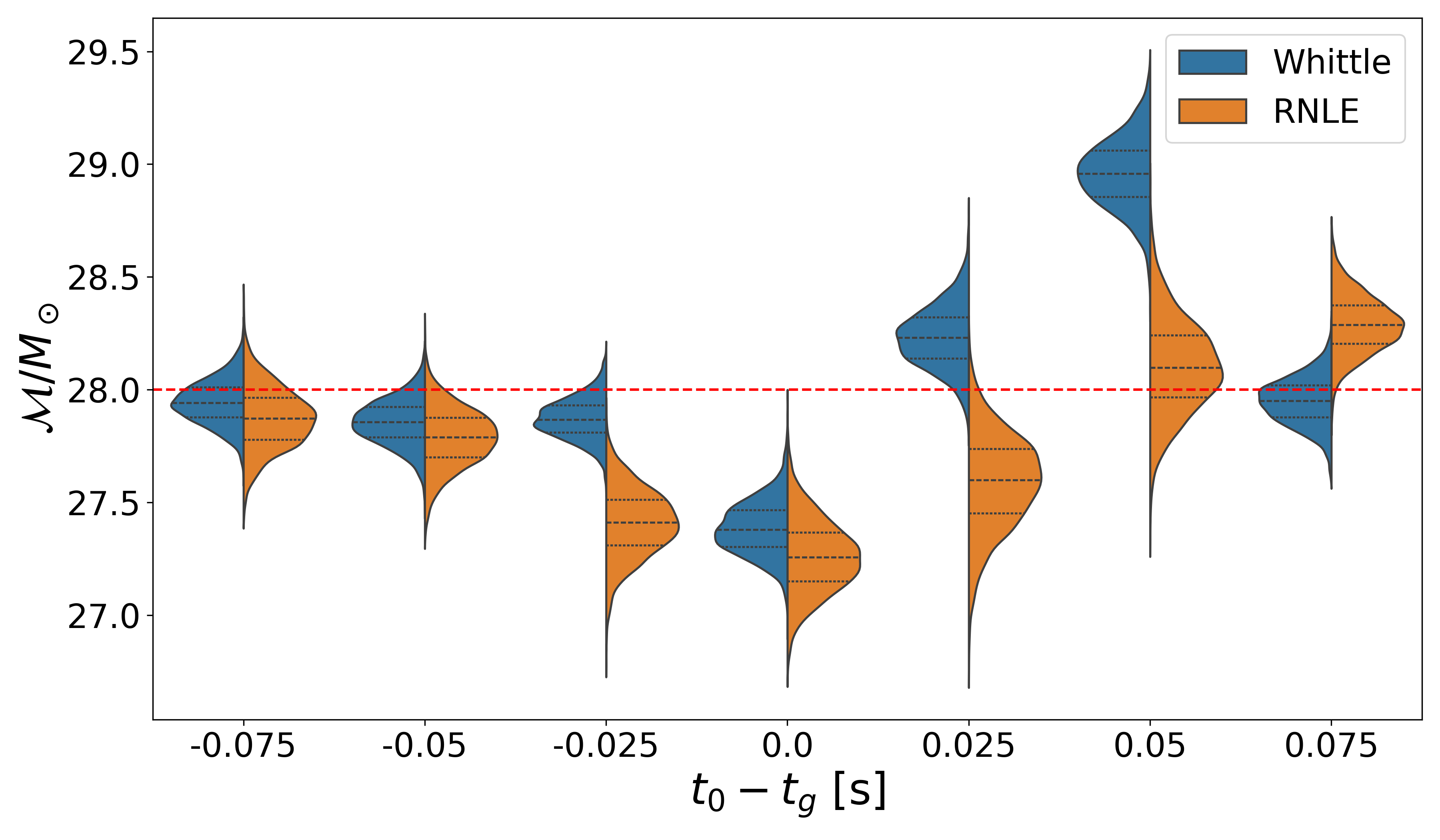}
        \caption{Results obtained with an \ac{RNLE} likelihood trained on the GD2.0 dataset containing all \ac{LHO} glitches from \ac{O2}, each taken one hundred times and repositioned using offsets sampled from $\mathcal{U}[-0.1,0.1]$.}
        \label{fig:blip_B}
    \end{subfigure}

    \begin{subfigure}{0.47\linewidth}
        \centering
        \includegraphics[width=\linewidth]{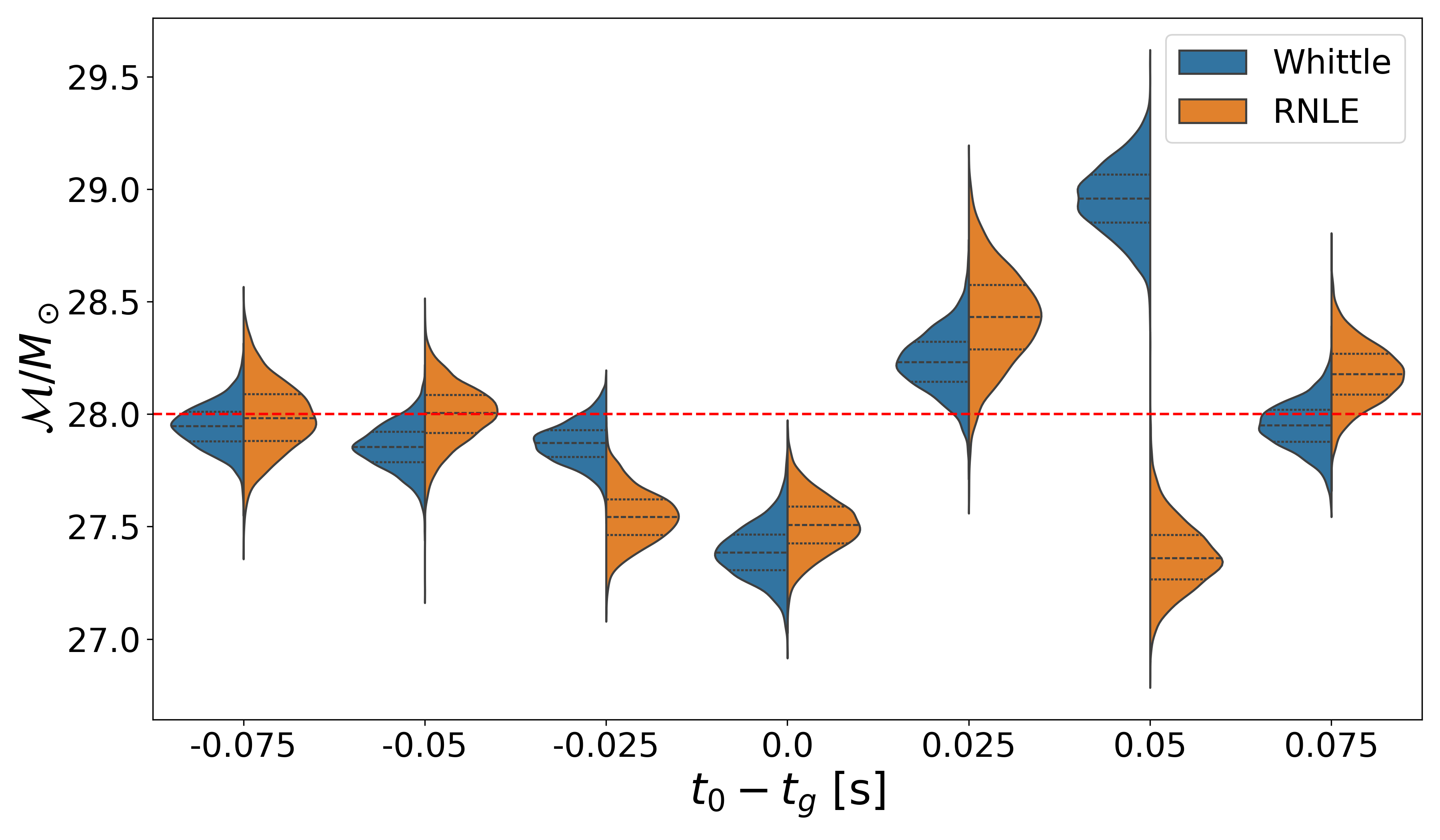}
        \caption{Results obtained with an \ac{RNLE} likelihood trained on the GD3.0 dataset containing all \ac{LHO} glitches from \ac{O2}, each taken ten times and positioned at discrete offsets.}
        \label{fig:blip_C}
    \end{subfigure}
    \hfill
    \begin{subfigure}{0.47\linewidth}
        \centering
        \includegraphics[width=\linewidth]{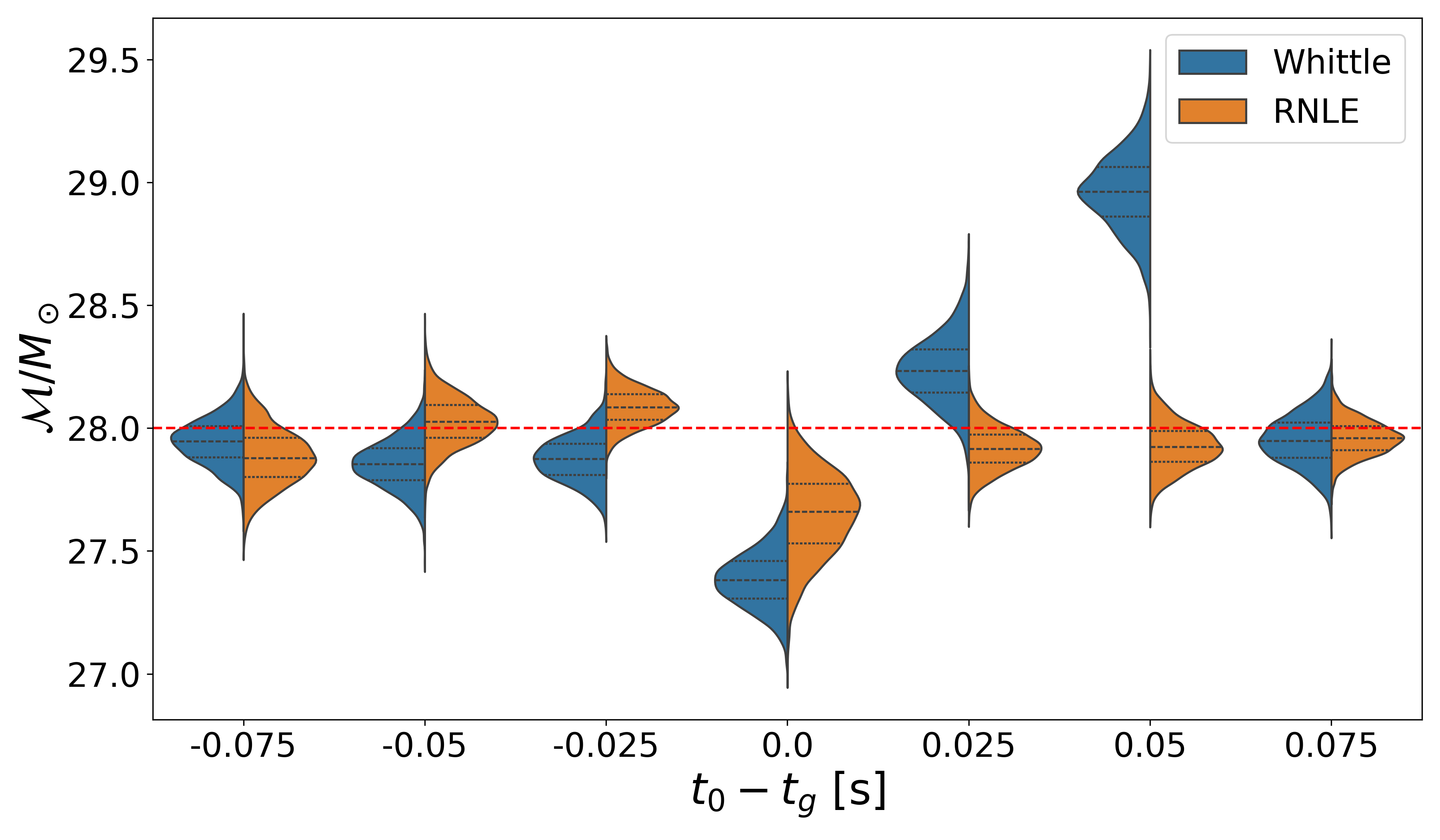}
        \caption{Results obtained with \ac{RNLE} likelihoods trained on the GD1 datasets (Table~\ref{tab:datasets}), which include all \ac{LHO} blip glitches from \ac{O2} always positioned at the offset time.}
        \label{fig:blip_D}
    \end{subfigure}

    \caption{
        Violin plots comparing Whittle (blue) and \ac{RNLE} (orange) results for various offsets between a blip glitch observed on 12 December 2016 and an injected \ac{GW} signal. Panels \textbf{A–D} correspond to the descriptions in the subcaptions.
    }
    \label{fig:blip_glitch_violin_plots}
\end{figure*}

Following the approach of {\hypersetup{hidelinks}\citeauthor{Ghonge:2023ksb}}, we conduct a standard set of seven analyses, varying the offset between the glitch and the injection. The \ac{SNR} of the injected \ac{BBH} signal is fixed at 12.88. Parameter estimation is carried out using both the Whittle likelihood and RNLE trained with different datasets.
We show the results using violin plots in Figure~\ref{fig:blip_glitch_violin_plots}. For each violin, the left side shows posterior distributions obtained with the Whittle likelihood, and the right side shows RNLE posteriors. The black dashed lines inside the violins represent the mean and 1-$\sigma$ credibility intervals for each posterior. The red dashed horizontal line represents the injection value. We again fix all parameters apart from the chirp mass. \\

For panel~\ref{fig:blip_A} in the top left, we use dataset T1, composed of 14300 s around the glitch, to train the conditional density estimator. We note that the most bias for the Whittle likelihood results is recovered for an offset of 0.0 and 0.05 s. Unfortunately, the RNLE likelihood is similarly biased for those offset values, while accurate results are recovered for every other offset. This is not unexpected, as the data around the blip glitch contains only a few other blip glitches, and we would not expect the estimator to capture their properties. \\
For panel~\ref{fig:blip_B} in the top right, we use dataset GD2.0 to train the conditional density estimator. This is composed of blip glitches, each taken 100 times and placed at a randomly sampled offset from the uniform distribution $\mathcal{U}[-0.1,0.1]$. While in this case the \ac{RNLE} likelihood recovers accurate results for the 0.05 s offset, i.e., the posterior distribution includes the true value, it recovers biased results for offsets for which the Whittle likelihood does not display any bias. \\
For panel~\ref{fig:blip_C} in the bottom left, we use dataset GD3.0, composed of blip glitches each taken 20 times and placed at discrete offsets coinciding with the offsets used for the observed data, to train the conditional density estimator. The resulting \ac{RNLE} likelihood recovers accurate posteriors for both the offsets 0.0 and 0.05 s. However, also in this case it recovers biased results for offset values for which the Whittle likelihood is not biased. \\
We attribute the oscillating behavior of these \ac{RNLE} likelihoods to the limited amount of training data and the non-convergence of the conditional density estimator.
To test this hypothesis, we create a new set of datasets, labeled as GD1 in Table~\ref{tab:datasets}. We create one dataset for each of the offset values used in our analysis. Each of them is composed of the same set of glitches from \ac{LHO} during \ac{O2} placed at the same offset, i.e., for GD1.0 all glitches are placed at a null offset. Using those datasets, we train seven different \ac{RNLE} likelihoods and use them to analyze the corresponding observed data realization, i.e., we use the likelihood obtained from GD1.0 to analyze the data with null offset. \\
Panel~\ref{fig:blip_D} shows the results obtained when each \ac{RNLE} likelihood trained on a given GD1 dataset is applied to the observed data realization with the matching offset. We recover accurate posteriors for all values of the offset, demonstrating that the approach is capable of learning an accurate likelihood even in the presence of non-Gaussian noise transients when the training and analysis conditions are closely matched. While this result validates the feasibility of the method, it is not a practical strategy for realistic analyses: training a separate likelihood for each specific glitch position would implicitly assume prior knowledge of the transient location, require a large number of independent trainings, and reduce the comparability and robustness of the inferred posteriors. This motivates the need for a single likelihood that can be applied consistently across different offsets and data realizations, minimizing its dependence on the training data and the estimator settings.

\section{Investigating training noise}\label{sec:training_noise}
As an initial step toward understanding how different sources of variability affect the performance of the \ac{RNLE} likelihood, we investigate the impact of training noise in the presence of blip glitches. Building on the approach described in Section~\ref{subsec:gw_1D_training_noise},
\begin{figure*}[htpb]
\centering

\begin{subfigure}{0.44\linewidth}
    \includegraphics[width=\linewidth]{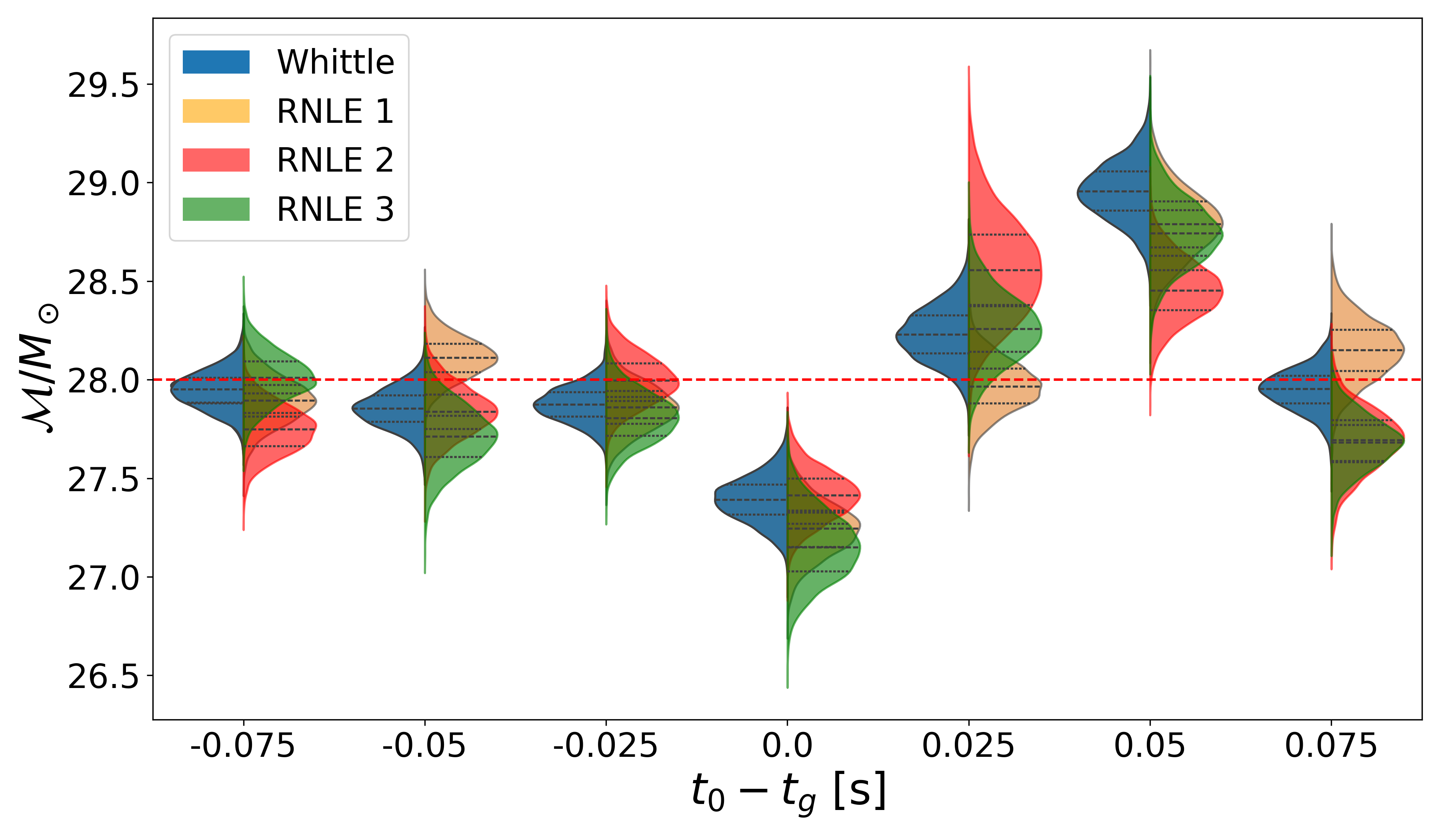}
    \caption*{(A1) RNLE posteriors — T1 dataset}
\end{subfigure}
\begin{subfigure}{0.44\linewidth}
    \includegraphics[width=\linewidth]{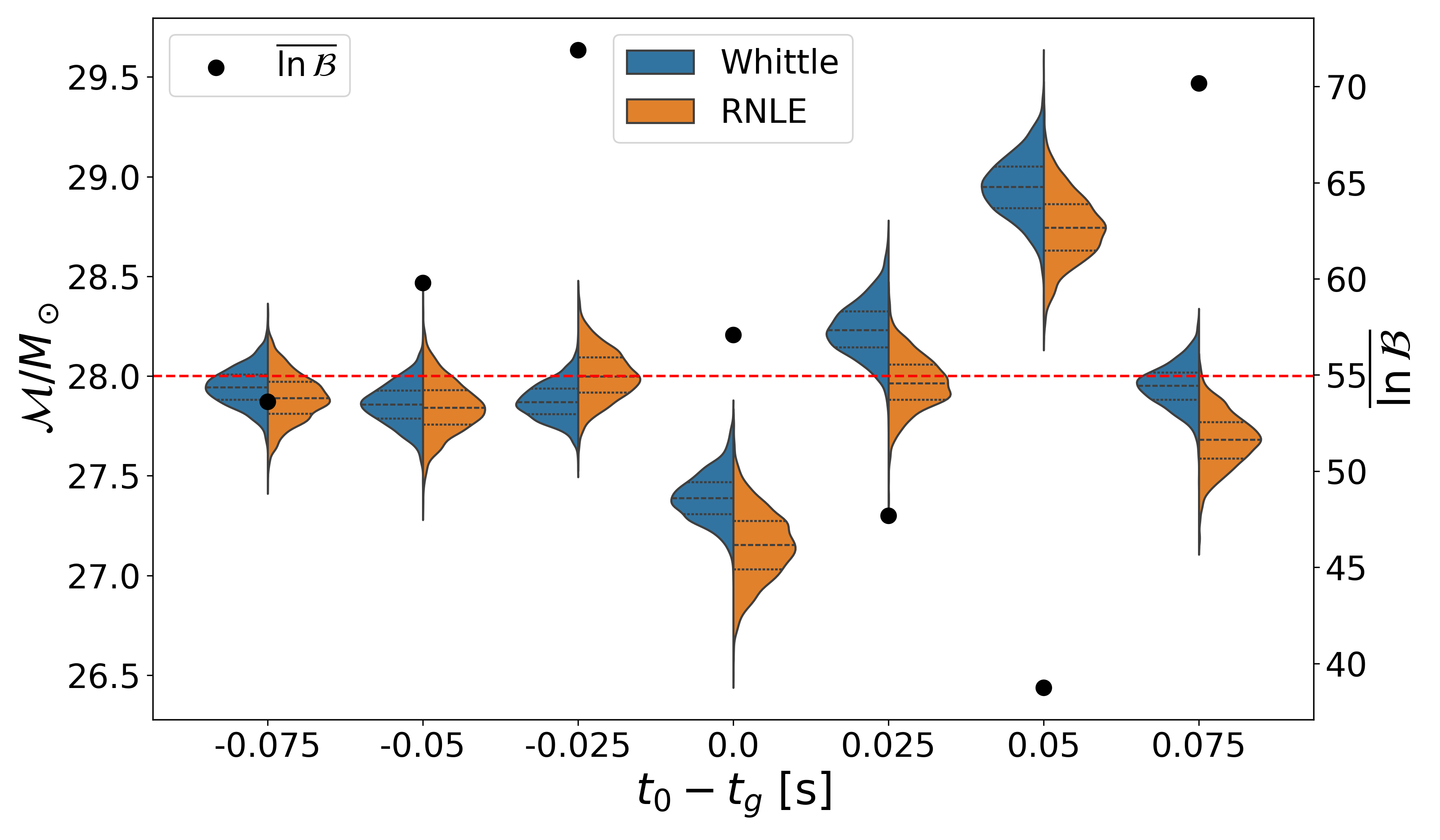}
    \caption*{(A2) Evidence–weighted posterior — T1 dataset}
\end{subfigure}

\begin{subfigure}{0.44\linewidth}
    \includegraphics[width=\linewidth]{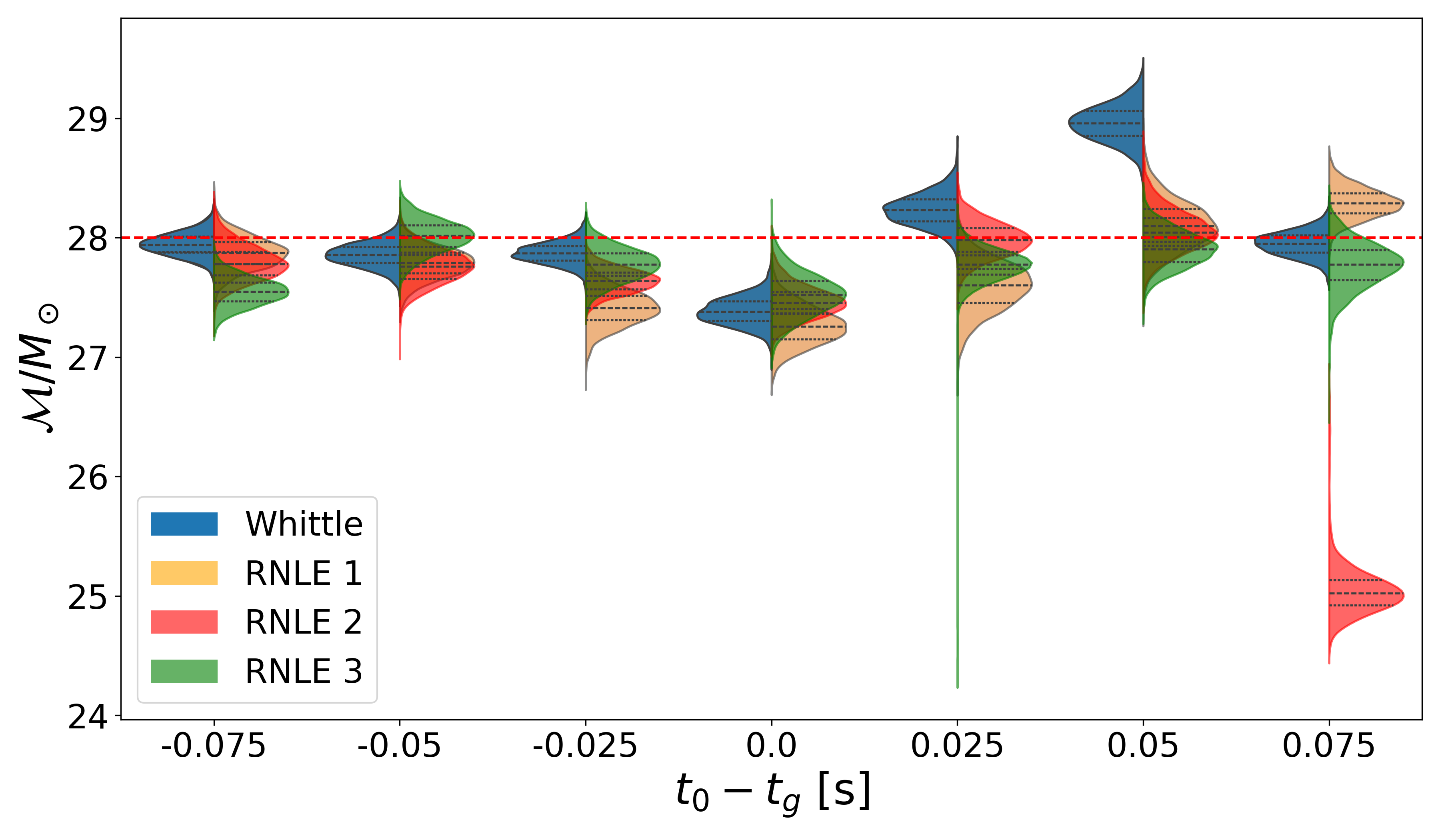}
    \caption*{(B1) RNLE posteriors — GD2.0}
\end{subfigure}
\begin{subfigure}{0.44\linewidth}
    \includegraphics[width=\linewidth]{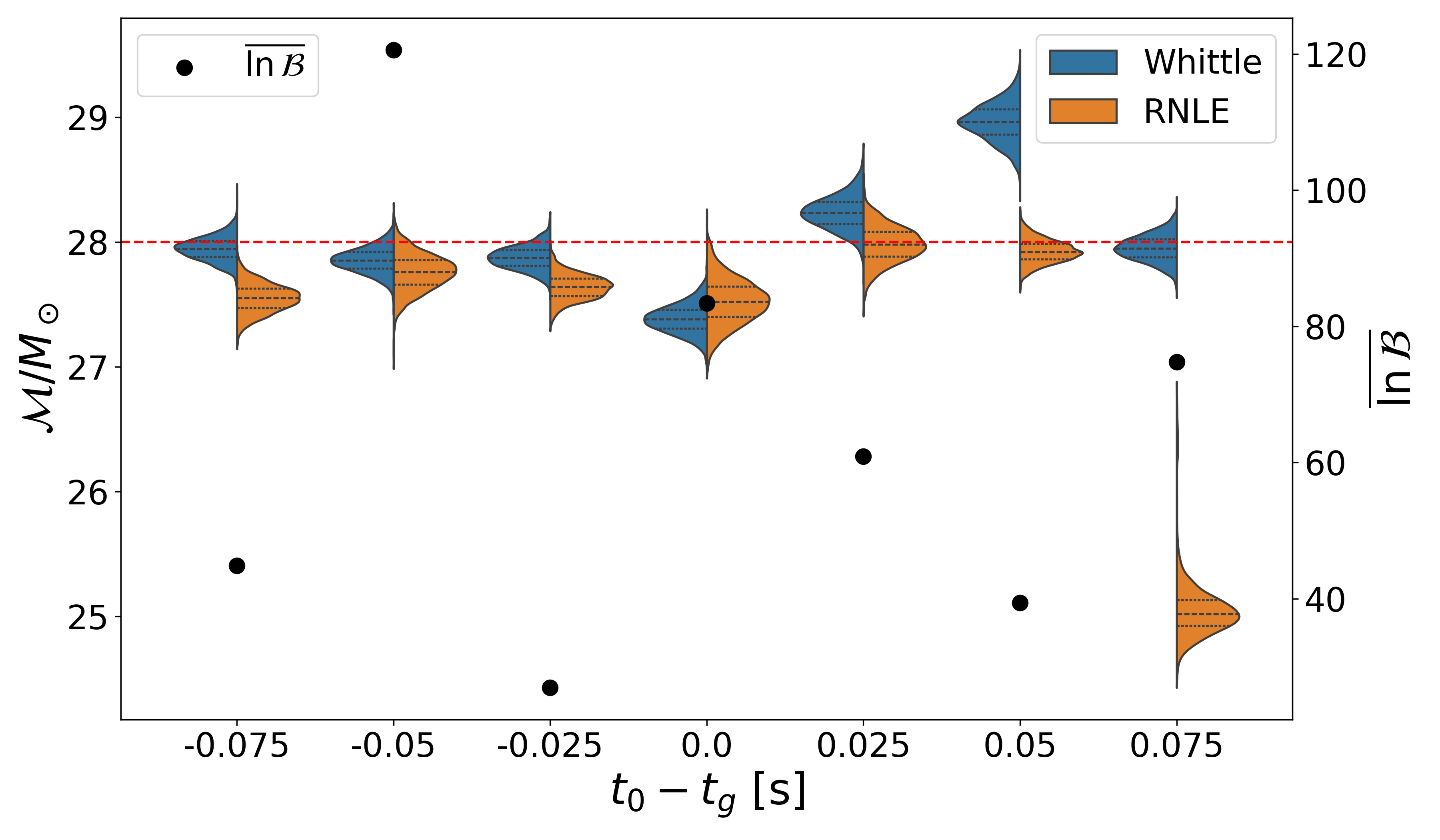}
    \caption*{(B2) Evidence–weighted posterior — GD2.0}
\end{subfigure}

\begin{subfigure}{0.44\linewidth}
    \includegraphics[width=\linewidth]{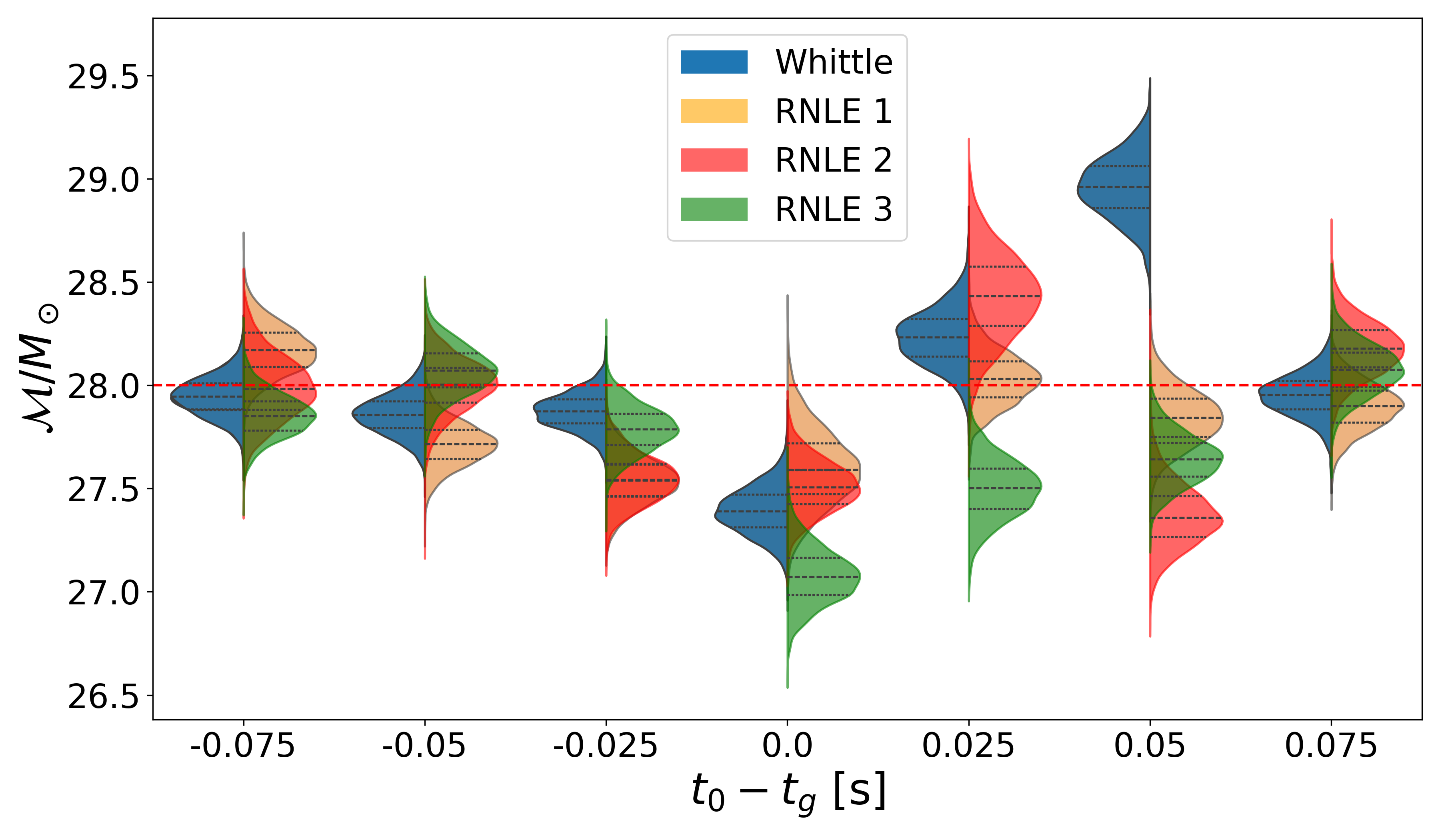}
    \caption*{(C1) RNLE posteriors — GD3.0}
\end{subfigure}
\begin{subfigure}{0.44\linewidth}
    \includegraphics[width=\linewidth]{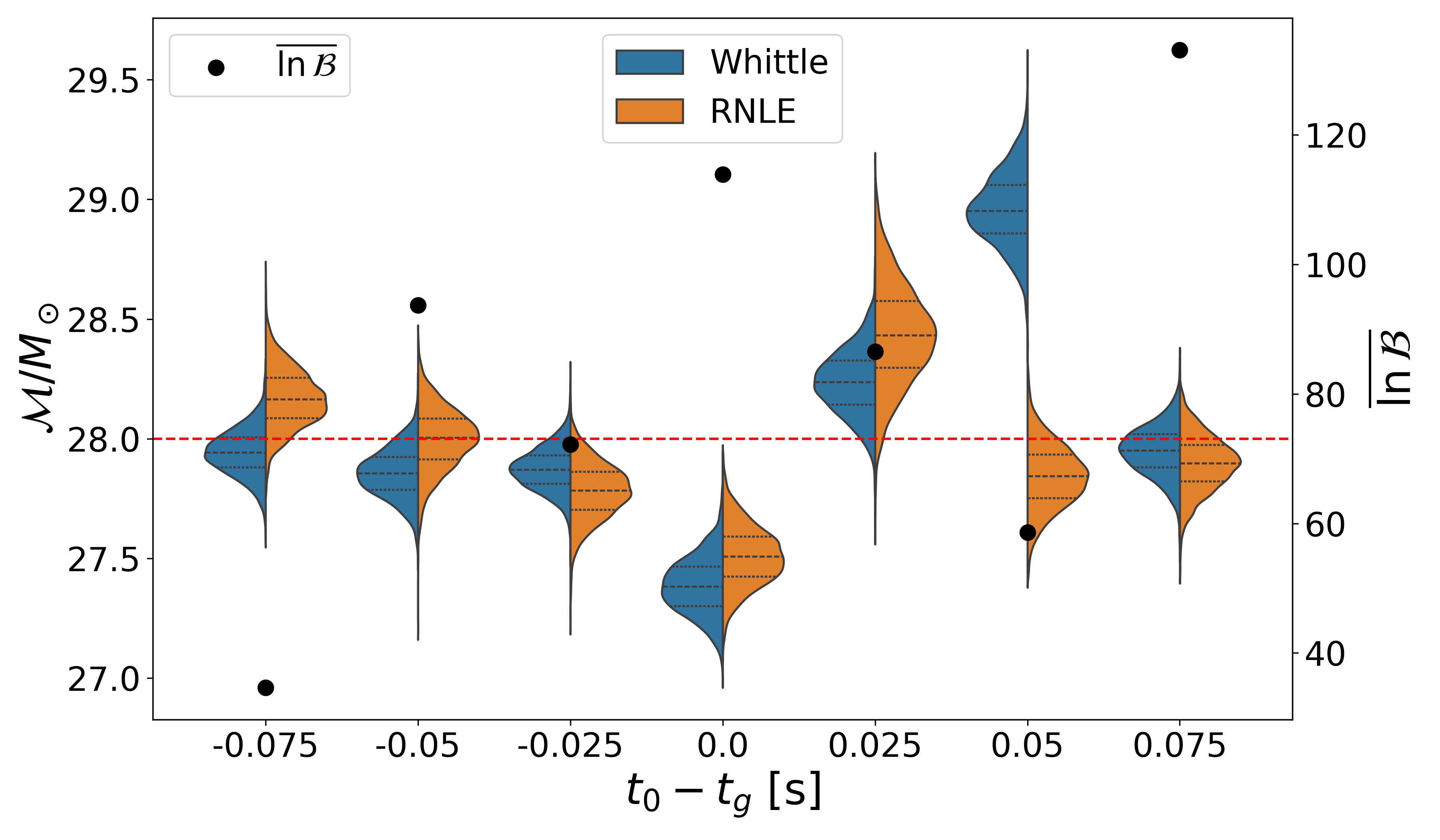}
    \caption*{(C2) Evidence–weighted posterior — GD3.0}
\end{subfigure}

\begin{subfigure}{0.44\linewidth}
    \includegraphics[width=\linewidth]{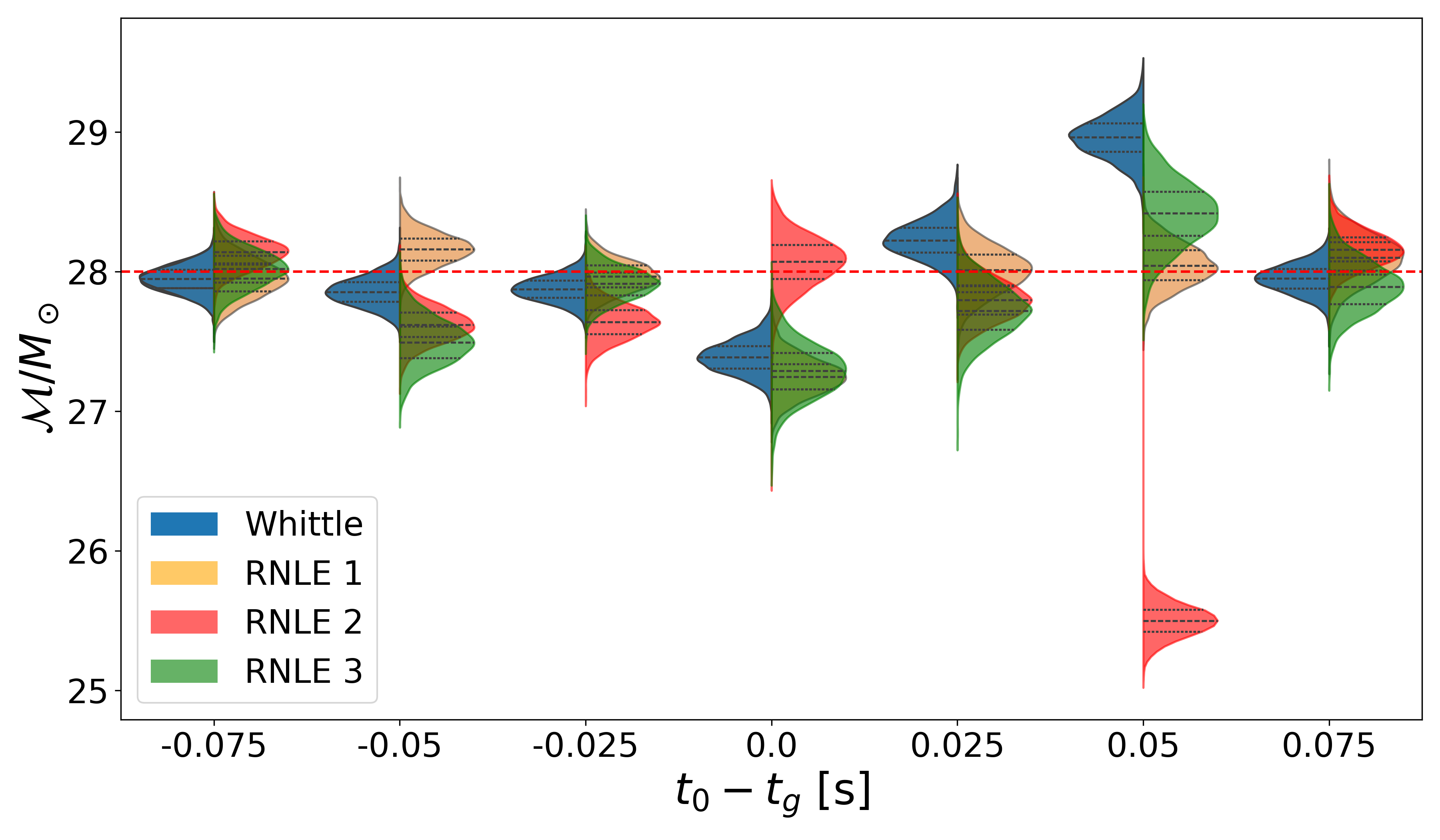}
    \caption*{(D1) RNLE posteriors — GD2.1}
\end{subfigure}
\begin{subfigure}{0.44\linewidth}
    \includegraphics[width=\linewidth]{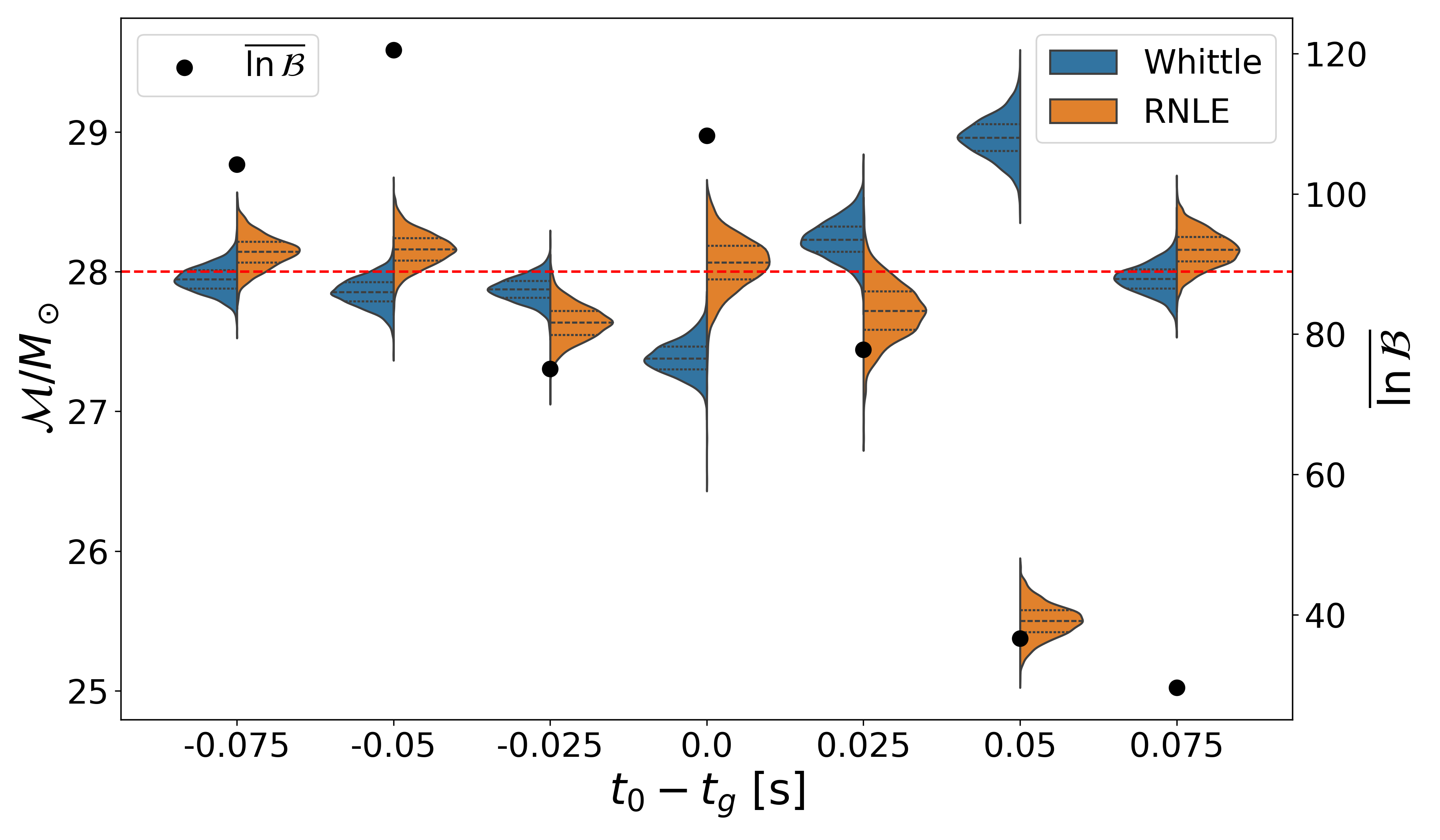}
    \caption*{(D2) Evidence–weighted posterior — GD2.1}
\end{subfigure}

\caption{
Comparison of posterior distributions obtained with the Whittle likelihood and multiple RNLE realizations across four different training datasets. Each row corresponds to a dataset:  
(A) T1, (B) GD2.0, (C) GD3.0, (D) GD2.1. The left columns show three RNLE realizations. The right columns show the evidence–weighted combination of the posteriors. The black dots represent the mean of the log-Bayes factors.
}
\label{fig:blip_glitch_violin_plots_repetitions}
\end{figure*}
where the stability of the conditional density estimator was assessed using ensembles of independently trained likelihoods, we observed that convergence of the learned likelihood corresponds to stabilization of the Bayesian evidence values across realizations (cf.~Figure~\ref{fig:benchmark_gw_violins}). Motivated by this insight, we now examine how variability in the \ac{RNLE} posteriors and evidence manifests when analyzing the blip glitch contaminated data. \\
To this end, we retrain the \ac{RNLE} conditional density estimator multiple times on an identical dataset, while changing the optimization seed of the conditional density estimator and the order of the training data realizations. After performing stochastic sampling, we quantify the variability among the resulting posterior distributions. This procedure isolates the intrinsic stochasticity introduced by the training process itself, analogous to the Gaussian-noise case discussed in Section~\ref{subsec:gw_1D_training_noise}. 
The left panels of Figure~\ref{fig:blip_glitch_violin_plots_repetitions} show violin plots analogous to those in Figure~\ref{fig:blip_glitch_violin_plots}, but now including three independent realizations of the \ac{RNLE} posteriors produced by retraining the likelihood. For this study, the GD1 datasets are excluded as we want to focus on obtaining a single likelihood to analyze all offset realizations. From top to bottom, the panels correspond to likelihoods trained with the T1, GD3.0, GD2.0, and GD2.1 datasets. The GD2.1 dataset extends GD2.0 by additionally incorporating blip glitches from Hanford during \ac{O3a}, selected using the same criteria as GD2.0, with each glitch used 100 times and offsets drawn from $\mathcal{U}[-0.1,0.1]$. \\
Across all datasets, the posteriors obtained from the different \ac{RNLE} realizations exhibit substantial variability. In particular, for GD3.0 and GD2.1, at every offset value, there exists at least one realization whose posterior samples include the injected parameter value at the 98\% credibility level, even for offsets where the Whittle likelihood yields biased estimates. This behavior is different from the results presented in Figure~\ref{fig:benchmark_gw_violins} for the simulated coloured Gaussian noise case, where independently trained likelihoods yielded consistent evidences and highly similar posterior distributions, exhibiting both accuracy and small uncertainties. The high-variability seen in the non-Gaussian scenario indicates that an ensemble of independently trained realizations enables a more reliable assessment of the variability induced by the training procedure itself, and that the mean log-Bayes can be an indicator of the robustness of the trained likelihood. 
Taken together, these findings suggest that the increased inter-realization variability observed here is driven by the complexity of the non-Gaussian noise features present in real data. The long-term objective is that increasing the amount of training data and improving model capacity will reduce this variability, approaching the level of stability observed in the Gaussian-noise case and yielding both more precise and more accurate parameter estimates.\\
To combine the different \ac{RNLE} realizations into a single posterior distribution, each posterior is weighted by its corresponding Bayesian evidence, using

\begin{equation}
    w_i = \frac{\mathcal{Z}_i}{\sum_j \mathcal{Z}_j},
    \label{eq:evidencee_weights}
\end{equation}
where $\mathcal{Z}_i$ is the evidence associated with the $i$-th posterior, and the sum runs over all posteriors being combined. Because all analyses employ identical priors, this procedure is equivalent to model averaging with Bayesian odds, as discussed by {\hypersetup{hidelinks}\citet{Thrane:2018qnx}}. Naively, one might hope that this approach will favor the most accurate posterior at each offset. \\
However, the right panels of Figure~\ref{fig:blip_glitch_violin_plots_repetitions} show the evidence-weighted posteriors corresponding to the left panels. Interestingly, for each offset, the combined posterior is almost entirely dominated by a single realization, with little to no mixing of posteriors. Moreover, the selected realization is not always the most accurate: in rows C and D, clearly biased posteriors receive the highest weight. This behavior, while seemingly counterintuitive, is consistent with the understanding from Section~\ref{subsec:gw_1D_training_noise} that evidence integrates the likelihood over the full parameter space and does not depend solely on the local posterior for a specific data realization. Consequently, a realization yielding an accurate posterior at a particular offset may not correspond to a higher evidence value. \\
The black dots in the right-hand panels further illustrate this behavior, representing the mean log-Bayes factor computed across the three posterior realizations at each offset. These results differ markedly from those shown in Fig.~\ref{fig:benchmark_gw_violins} for the analogous study using simulated colored Gaussian noise. In that case, the mean evidence was observed to approach zero as the number of simulations increased, indicating that effective mixing would occur if the weighting scheme introduced in this section were applied. By contrast, in Fig.~\ref{fig:blip_glitch_violin_plots_repetitions} the mean evidence remains consistently large, exceeding 40 across all offsets. Such values correspond to substantial variability among the inferred posterior distributions.
Overall, the pronounced variance in the evidence estimates across all datasets suggests that the number of simulations employed is insufficient to ensure full convergence of the conditional density estimator. A systematic investigation of convergence properties and alternative training strategies is therefore deferred to future work. In this context, the ensemble weighting approach based on the evidence, introduced here, provides a promising diagnostic tool to assess convergence and to mitigate the stochasticity inherent in the training procedure of the conditional density estimator.

\section{Conclusion \label{sec:conclusion}}

We have presented the development and application of a novel \ac{SBI} algorithm for gravitational-wave parameter estimation based on a modified neural likelihood estimation scheme, termed \ac{RNLE}. The method exploits both the \texttt{sbi} implementation of \ac{NLE} and the additive structure of gravitational-wave signals and detector noise to directly learn the likelihood of the noise contribution. A key feature of \ac{RNLE} is that the conditional density estimator can be trained exclusively on background noise, without requiring gravitational-wave signal injections. This substantially reduces the dimensionality of the parameter space explored during training and enables the modeling of non-Gaussian likelihoods associated with glitch-contaminated data. The \ac{RNLE} algorithm is implemented within the \texttt{Bilby} framework, allowing seamless integration with standard stochastic samplers.

The performance of the algorithm was validated through a sequence of controlled studies using a sine-Gaussian toy model, simulated gravitational-wave detector noise, and real interferometer data. These investigations included injection campaigns benchmarking \ac{RNLE} against the Whittle likelihood, with a systematic exploration of its dependence on the analyzed data duration and on the size and composition of the training dataset. Under approximately Gaussian noise conditions, \ac{RNLE} accurately learns the likelihood function and recovers posterior distributions consistent with those obtained using the Whittle likelihood. When applied to real detector data, this agreement persists whenever the assumptions underlying the Whittle likelihood are satisfied. Crucially, in the presence of loud non-Gaussian noise transients, \ac{RNLE} outperforms the Whittle likelihood, recovering posterior distributions consistent with the injected true values, while the Whittle likelihood leads to biased posteriors that do not include the true parameters. Analyses of loud glitches from the \ac{O3b} run, and a blip glitch from \ac{O2} demonstrate that glitch morphology, duration, and temporal overlap with the gravitational-wave signal play a critical role in determining likelihood accuracy. These results provide clear practical guidance: parameter estimation in the presence of glitches benefits from training datasets that closely resemble the data being analyzed, i.e., same \texttt{GravitySpy} glitch class, same frequency range, and similar relative timing between the signal and noise transients.

Beyond accuracy, we investigated sources of variability intrinsic to the \ac{RNLE} likelihood estimation procedure. By retraining the conditional density estimator multiple times on identical datasets, we quantified the stochastic variability induced by training noise. We observed substantial variability among posterior realizations in the presence of non-Gaussian noise. In contrast, for Gaussian noise, increasing the number of simulations led to consistent likelihood estimates across retraining runs.
We propose an evidence-weighting scheme to combine multiple realizations into a single posterior; however, the Bayesian evidence was found to often strongly favor a single realization, sometimes even a biased one, reflecting its sensitivity to global likelihood structure rather than local accuracy near the true parameters. Comparing multiple likelihood realizations estimated from the same dataset provides a diagnostic for assessing the reliability of the RNLE results. In the Gaussian noise case, the evidence of the RNLE realizations converged when increasing the size of the training dataset. Convergence was not observed in the non-Gaussian noise analysis, hinting at an insufficient training dataset size to capture the relevant non-Gaussian features of the noise.
This behavior underscores the importance of careful training dataset construction and highlights the need for stability-aware inference strategies when deploying neural likelihood estimators.

Overall, \ac{RNLE} provides a promising pathway toward accurate gravitational-wave parameter estimation in the presence of non-Gaussian noise transients, emphasizing likelihood-level modeling rather than post hoc glitch mitigation. This perspective is further motivated by recent work showing that glitch subtraction approaches generally leave residuals that can bias inference, suggesting that joint modeling of the signal and noise contributions is fundamentally more robust \cite{Udall:2025bts}. In this context, \ac{RNLE} offers a flexible framework for incorporating complex noise behavior directly into the inference process.

Future efforts will focus on enabling more systematic exploration of training dataset design and likelihood stability~\cite{Lyu:2025vqk}, with the primary goal of improving the reliability of parameter estimation accuracy. Methodological and computational improvements, such as faster training and likelihood evaluation, will facilitate larger-scale studies of training strategies and noise conditions~\cite{Dirmeier:2024sbi}, allowing the applicability and limitations of \ac{RNLE} to be more thoroughly characterized across a broad range of gravitational-wave analyses.

\section*{Code availability}
The notebooks and scripts to reproduce the results and plots of this study are publicly available on the Zenodo repository~\cite{emma_rnle_2026} and can be used with the \texttt{sbilby} package available on pypi~\cite{emma_sbilby_2026}.

\begin{acknowledgments}
We would like to thank Alessio Spurio Mancini for the useful discussions during the development of the project and Julianna Ostrovska for the work done on NLE in the early stages of the project. The authors are grateful for computational resources provided by the LIGO Laboratory and supported by National Science Foundation Grants PHY-0757058 and PHY-0823459. This work is supported by the Science and Technology Facilities Council (STFC) grant UKRI2488. The authors are also grateful for computational resources provided by Cardiff University and supported by STFC grants ST/I006285/1 and ST/V005618/1.
This research has made use of data or software obtained from the Gravitational Wave Open Science Center (gwosc.org), a service of the LIGO Scientific Collaboration, the Virgo Collaboration, and KAGRA. This material is based upon work supported by NSF's LIGO Laboratory which is a major facility fully funded by the National Science Foundation, as well as the Science and Technology Facilities Council (STFC) of the United Kingdom, the Max-Planck-Society (MPS), and the State of Niedersachsen/Germany for support of the construction of Advanced LIGO and construction and operation of the GEO600 detector. Additional support for Advanced LIGO was provided by the Australian Research Council. Virgo is funded, through the European Gravitational Observatory (EGO), by the French Centre National de Recherche Scientifique (CNRS), the Italian Istituto Nazionale di Fisica Nucleare (INFN) and the Dutch Nikhef, with contributions by institutions from Belgium, Germany, Greece, Hungary, Ireland, Japan, Monaco, Poland, Portugal, Spain. KAGRA is supported by Ministry of Education, Culture, Sports, Science and Technology (MEXT), Japan Society for the Promotion of Science (JSPS) in Japan; National Research Foundation (NRF) and Ministry of Science and ICT (MSIT) in Korea; Academia Sinica (AS) and National Science and Technology Council (NSTC) in Taiwan.

\end{acknowledgments}

\bibliography{reference.bib}
\bibliographystyle{apsrev4-2} 

\appendix
\onecolumngrid
\section{Tables of parameter values and priors}~\label{app:tables}

\begin{center}
\setlength\tabcolsep{7pt}
\begin{table}[htp!]
\begin{tabular}{ccccc}
Parameter & $\alpha$ & $\mathrm{f}$ &  $A$  & $\sigma$ \\
\hline
Prior & $\mathcal{U}[1,3]$ & $\mathcal{U}[1,3]$ & $\mathcal{U}[9,11]$ & $\mathcal{U}[0,2]$ \\
\hline
\end{tabular}
\caption{ Priors on the parameters of the sine-Gaussian toy model.} 
\label{tab:priors_toy_model}
\end{table}
\end{center}

\begin{table*}[htp!]
\caption{\label{tab:simulation} Parameter values for simulated gravitational wave signals corresponding to precessing and aligned-spins \ac{BBH} mergers. Values are given for the chirp mass ($\mathcal{M}$), mass ratio (q), luminosity distance (D$_L$), dimensionless spins (a$_1$ and a$_2$), tilt angles ($\theta_1$ and $\theta_2$), phase angles ($\phi_{12}$ and $\phi_{jl}$), geocentric time ($\mathrm{t_{geocent}}$), angular parameter ($\theta_{jn}$), polarization angle ($\psi$), and overall phase ($\phi$). These values serve as the ground truth for the simulated signals. Right to each parameter value we report the \ac{JS} divergence value between the posterior distribution obtained with the Whittle and \ac{RNLE} likelihood.}
\begin{ruledtabular}
\begin{tabular}{ccccccc}
 Parameter& 10D Precessing BBH & \ac{JS} values & 15D Precessing BBH  & \ac{JS} values & 11D Aligned-spin BBH & \ac{JS} values
 \\   \hline 
 $\mathcal{M}$& 28 & 0.002 & 28 &0.008 & 28 & 0.005 \\
 q& 0.82 & 0.003 & 0.82 & 0.010 & 0.94 & 0.001 \\
 D$_L$& 1400 & 0.001& 1400& 0.003 & 1400 & 0.002 
 \\
  $\chi_1$& - & - & -& - & 0.1 & 0.008\\
 $\chi_2$ & - &- & -& - & 0.1& 0.006\\
 a$_1$&0.32 &0.001 &0.32&0.024 & - & - \\
 a$_2$&0.44 & 0.001 &0.44& 0.027 & - & - \\
 $\theta_1$&0.5& 0.001&0.5 & 0.004 &0.0 & -\\
 $\theta_2$& 1.0&0.001 &1.0 & 0.025 &0.0 & -\\
 $\phi _{12}$&0.0& -&1.7& 0.002 &0.0& - \\
 $\phi _{jl}$& 0.0& -&0.3 & 0.001 &0.0& - \\
 DEC&-1.21 & - & 0.7219&0.007 & 0.7219 & 0.060 \\
 RA&1.375& - &1.375 & 0.012&0.1904 & 0.040 \\
 $\mathrm{t_{geocent}}$ & 1126259642.0 & 0.002 & 1126259642.0 & 0.001 & 1126259642.0 & 0.010 \\
 $\theta _{jn}$&0.4& 0.001&0.4& 0.008 &0.4 & 0.060 \\
 $\psi$&2.66& 0.001&2.66 & 0.003 &2.66 & 0.001\\
  $\phi$&1.3& 0.001&1.3& 0.002 & 1.3 & 0.001 \\
\end{tabular}
\end{ruledtabular}
\end{table*}

\begin{table*}[htp!]
\caption{\label{tab:prior_gw}Prior distributions employed in the parameter estimation of simulated gravitational wave signals arising from precessing and spin-aligned \ac{BBH} mergers. The symbol $\mathcal{U}$ denotes a uniform prior, and $\lambda^2$ represents a power law prior within the specified ranges. The table details the priors for chirp mass ($\mathcal{M}$), mass ratio (q), dimensionless spins ($\chi_1$ and $\chi_2$), tilt angles ($\theta_1$ and $\theta_2$), declination (DEC), luminosity distance (D$_L$), right ascension (RA), the geocentric time, and various angle parameters.}
\begin{ruledtabular}
\begin{tabular}{ccc}
 Parameter& Precessing BBH & Aligned-spin BBH
 \\   \hline \vspace{-0.2cm} \\
 $\mathcal{M}$& $\mathcal{U}$ [24,32] &$\mathcal{U}$ [24,32] \\
 q& $\mathcal{U}$ [0.125,1]& $\mathcal{U}$ [0.125,1] \\
 $\chi_1$& - & $\mathcal{U}$ [0,0.99] \\
 $\chi_2$& - & $\mathcal{U}$ [0,0.99] \\
 $\mathrm{a}_1$& $\mathcal{U}$ [0,0.99] & - \\
 $\mathrm{a}_2$& $\mathcal{U}$ [0,0.99] & - \\
  $\theta_1$& $\sin$& - \\ 
  $\theta_2$& $\sin$& - \\
  $\phi_{12}$& $\mathcal{U}$ [0,2$\pi$]& - \\
  $\phi_{jl}$& $\mathcal{U}$ [0,2$\pi$]& - \\
 D$_L$& $\lambda^2$ [100,10000] & $\lambda^2$ [100,10000]  \\
 DEC& $\cos$& $\cos$ \\
 RA&$\mathcal{U}$ [0,2$\pi$] & $\mathcal{U}$ [0,2$\pi$] \\
 $\mathrm{t_{geocent}}$ & $\mathcal{U}$ [$\mathrm{t_{trigger}}-0.1$,$\mathrm{t_{trigger}}+0.1$] & $\mathcal{U}$ [$\mathrm{t_{trigger}}-0.1$,$\mathrm{t_{trigger}}+0.1$] \\
 $\theta_{jn}$&$\sin$ &$\sin$ \\
 $\psi$& $\mathcal{U}$ [0,$\pi$]&  $\mathcal{U}$ [0,$\pi$] \\
 $\phi$& $\mathcal{U}$ [0,2$\pi$]& $\mathcal{U}$ [0,2$\pi$] 
\end{tabular}
\end{ruledtabular}
\end{table*}
\newpage
\section{PSD windowing comparison}~\label{app:psd_window}

\begin{figure*}[htp!]
    \centering
    \includegraphics[width=\linewidth]{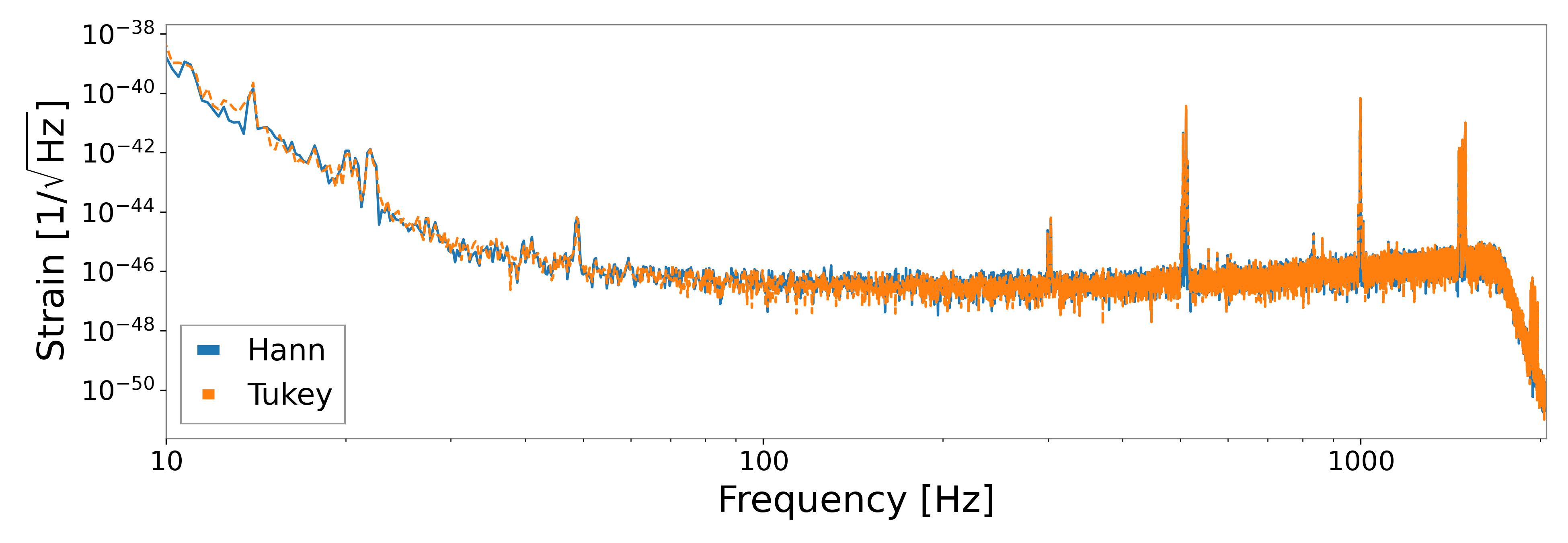}
    \caption{Comparison of a \ac{PSD} computed using the same data and method, using a Hann window, in blue, and a Tukey window, in orange.  }
    \label{fig:app_psd}
\end{figure*}

\end{document}